\documentclass[prd,twocolumn,aps,superscriptaddress,showpacs,preprintnumbers,nofootinbib]{revtex4}
\usepackage{varioref,exscale,latexsym,amsmath,amssymb}
\usepackage{graphicx}
\usepackage{bm}
\usepackage{slashed}
\usepackage{ulem}
\usepackage{epsfig}
\usepackage{color}

\newcommand{\Tr}{\mathrm{Tr}}
\newcommand{\STr}{\mathrm{STr}}
\newcommand{\vev}[1]{\langle #1 \rangle}

\begin{document}
%\setlength{\parskip}{0pt}
%%%%%%%%%%%%%%%%%%%%%%%%%%%%%%%%%%%%%%%%%%%%%%%%%%%%%%%%%%%%%%%%%%%%%%%%%%%%%%%%%%%%%%%%%%%%
%%%%%%%%%%%%%%%%%%%%%%%%%%%%%%%%%%%%%%%%%%%%%%%%%%%%%%%%%%%%%%%%%%%%%%%%%%%%%%%%%%%%%%%%%%%%
\title{The Higgs Mass, Superconnections and the TeV-scale Left-Right Symmetric Model}

\author{Ufuk Aydemir}\email{ufuk.aydemir@physics.uu.se}
\affiliation{Department of Physics and Astronomy, Uppsala University, Uppsala 751 20, Sweden}

\author{Djordje Minic}\email{dminic@vt.edu}
\affiliation{Department of Physics, Virginia Tech, Blacksburg, VA 24061, U.S.A.}

\author{Chen Sun}\email{sunchen@vt.edu}
\affiliation{Department of Physics, Virginia Tech, Blacksburg, VA 24061, U.S.A.}

\author{Tatsu Takeuchi}\email{takeuchi@vt.edu}
\affiliation{Department of Physics, Virginia Tech, Blacksburg, VA 24061, U.S.A.}
\affiliation{Kavli Institute for the Physics and Mathematics of the Universe (WPI),
The University of Tokyo, Kashiwa-shi, Chiba-ken 277-8583, Japan}

\date{\today}

\begin{abstract}
We discuss the physical implications of formulating the Standard Model (SM) in terms of the superconnection formalism involving the superalgebra $su(2/1)$. 
In particular, we discuss the prediction of the Higgs mass according to the formalism and point out
that it is $\sim\!170$~GeV, in clear disagreement with experiment.
To remedy this problem, we extend the formalism to the superalgebra $su(2/2)$,
which extends the SM to the left-right symmetric model (LRSM) and accommodates a $\sim\!126$ GeV Higgs.
Both the SM in the $su(2/1)$ case and the LRSM in the $su(2/2)$ case are argued to emerge at $\sim\!4$ TeV
from an underlying theory in which the spacetime geometry is modified by the addition of a discrete 
extra dimension.  
The formulation of the exterior derivative in this model space suggests a deep connection between the modified geometry, which can be described in the language of non-commutative geometry (NCG), and the spontaneous breaking of the gauge symmetries.
The implication is that spontaneous symmetry breaking could actually be geometric/quantum gravitational in nature.
The non-decoupling phenomenon seen in the Higgs sector can then be reinterpreted in a new light as due to
the mixing of low energy (SM) physics and high energy physics associated with quantum gravity, such as string theory. 
The phenomenology of a TeV scale LRSM is also discussed, and we argue that some exciting
discoveries may await us at the LHC, and other near-future experiments.
\end{abstract}

\pacs{12.90.+b,14.80.Bn}

\preprint{IPMU14-0307}

\maketitle
%%%%%%%%%%%%%%%%%%%%%%%%%%%%%%%%%%%%%%%%%%%%%%%%%%%%%%%%%%%%%%%%%%%%%%%%%%%%%%%%%%%%%%%%%%%%
%%%%%%%%%%%%%%%%%%%%%%%%%%%%%%%%%%%%%%%%%%%%%%%%%%%%%%%%%%%%%%%%%%%%%%%%%%%%%%%%%%%%%%%%%%%%
\section{Introduction and Overview}

\subsection{The Higgs Mass and Beyond the Standard Model Physics}

The Standard Model (SM) of particle physics is a phenomenally successful 
phenomenological theory
whose last building block associated with the Higgs sector has finally been
detected \cite{atlas:2012gk,cms:2012gu}.
The discovery of the Higgs boson with a mass of $\sim$126 GeV
is made even more significant by the lack of discovery (so far) 
of any new particle associated with supersymmetry (SUSY), technicolor, or any other
Beyond the Standard Model (BSM) scenario which have been proposed as 
solutions to the hierarchy problem.
This apparent failure of existing BSM models
may simply be an indication that, while one of the approaches is actually correct,
the corresponding model is not sophisticated enough to accurately represent nature, 
and that further refinements would eventually lead to a successful theory.
Another point of view may be that the Higgs mass and the lack of new particles
are pointing to the limitations of existing BSM paradigms such as SUSY and
technicolor, and that new ideas to guide BSM model building should be 
rigorously searched for.
Since no stone should be left unturned, 
this search should include reassessments of old ideas as well.

%%%%%%%%%%%%%%%%%%%%%%%%%%%%%%%%%%%%%%%%%%%%%%%%%%%%%%%%%%%%%%%%%%%%%%%%%%%%%%%%%%%%%%%%%%%%
\subsection{The Superconnection Formalism}

In a previous paper \cite{Aydemir:2013zua}, 
we have investigated the possibility of reviving 
the superconnection formalism first discussed in 1979 by 
Ne'eman \cite{Neeman:1979wp}, Fairlie \cite{Fairlie:1979zy,Fairlie:1979at}, and
others \cite{Squires:1979je,Taylor:1979sm,Dondi:1979ib}. The original observation 
of Ne'eman was that the $SU(2)_L\times U(1)_Y$ 
gauge fields and the Higgs doublet in the SM could be embedded
into a single $su(2/1)$ superconnection \cite{Quillen:1985,Sternberg:2012} 
with the $SU(2)_L\times U(1)_Y$ gauge fields
constituting the even part of the superconnection and the Higgs doublet $\phi$
constituting the odd part, to wit:
\begin{equation}
\mathcal{J}\;=\; i
  \begin{bmatrix}
    \mathcal{W}-\frac{1}{\sqrt{3}}B\cdot \mathbf{1}_{2\times 2} & \sqrt{2}\,\phi \\
    \sqrt{2}\,\phi^\dagger & -\frac{2}{\sqrt{3}}B \\
  \end{bmatrix}
\,,
\label{Jdef}
\end{equation}
where $\mathcal{W}=W_i\tau_i$.
This embedding predicts $\sin^2\theta_W=1/4$
as well as the Higgs quartic coupling,
the latter leading to a prediction of the Higgs mass \cite{Ne'eman:1986bn,Hwang:1995wk,Roepstorff:1998vh}.
The leptons and quarks could also be embedded into 
irreducible representations of $su(2/1)$ \cite{Corwin:1974fi,Scheunert:1976wi,Bars:1982ep,Bars:1984rb}, 
thereby fixing their electroweak quantum numbers in a natural fashion.
Fairlie started from a 6-dimensional gauge-Higgs unified theory reduced to
four dimensions and arrived at a similar observation.\footnote{%
See also Ref.~\cite{Manton:1979kb}.}
Subsequently,
suggestions have been made to incorporate QCD into the formalism  by extending the superalgebra to
$su(5/1)$ \cite{Taylor:1979pk,Dondi:1979pb,Neeman:1980dg}.

Though the appearance of the $su(2/1)$ superconnection suggested an underlying `internal' $SU(2/1)$ supersymmetry, gauging this supersymmetry to obtain the superconnection proved problematic as discussed in Refs.~\cite{Taylor:1979vx,ThierryMieg:1982nn}.
For instance, the Higgs doublet is a boson whereas an $SU(2/1)$ supersymmetry would demand the off-diagonal scalar components of the superconnection be fermionic with the wrong spin-statistics.
Interpreting these degrees of freedom as ghosts would render the model non-unitary, and though attempts have been made to deal with this problem \cite{Neeman:1980wr,ThierryMieg:1981kb} the issue has never been completely resolved. 
It is also clear that the quarks and leptons placed in $SU(2/1)$ representations cannot all be fermions \cite{Neeman:1980wr,Naka:1990sp}.
The $SU(2/1)$ supersymmetry must also be broken by hand to give the gauge boson kinetic terms the correct signs \cite{Ecclestone:1979fc}.
Due to these, and various other problems, interest in the approach waned.

%%%%%%%%%%%%%%%%%%%%%%%%%%%%%%%%%%%%%%%%%%%%%%%%%%%%%%%%%%%%%%%%%%%%%%%%%%%%%%%%%%%%%%%%%%%%
\subsection{Connection to Noncommutative Geometry}

It was subsequently recognized, however, that the appearance of a superconnection does
not necessarily require the involvement of the familiar boson$\leftrightarrow$fermion supersymmetry.
This development follows the 1990 paper of Connes and Lott \cite{Connes:1990qp}
who constructed a new description of the SM using the framework of
noncommutative geometry (NCG)
in which the Higgs doublet appears as part of the Yang-Mills field (i.e. connection)
in a spacetime with a modified geometry.
The full Yang-Mills field in this approach was described by a superconnection,
the off-diagonal elements of which were required to be bosonic.

The NGC-superconnection approach to the SM was studied by many authors and a vast
literature on the subject exists, \textit{e.g.}
Refs.~\cite{Coquereaux:1990ev,Haussling:1991ns,Coquereaux:1991ds,DuboisViolette:1988ir,DuboisViolette:1988vq,DuboisViolette:1988ps,DuboisViolette:1989at,Madore:1999bi,Madore:2000en,Ne'eman:1990nr,Hwang:1995xc,Wulkenhaar:1996at,Schucker:1993xj,Iochum:1996ph,Carminati:1996em,Kastler:2000px,Morita:1993kp,Morita:1993zv,Morita:1993dn,Okumura:1994ck,Sogami:1995hi,Naka:1994fd,Alvarez:1995wb,Varilly:1993ib,Martin:1996wh,Varilly:1997qg}
to give just a representative list.\footnote{%
See Ref.~\cite{Scheck:2002rs} for a collection of lectures from 1999 by various authors.
}
Though these works differ from each other in detail, the
basic premise is the same.
The models are all of the Kaluza-Klein type in which the extra dimension is discrete and consists
of only two points.
In other words, the model spacetime consists of two $3+1$ dimensional `branes.'
In such a setup, the connection must be generalized to connect not
just points within each brane, but also to bridge the gap between the two.
If the left-handed fermions live on one brane and the right-handed fermions on the other,
then the connections within each brane, \textit{i.e.} the even part of the superconnection, 
will involve the usual SM gauge fields which couple to fermions of that chirality. 
In contrast, the connection across the gap, \textit{i.e.} the odd part of the
superconnection, connects fermions of opposite chirality and can be identified with the Higgs
doublet.

In this approach,
both the even and odd parts of the superconnection are bosonic, 
the $\mathbb{Z}_2$-grading of the superalgebra
resulting not from fermionic degrees of freedom but from the existence of the two
`branes' (on which the chirality $\gamma_5$ provides the $\mathbb{Z}_2$ grading operator), 
and the definition of the generalized exterior derivative $d$ in the discrete
direction.\footnote{%
Since $d^2=0$, the exterior derivative is intrinsically `fermionic.'}
That is, the superconnection emerges from the `geometry' of the discrete extra dimension.

In algebraic geometry, the geometric properties of a manifold $M$ are studied via
the algebraic properties of the commutative algebra of smooth functions $C^\infty(M)$ defined on it.
If this algebra is allowed to be noncommutative in general, one has a NCG \cite{Connes:1990qp,Chamseddine:1991qh,Connes:1994yd,Chamseddine:1996zu,Connes:2006qv,Chamseddine:2007hz,Chamseddine:2010ud,Chamseddine:2012sw,Chamseddine:2013rta}.
In the discrete extra dimension case, one usually starts with the algebra 
$\mathcal{A} = C^\infty(M)\otimes( \mathbb{C}\oplus \mathbb{H})$,
and the fermions on the branes are required to lie in representations of this algebra.
Gauge transformations correspond to the unitary inner automorphisms of the algebra,\footnote{%
The unitary condition renders the resulting gauge theory anomaly free \cite{Alvarez:1995wb}.
}
which in this case is $U(1)\times SU(2)$.
The exterior derivative $d$ is defined via
\begin{eqnarray}
\label{extder}
d\alpha\;=\; [D,\alpha]_s\;,\qquad
\alpha\in\mathcal{A}\;,
\end{eqnarray}
where $[\cdot,\cdot]_s$ is the super-commutator, and the operator $D$
includes the usual exterior derivative acting on the $C^\infty(M)$ part of the
algebra, as well as a `matrix derivative' \cite{Coquereaux:1990ev,Haussling:1991ns,Hwang:1995xc}
which acts on the $\mathbb{C}\oplus\mathbb{H}$
part.
QCD can be included in the model by extending the algebra to 
$\mathcal{A} = C^\infty(M)\otimes( \mathbb{C}\oplus \mathbb{H}\oplus M_3(\mathbb{C}))$,
where $M_3(\mathbb{C})$ is the algebra of $3\times 3$ matrices with elements in $\mathbb{C}$.
Indeed, Connes et al. have shown that the entire SM can be rewritten in the NCG language \cite{Connes:1990qp,Chamseddine:2007hz}.

The extra-discrete-dimension interpretation of the
superconnection model also solves the problem that the
prediction $\sin^2\theta_W=1/4$ is not stable under renormalization group
running and can only be imposed at one scale \cite{Alvarez:1993vk,Alvarez:1994jr}.
That scale can be interpreted as the scale at which 
the SM with $\sin^2\theta_W=1/4$ emerges from the
underlying discrete extra dimension model.
The same scale should also characterize the separation of the two `branes' in the
discrete direction.
Given the current experimental knowledge of the SM, this
scale turns out to be $\sim 4$~TeV \cite{Aydemir:2013zua},
suggesting a phenomenology that could potentially be explored at the LHC,
as well as the existence of a new fundamental scale of nature at those energies.
We will have more to say about this later.

These developments notwithstanding, 
a definitive recipe for constructing a NCG
Kaluza-Klein model for a given algebra still seems to be in the works.
Different authors use different definitions of the
exterior derivative $d$, which, naturally, lead to different Higgs sectors and different predictions.
In the Spectral SM of Connes et al. \cite{Chamseddine:1996zu,Connes:2006qv,Chamseddine:2007hz,Chamseddine:2010ud,Chamseddine:2012sw,Chamseddine:2013rta}, for instance, the prediction for the
$U(1)\times SU(2)\times SU(3)$ gauge couplings are of the $SO(10)$ GUT type,
pushing up the scale of emergence to the GUT scale.
The Spectral SM is not particularly predictive either:
the fermionic masses and mixings must all be put in by hand into the operator $D$.
Thus, the NCG-superconnection approach still has much to be desired and
further development is called for.

Despite the still incomplete nature of the NCG-superconnection approach,
one can still make predictions and assessments based on the SM which we assume to emerge from
it at the emergence scale.
We have already commented on the fact that 
the prediction $\sin^2\theta_W=1/4$ leads to an emergence scale of
$\sim$4~TeV.
The $su(2/1)$ superconnection also predicts the Higgs quartic coupling at that scale, 
from which in turn one can predict the Higgs boson mass to be $\sim$170 GeV.
As discussed in Ref.~\cite{Aydemir:2013zua},
lowering this prediction down to $\sim$126~GeV requires
the introduction of extra scalar degrees of freedom
which modify the renormalization group equations (RGE) of the Higgs
couplings. 
Those degrees of freedom would be available, for instance, if the
$su(2/1)$ superconnection were extended to $su(2/2)$.
The extra-discrete-dimensional $su(2/2)$ model shares the same
prediction for $\sin^2\theta_W$ as the $su(2/1)$ version,
and therefore the same scale ($\sim$4 TeV) at which 
an effective $SU(2)_L\times SU(2)_R\times U(1)_{B-L}$ gauge theory can be expected to emerge.
Thus, explaining the Higgs mass within the NCG-superconnection
formalism seems to demand an extension of the SM gauge group.

Curiously,  
Connes et al.'s Spectral SM with a GUT emergence scale 
also predicts the Higgs mass to be $\sim$170 GeV.
Lowering this to $\sim$126 GeV 
requires the introduction of extra scalar degrees of freedom
as discussed above \cite{Chamseddine:2012sw,Chamseddine:2013rta}.
See also Refs.~\cite{Devastato:2013oqa,Devastato:2014xga}.
Here too, the Higgs mass seems to suggest that
the SM gauge group needs to be extended to 
$SU(2)_L\times SU(2)_R\times U(1)_{B-L}$, or including the QCD sector, to
$SU(2)_L\times SU(2)_R\times SU(4)$.

Thus, the NCG-superconnection formalism 
already requires the extension of the SM gauge group to
that of the left-right symmetric model (LRSM), or that of Pati-Salam \cite{Pati:1974yy}.\footnote{%
Coincidentally, the analysis of possible string compactifications
by Dienes \cite{dienes} also finds frequent
occurrence of the Pati-Salam group.}
In this paper, we will take a look at some of the phenomenological consequences
of a NCG-superconnection motivated LRSM,
in anticipation of the start of the upgraded LHC program in 2015,
and various experiments at the intensity frontier which will be able to constrain 
new physics via rare decay processes.

This paper is organized as follows.
In section~\ref{su21}, we first review the $su(2/1)$ superconnection approach to the SM. 
We follow the bottom-up approach of Ne'eman et al. \cite{Neeman:1979wp,Ne'eman:1990nr,Hwang:1995wk}, Coquereaux et al. \cite{Coquereaux:1990ev},
and Haussling et al. \cite{Haussling:1991ns}, in which we start with
the superconnection and build up the theory around it.
This review goes into some pedagogical detail,
and also shows where the Higgs mass prediction of $\sim$170~GeV comes from.
In section~\ref{su22}, we extend the formalism developed in section~\ref{su21} to
the $su(2/2)$ superconnection into which the LRSM gauge group
$SU(2)_L\times SU(2)_R\times U(1)_{B-L}$ is embedded.
Again, the model is reviewed in some detail to clearly present the
assumptions that go into its construction, and the resulting predictions including that of the Higgs mass.
Section~\ref{Fermions} discusses how fermion masses and mixings can be
incorporated into the model. 
Section~\ref{PHENO} 
discusses whether the new particles predicted by the $su(2/2)$ superconnection motivated LRSM 
are accessible at the LHC and other experiments.
Section~\ref{Summary} concludes with a summary of what was discovered, 
review of remaining questions, and some speculation on what all this could mean.
The review of the Spectral SM of Connes et al. is relegated to a
subsequent paper \cite{AMST2}.

%%%%%%%%%%%%%%%%%%%%%%%%%%%%%%%%%%%%%%%%%%%%%%%%%%%%%%%%%%%%%%%%%%%%%%%%%%%%%%%%%%%%%%%%%%%%
%%%%%%%%%%%%%%%%%%%%%%%%%%%%%%%%%%%%%%%%%%%%%%%%%%%%%%%%%%%%%%%%%%%%%%%%%%%%%%%%%%%%%%%%%%%%
\section{The $\bm{su(2/1)}$ superconnection formalism of the Standard Model}\label{su21}

We begin by reviewing the $su(2/1)$ superconnection formalism of Ne'eman et al. \cite{Neeman:1979wp,Ne'eman:1990nr,Hwang:1995wk}, supplemented by the
matrix derivative of Coquereaux et al. \cite{Coquereaux:1990ev}
and Haussling et al. \cite{Haussling:1991ns}, and some of our own observations.
This will be done in some detail to dispel many misconceptions that exist
concerning the formalism, while at the same time expose its weaknesses.
For a pedagogical introduction to superconnections, we point the reader to
Ref.~\cite{Sternberg:2012} by Sternberg.

%%%%%%%%%%%%%%%%%%%%%%%%%%%%%%%%%%%%%%%%%%%%%%%%%%%%%%%%%%%%%%%%%%%%%%%%%%%%%%%%%%%%%%%%%%%%%
\subsection{Superalgebras}

Let $K$ be a field such as $\mathbb{R}$ or $\mathbb{C}$.
A superalgebra $A$ over $K$ is a vector space over $K$ with
a direct sum decomposition
\begin{equation}
A \;=\; A_0 \oplus A_1\;,
\end{equation}
together with a bilinear multiplication $A\times A\rightarrow A$ such that
\begin{equation}
A_i \cdot A_j \;\subseteq\; A_{(i+j)\,\mathrm{mod}\,2}
\;.
\end{equation}
The subscripts $0$ and $1$ of $A_0$ and $A_1$ are known as the `grading' 
of each space and its elements.
The above relation indicates that when two elements of $A$ are multiplied
together, the gradings of the elements add as elements of the group $\mathbb{Z}_2$.
Consequently, superalgebras are also known as $\mathbb{Z}_2$-graded algebras.
If we call the elements of $A_0$ and $A_1$ respectively `even' and `odd,'
then $A_i\cdot A_j \subseteq A_{(i+j)\,\mathrm{mod}\,2}$ means that
\begin{eqnarray}
\mathrm{even}\cdot\mathrm{even} & = & \mathrm{even}\;,\cr
\mathrm{even}\cdot\mathrm{odd} & = & \mathrm{odd}\;,\cr
\mathrm{odd}\cdot\mathrm{even} & = & \mathrm{odd}\;,\cr
\mathrm{odd}\cdot\mathrm{odd} & = & \mathrm{even}\;.
\label{EvenOdd}
\end{eqnarray}
Some texts use the symbols $+$ and $-$ instead of $0$ and $1$ for the $\mathbb{Z}_2$
gradings 
\begin{equation}
A \;=\; A_+ \oplus A_- \;,
\end{equation}
so that
\begin{equation}
A_i\cdot A_j \;\subseteq\; A_{ij}\;,
\label{SAproductpm}
\end{equation}
in which case Eq.~(\ref{EvenOdd}) can also be written
\begin{eqnarray}
\mathrm{+}\cdot\mathrm{+} & = & \mathrm{+}\;,\cr
\mathrm{+}\cdot\mathrm{-} & = & \mathrm{-}\;,\cr
\mathrm{-}\cdot\mathrm{+} & = & \mathrm{-}\;,\cr
\mathrm{-}\cdot\mathrm{-} & = & \mathrm{+}\;,
\label{PlusMinus}
\end{eqnarray}
and the analogy with regular multiplication is manifest.
In this text, however, we will stick to $0$ and $1$
for notational convenience.

%%%%%%%%%%%%%%%%%%%%%%%%%%%%%%%%%%%%%%%%%%%%%%%%%%%%%%%%%%%%%%%%%%%%%%%%%%%%%%%%%%%%%%%%%%%%%
\subsection{The Commutative Superalgebra of Differential Forms}

Consider the vector space of differential forms $\Omega(M)$ on the manifold $M$,
which decomposes as:
\begin{equation}
\Omega(M) \;=\; \Omega_0(M)\oplus\Omega_1(M)\;, 
\end{equation}
where
\begin{eqnarray}
\Omega_0(M) & = & \bigoplus_{n=\mathrm{even}}\Omega^n(M) 
%\,=\, \Omega^0(M) \oplus \Omega^2(M) \oplus \Omega^4(M) \oplus \cdots 
\;,\cr
\Omega_1(M) & = & \bigoplus_{n=\mathrm{odd}}\Omega^n(M)  
%\,=\, \Omega^1(M) \oplus \Omega^3(M) \oplus \Omega^5(M) \oplus \cdots 
\;.
%\cr & &
\end{eqnarray}
Here, $\Omega^n(M)$ is the vector space of $n$-forms on $M$.
$\Omega_0(M)$ is the vector space of even-order differential forms, while
$\Omega_1(M)$ is the vector space of odd-order differential forms.
$\Omega(M)=\Omega_0(M)\oplus \Omega_1(M)$ is a superalgebra under the wedge product $\wedge$ since, clearly,
\begin{equation}
\Omega^i(M) \wedge \Omega^j(M) \;\subseteq\; \Omega^{i+j}(M)
\end{equation}
implies
\begin{equation}
\Omega_i(M) \wedge \Omega_j(M) \;\subseteq\; \Omega_{(i+j)\,\mathrm{mod}\,2}(M)
\;.
\end{equation}
Furthermore, for any $a,b\in\Omega(M)$ with definite gradings $|a|$ and $|b|$, we have
\begin{equation}
a\wedge b \;=\; (-1)^{|a||b|}\,b\wedge a\;,
\end{equation}
that is, 
\begin{equation}
a\wedge b-(-1)^{|a||b|}\,b\wedge a\;=\;0\;.
\label{superwedgecommutator}
\end{equation} 
For generic superalgebras, when
\begin{equation}
a\cdot b -(-1)^{|a||b|}\,b\cdot a \;=\; 0\;,
\label{commutativeSA}
\end{equation}
the superalgebra is said to be commutative.
Thus, $\Omega(M)$ is a commutative superalgebra.

%%%%%%%%%%%%%%%%%%%%%%%%%%%%%%%%%%%%%%%%%%%%%%%%%%%%%%%%%%%%%%%%%%%%%%%%%%%%%%%%%%%%%%%%%%%%
\subsection{The Lie Superalgebra $\bm{su(2/1)}$}

%\subsubsection{Generics}

Formally, a Lie superalgebra is a superalgebra whose
product $a\cdot b$ satisfies the relations
\begin{eqnarray}
a\cdot b & = & -(-1)^{|a||b|}\,b\cdot a\;,\cr
a\cdot (b\cdot c) & = &
(a\cdot b)\cdot c + (-1)^{|a||b|}\,b\cdot (a\cdot c) \;.\cr
& &
\label{SuperLie}
\end{eqnarray}
Elements of the real Lie superalgebra $su(N/M)$ are represented by
$(N+M)\times(N+M)$ supertraceless Hermitian matrices of the form \cite{Bars:1982ep,Bars:1984rb}
\begin{eqnarray}\label{supermatrix}
\mathcal{H}
& = &
  \begin{bmatrix}
    H^{(N)}          & \theta  \\
    \theta^{\dagger} & H^{(M)} \\
  \end{bmatrix}
\;=\;
 \underbrace{
  \begin{bmatrix}
    H^{(N)}          & 0       \\
    0                & H^{(M)} \\
  \end{bmatrix}
 }_{\displaystyle \mathcal{H}_0}
\;+\;
 \underbrace{
  \begin{bmatrix}
    0^{\phantom{\dagger}} & \theta  \\
    \theta^{\dagger}      & 0       \\
  \end{bmatrix}
 }_{\displaystyle \mathcal{H}_1}
\;,
\cr
& &
\end{eqnarray}
where $H^{(N)}$ and $H^{(M)}$ are, respectively, $N\times N$ and $M\times M$ Hermitian matrices
and constitute the even (grading 0) part of the superalgebra, 
while $\theta$ ($\theta^\dagger$) is an $N\times M$ ($M\times N$) matrix and constitutes
the odd (grading 1) part.
The `supertrace' of $\mathcal{H}$ is defined as
\begin{eqnarray}
\STr\,\mathcal{H}\;=\;\Tr\,H^{(N)}-\Tr\,H^{(M)}\;,
\end{eqnarray}
and the elements of $su(N/M)$ all have vanishing supertrace.
Note that the traceless parts of $H^{(N)}$ and $H^{(M)}$ 
respectively generate $SU(N)$ and $SU(M)$, while the non-vanishing trace part generates $U(1)$. 
Therefore, the even part of the $su(N/M)$ superalgebra generates $SU(N)\times SU(M)\times U(1)$
upon exponentiation.

The product of $X,Y\in su(N/M)$
in the matrix representation is given by 
\begin{eqnarray}
\dfrac{1}{i}[\,X,\,Y\,] & & \mbox{if $|X||Y|=0$}\;, \cr
\{\,X,\,Y\,\}           & & \mbox{if $|X||Y|=1$}\;,
\end{eqnarray}
where $[*,*]$ and $\{*,*\}$ respectively denote the 
standard commutator and anti-commutator between two matrices.
Note that the factor of $i^{-1}$ for the $|X||Y|=0$ case
is necessary to render the product Hermitian.
Ref.~\cite{Haussling:1991ns} denotes the two cases collectively as
\begin{equation}
\dfrac{1}{i}[\,X,\,Y\,]_s
\label{LieSAproduct}
\end{equation}
where $[X,Y]_s$ is the `supercommutator.' 
Given the even-odd decompositions $X=X_0+X_1$ and $Y=Y_0+Y_1$,
it is defined as \cite{Haussling:1991ns}
\begin{eqnarray}
\lefteqn{[\,X,\,Y\,]_s} \cr
& = &
[\,X_0+X_1,\,Y_0+Y_1\,]_s
\cr
& = &
  [\,X_0,\,Y_0\,] 
+ [\,X_0,\,Y_1\,]
+ [\,X_1,\,Y_0\,]
+ i\{\,X_1,\,Y_1\,\}
\;.
\cr
& & 
\label{supercommutator1}
\end{eqnarray}
In the literature, the supercommutator is also defined as
\begin{equation}
[\,X,\,Y\,]_s \;=\; XY -(-1)^{|X||Y|}\,YX\;,
\label{supercommutator2A}
\end{equation}
which when written out explicitly reads
\begin{eqnarray}
\lefteqn{[\,X,\,Y\,]_s} \cr
& = &
[\,X_0+X_1,\,Y_0+Y_1\,]_s
\cr
& = &
  [\,X_0,\,Y_0\,] 
+ [\,X_0,\,Y_1\,]
+ [\,X_1,\,Y_0\,]
+ \{\,X_1,\,Y_1\,\}
\;.
\cr
& & 
\label{supercommutator2B}
\end{eqnarray}
Though we will be using the first definition to express multiplication in Lie superalgebras,
we will also have a use for the latter definition later in the text, so we
request the reader to keep in mind that the $i$ in front of the
anti-commutator terms may or may not be there depending on the context.
It is straightforward to check that both definitions of the supercommutator
satisfy Eq.~(\ref{SuperLie}), that is:
\begin{eqnarray}
[X,Y]_s & = & -(-1)^{|X||Y|}\,[Y,X]_s\;, \cr
[X,[Y,Z]_s]_s & = &
[[X,Y]_s,Z]_s + (-1)^{|X||Y|}\,[Y,[X,Z]_s]_s \;. \cr
& &
\label{SuperLieConditions}
\end{eqnarray}
%

%%%%%%%%%%%%%%%%%%%%%%%%%%%%%%%%%%%%%%%%%%%%%%%%%%%%%%%%%%%%%%%%%%%%%%%%%%%%%%%%%%%%%%%%%%%%
%\subsubsection{$\bm{su(2/1)}$}

Let us look at a specific case. 
The Lie superalgebra $su(2/1)$ is the algebra of $3\times3$ 
supertraceless Hermitian matrices,
the basis for which can be chosen as
\begin{eqnarray}\label{generators1}
\lambda_1^s 
\!& = &\!
  \begin{bmatrix}
    0 & 1 & 0 \\
    1 & 0 & 0 \\
    0 & 0 & 0 \\
  \end{bmatrix}
,\;
\lambda_2^s
\,=
  \begin{bmatrix}
    0 & -i & 0 \\
    i &  0 & 0 \\
    0 &  0 & 0 \\
  \end{bmatrix}
,\;
\lambda_3^s
\,=
  \begin{bmatrix}
    1 &  0 & 0 \\
    0 & -1 & 0 \\
    0 &  0 & 0 \\
  \end{bmatrix}
,\nonumber\\
\lambda_4^s
\!& = &\!
  \begin{bmatrix}
    0 & 0 & 1 \\
    0 & 0 & 0 \\
    1 & 0 & 0 \\
  \end{bmatrix}
,\;
\lambda_5^s
\,=
  \begin{bmatrix}
    0 & 0 & -i \\
    0 & 0 &  0 \\
    i & 0 &  0 \\
  \end{bmatrix}
,\;
\lambda_6^s
\,=
  \begin{bmatrix}
    0 & 0 & 0 \\
    0 & 0 & 1 \\
    0 & 1 & 0 \\
  \end{bmatrix}
,\nonumber\\
\lambda_7^s
\!& = &\!
  \begin{bmatrix}
    0 & 0 &  0 \\
    0 & 0 & -i \\
    0 & i &  0 \\
  \end{bmatrix}
,\;
\lambda_8^s
\,= \frac{1}{\sqrt{3}}
  \begin{bmatrix}
    -1 &  0 &  0 \\
     0 & -1 &  0 \\
     0 &  0 & -2 \\
  \end{bmatrix}
  \;.
\end{eqnarray}
These are the usual $su(3)$ Gell-mann matrices except for the eighth ($\lambda_8^s$) due to the requirement of vanishing supertrace.
Of these, $\lambda_1^s,\lambda_2^s,\lambda_3^s,\lambda_8^s$ span the even part of the superalgebra while $\lambda_4^s,\lambda_5^s,\lambda_6^s,\lambda_7^s$ span the odd part. 
They close under commutation and anti-commutation relations as \cite{Neeman:1979wp}
\begin{eqnarray}\label{gradedrelations}
\dfrac{1}{i}
\left[\lambda_i^s,\,\lambda_j^s\right] & = & 2\,f_{ijk}^{\phantom{s}}\lambda_k^s\,,
%\qquad i,j=1,2,3,
\nonumber\\
\left[\lambda_i^s,\,\lambda_8^s\right] & = & 0\,,\nonumber\\  
\dfrac{1}{i}
\left[\lambda_i^s,\,\lambda_m^s\right] & = & 2\,f_{iml}^{\phantom{s}}\lambda_l^s\,,
%\qquad m,n=4,5,6,7,
\nonumber\\
\dfrac{1}{i}
\left[\lambda_8^s,\,\lambda_m^s\right] & = &\frac{2}{3}f_{8ml}^{\phantom{s}}\lambda_l^s\,,\nonumber\\
\left\{\lambda_m^s,\,\lambda_n^s\right\} & = & 2\,d_{mnk}^{\phantom{s}}\lambda_k^s-\sqrt{3}\,\delta_{mn}\lambda_8^s\,,
\end{eqnarray}
where $i,j,k$ denote the even indices $1,2,3,8$ and $m,n,l$ denote the odd indices $4,5,6,7$. 
The $f$'s and the $d$'s are the same as the $su(3)$ structure constants defined in Ref.~\cite{Itzykson:1980rh}.
Note that the odd matrices close into the even ones under anti-commutation (instead of commutation), which is the main difference from the $su(3)$ case.
Note also that we have chosen to normalize the above matrices, including $\lambda_8^s$, 
in the usual way
\begin{equation}
\Tr(\lambda^s_a\lambda^s_b)\;=\;2\delta_{ab}\;,
\label{superlambda-normalization}
\end{equation}
and not via the supertrace.

%%%%%%%%%%%%%%%%%%%%%%%%%%%%%%%%%%%%%%%%%%%%%%%%%%%%%%%%%%%%%%%%%%%%%%%%%%%%%%%%%%%%%%%%%%%%%
%%%%%%%%%%%%%%%%%%%%%%%%%%%%%%%%%%%%%%%%%%%%%%%%%%%%%%%%%%%%%%%%%%%%%%%%%%%%%%%%%%%%%%%%%%%%%
\subsection{Tensor Product of Superalgebras}

If $A$ and $B$ are superalgebras, then the tensor product $A\otimes B$ is 
also a superalgebra under the multiplication
\begin{equation}
(a\otimes b)\cdot(a'\otimes b') \;\equiv\;
(-1)^{|b||a'|}(a\cdot a')\otimes(b\cdot b')\;,
\label{TensoredSAproduct}
\end{equation}
where $a,a'\in A$ and $b,b'\in B$.
In constructing this product, elements of $A$ and $B$ are assumed
to (super)commute, cf. Eq.~(\ref{commutativeSA}).
The grading of the element $a\otimes b\in A\otimes B$ 
is given by
\begin{equation}
|a\otimes b| \;=\; |a|+|b| \mod 2\;,
\end{equation}
and the even-odd decomposition $A\otimes B=(A\otimes B)_0 \oplus (A\otimes B)_1$ is 
\begin{eqnarray}
(A\otimes B)_0 & = & (A_0\otimes B_0) \oplus (A_1\otimes B_1)\;,\cr
(A\otimes B)_1 & = & (A_0\otimes B_1) \oplus (A_1\otimes B_0)\;,
\end{eqnarray}
where $A=A_0\oplus A_1$ and $B=B_0\oplus B_1$.

In particular, 
the tensor product of a commutative superalgebra of differential forms 
$\Omega(M)$ and a Lie superalgebra $L$ is again a Lie superalgebra with product
\begin{equation}
[\,a\otimes X,\,b\otimes Y\,]_S \;=\;
(-1)^{|X||b|}(a\wedge b)\otimes [\,X,\,Y\,]_s
\;,
\label{OmegaLproduct}
\end{equation}
where $a,b\in \Omega(M)$ and $X,Y\in L$.
The tensor product $\Omega(M)\otimes L$ is the space of
$L$ valued differential forms.

%%%%%%%%%%%%%%%%%%%%%%%%%%%%%%%%%%%%%%%%%%%%%%%%%%%%%%%%%%%%%%%%%%%%%%%%%%%%%%%%%%%%%%%%%%%%%
%%%%%%%%%%%%%%%%%%%%%%%%%%%%%%%%%%%%%%%%%%%%%%%%%%%%%%%%%%%%%%%%%%%%%%%%%%%%%%%%%%%%%%%%%%%%%
\subsection{Superconnection}

Just as the gauge connection in QCD is given by
$G=i\sum_{a=1}^{8} G_a \lambda_a$, where $G_a=G_a^\mu dx_\mu$ are one-forms
corresponding to the gluon fields, 
we construct the $su(2/1)$ superconnection $\mathcal{J}$ using the $\lambda^s$ matrices as\footnote{%
We take the elements of $su(2/1)$ to be Hermitian, but the superconnection $\mathcal{J}$
and the supercurvature $\mathcal{F}$, to be defined in section~\ref{supercurvature21}, are taken to
be anti-Hermitian.
}
\begin{eqnarray}
\mathcal{J} 
& = & i\sum_{a=1}^{8} J_a^{\phantom{s}} \lambda^s_a
\cr
& = & i\sum_{i=1,2,3,8} J_i^{\phantom{s}} \lambda^s_i 
    + i\sum_{m=4,5,6,7} J_m^{\phantom{s}} \lambda^s_m 
\;.
\end{eqnarray}
For the terms multiplying the even $su(2/1)$ matrices,
we make the identifications $J_i=W_i$ $(i=1,2,3)$ and $J_8=B$, where 
$W_i = W_{i}^{\mu} dx_\mu$ and $B=B^\mu dx_\mu$ are respectively the one-form fields corresponding to the $SU(2)_L$ and $U(1)_Y$ gauge fields. 
The terms multiplying the odd $su(2/1)$ matrices are 
identified with zero-form fields corresponding to the Higgs doublet:
\begin{eqnarray}
J_4\mp iJ_5 & = & \sqrt{2}\,\phi^\pm\;, \\
J_6  - iJ_7 & = & \sqrt{2}\,\phi^0\;,   \\
J_6  + iJ_7 & = & \sqrt{2}\,\phi^{0\ast}\;.
\end{eqnarray}
Then, the superconnection can be written as
\begin{equation}
\mathcal{J} \;=\; i
\begin{bmatrix}
   \mathcal{W}-\frac{1}{\sqrt{3}}B\cdot \mathbf{1}_{2\times 2} & \sqrt{2}\phi \\
    \sqrt{2}\phi^\dagger & -\frac{2}{\sqrt{3}}B 
\end{bmatrix}
\;,
\end{equation}
where, $\mathcal{W}=W_i\,\tau_i$ (where $\tau_i$ are the Pauli matrices), 
and 
\begin{equation}
\phi \;=\; \left[
\begin{array}{c}                                                                                                                            \phi^+  \\ \phi^0                                                                                                                                                                                                                                \end{array}
\right]\;.
\end{equation}

Note that the superconnection $\mathcal{J}$ is an odd element of
$\Omega(M)\otimes su(2/1)$, where $M$ is the $(3+1)$ dimensional spacetime manifold.
Though $\phi$ by itself is a zero-form, the superconnection $\mathcal{J}$ as a whole is actually a generalized one-form,
the odd grading of $\lambda^s_{m}$ ($m=4,5,6,7$)
supplying the extra grading associated with every application of the
exterior derivative operator.

Note also that the 1-forms $W_i=W_i^\mu dx_\mu$ and $B=B^\mu dx_\mu$ are dimensionless, 
so the 0-form $\phi$ which appears together with them in the superconnection must also be
dimensionless.  To give $\phi$ its usual mass dimension of one, some authors replace
$\phi$ with $\phi/\mu$, where $\mu$ is a mass scale.
However, for notational simplicity we will not do this.
We request the reader to assume that, not just $\phi$, but all dimensionful quantities are multiplied by the appropriate (but invisible) powers of $\mu$ to make them dimensionless, \textit{e.g.}
$B^\nu \rightarrow B^\nu/\mu$, $dx^\nu \rightarrow \mu\,dx^\nu$.
In particular, the Hodge dual should not change the dimension of the operand:
$*1 =  \mu^4 d^4x$, $*(\mu\,dx^\nu)=\frac{1}{6}\mu^3\,\varepsilon^{\kappa\lambda\mu\nu}dx_\kappa\wedge dx_\lambda \wedge dx_\mu$, etc.
Once all the dust has settled, the powers of $\mu$ will disappear from the final expression for
the action, and we will then be free to think of all quantities to have their usual dimensions.

As stated in the introduction, we are considering a model space consisting of 
two $3+1$ dimensional branes separated by a gap.
We interpret the even part of the superconnection $\mathcal{J}$ as
connecting points within the two $3+1$ dimensional branes, 
the 1-from $\mathcal{W}-\frac{1}{\sqrt{3}}B\cdot \mathbf{1}_{2\times 2}$ acting on the
left-handed brane, and the 1-from $-\frac{2}{\sqrt{3}}B$ acting on the right.
The 0-form $\sqrt{2}\phi$ connects the left-handed brane to the right, and
$\sqrt{2}\phi^\dagger$ the right-handed brane to the left.

%%%%%%%%%%%%%%%%%%%%%%%%%%%%%%%%%%%%%%%%%%%%%%%%%%%%%%%%%%%%%%%%%%%%%%%%%%%%%%%%%%%%%%%%%%%%%
%%%%%%%%%%%%%%%%%%%%%%%%%%%%%%%%%%%%%%%%%%%%%%%%%%%%%%%%%%%%%%%%%%%%%%%%%%%%%%%%%%%%%%%%%%%%%
\subsection{Supercurvature}\label{supercurvature21}

%%%%%%%%%%%%%%%%%%%%%%%%%%%%%%%%%%%%%%%%%%%%%%%%%%%%%%%%%%%%%%%%%%%%%%%%%%%%%%%%%%%%%%%%%%%%%
%%%%%%%%%%%%%%%%%%%%%%%%%%%%%%%%%%%%%%%%%%%%%%%%%%%%%%%%%%%%%%%%%%%%%%%%%%%%%%%%%%%%%%%%%%%%%
\subsubsection{Extension of the Exterior Derivative}

In usual differential geometry
the curvature of the connection $\omega$ is given by $(d\omega)+\omega\wedge\omega$,
and in QCD the curvature of the gauge connection $G$ is given by
$F_G = (dG) + \frac{1}{2}[G,G]$.
We would like to calculate the supercurvature from the superconnection $\mathcal{J}$ via
the analogous expression
\begin{eqnarray}
\label{supercurvatureF}
\mathcal{F} \;=\; 
(\mathbf{d}_S\,\mathcal{J})+\dfrac{1}{2}\left[\,\mathcal{J},\,\mathcal{J}\,\right]_S
\;,
\end{eqnarray}
where $\mathbf{d}_S$ is the extension of the usual exterior derivative operator $d$ 
to the superalgebra $\Omega(M)\otimes su(2/1)$.
Let us define what it is.

The exterior derivative operator $d=dx^\mu\wedge\partial_\mu$ is a map from
$\Omega^i(M)$ to $\Omega^{i+1}(M)$:
\begin{equation}
\Omega^i(M) \;\stackrel{d}{\longrightarrow}\; \Omega^{i+1}(M)\;,
\end{equation}
or in terms of the $\mathbb{Z}_2$-grading decomposition $\Omega(M)=\Omega_0(M)+\Omega_1(M)$,
it maps from one grading to the other:
\begin{equation}
\Omega_0(M)\;\stackrel{d}{\longleftrightarrow}\;\Omega_1(M)\;.
\end{equation}
Since it changes the $\mathbb{Z}_2$ grading of differential forms by 1, it carries the grading of 1 itself.
Its characteristic properties are 
that it satisfies the super-Leibniz rule
\begin{equation}
d(a\wedge b) \;=\; (da) \wedge b +(-1)^{|a|}a\wedge (db)
\;.
\label{superLeibniz1}
\end{equation}
and that it is nilpotent
\begin{equation}
d^2 \;=\; 0\;.
\label{nilpotent0}
\end{equation}

The extension $\mathbf{d}_S$ operating on
$\Omega(M)\otimes su(2/1)$ should also be a grading-switching operator
\begin{equation}
[\Omega(M)\otimes su(2/1)]_0 
\;\stackrel{\mathbf{d}_S}{\longleftrightarrow} \;
[\Omega(M)\otimes su(2/1)]_1
\;,
\end{equation}
and should possess the same properties of obeying the super-Leibniz rule
and nilpotency.
To this end, let us write
\begin{equation}
\mathbf{d}_S \;=\; \mathbf{d} + \mathbf{d}_M\;,
\end{equation}
where
\begin{eqnarray}
\Omega_0(M)\otimes su(2/1)
& \stackrel{\mathbf{d}}{\longleftrightarrow} &
\Omega_1(M)\otimes su(2/1)
\;,
\cr
\Omega(M)\otimes su(2/1)_0
& \stackrel{\mathbf{d}_M}{\longleftrightarrow} &
\Omega(M)\otimes su(2/1)_1
\;,
\end{eqnarray}
that is, $\mathbf{d}$ switches the grading of the $\Omega(M)$ part
while $\mathbf{d}_M$ switches the grading of the $su(2/1)$ part,
and consider the two operators separately.
Since the operators themselves have grading 1 in $\Omega(M)\otimes su(2/1)$,
they should anti-commute:
\begin{equation}
\mathbf{d}\mathbf{d}_M + \mathbf{d}_M\mathbf{d} \;=\; 0\;.
\end{equation}
From the model building perspective,
the $\mathbf{d}$ operator generates translations
within each of the two $3+1$ dimensional branes while the
`matrix derivative' $\mathbf{d}_M$ \cite{Coquereaux:1990ev,Haussling:1991ns}
accounts for transitions between the two branes.

%%%%%%%%%%%%%%%%%%%%%%%%%%%%%%%%%%%%%%%%%%%%%%%%%%%%%%%%%%%%%%%%%%%%%%%%%%%%%%%%%%%%%%%%%%%%%
\subsubsection{The operator $\mathbf{d}$.}

We define 
the action of the operator $\mathbf{d}$ on $a\otimes X\in \Omega(M)\otimes su(2/1)$ by
\begin{equation}
\mathbf{d}(a\otimes X)
\;=\; (da)\otimes X\;.
\label{ddef}
\end{equation}
It is straightforward to show that $\mathbf{d}$ satisfies the
super-Leibniz rule given by
\begin{eqnarray}
\lefteqn{
\mathbf{d}\Bigl(\bigl[\,a\otimes X,\,b\otimes Y\,\bigr]_S\Bigr) 
}
\cr
& = &
  \bigl[\,\mathbf{d}(a\otimes X),\,b\otimes Y\,\bigr]_S
\cr
& & 
+ (-1)^{|a|+|X|}\bigl[\,a\otimes X,\,\mathbf{d}(b\otimes Y)\bigr]_S
\;.
%\cr
%& &
\label{superLeibniz2}
\end{eqnarray}
Nilpotency $\mathbf{d}^2(a\otimes X)=0$ also follows immediately from $(d^2a)=0$.
From Eqs.~(\ref{superLeibniz1}) and (\ref{ddef}), we infer
\begin{eqnarray}
\mathbf{d}(a\otimes X)
& = & \left[ da -(-1)^{|a|}ad \right]\otimes X \cr
& = & da \otimes X -(-1)^{|a|} ad\otimes X \cr
& = & da \otimes X -(-1)^{|a|+|X|} a\otimes X d 
\;,
\end{eqnarray}
or using the second definition of the supercommutator, Eq.~(\ref{supercommutator2B}),
we can write
\begin{equation}
\mathbf{d}(a\otimes X) \;=\; \bigl[\,\mathbf{d},\,a\otimes X\,\bigr]_S\;.
\end{equation}
%

%%%%%%%%%%%%%%%%%%%%%%%%%%%%%%%%%%%%%%%%%%%%%%%%%%%%%%%%%%%%%%%%%%%%%%%%%%%%%%%%%%%%%%%%%%%%%
\subsubsection{The Matrix Derivative $\mathbf{d}_M$}

Let us first find an operator $d_M$ which acts on $su(2/1)$ such that
\begin{equation}
su(2/1)_0 \;\stackrel{d_M}{\longleftrightarrow}\; su(2/1)_1 
\end{equation}
with the required properties.
For $X,Y\in su(2/1)$,
the super-Leibniz rule demands
\begin{equation}
(d_M[\,X,\,Y\,]_s) 
\;=\; [\,(d_M X),\,Y\,]_s 
+(-1)^{|X|} [\,X,\,(d_M Y)\,]_s
\;.
\end{equation}
Comparing with the second line of Eq.~(\ref{SuperLieConditions}), we see that
such an operator can be realized as \footnote{%
The supercommutator that appears here is that of the first definition, Eq.~(\ref{supercommutator1}).
}
\begin{equation}
(d_M X) \;=\; i\,[\,\eta,\,X\,]_s\;,
\label{Mderivative}
\end{equation}
where $\eta$ is any odd element of $su(2/1)$.
It is clear that this operator switches the grading of $X$.

Nilpotency is more difficult to realize and how it is treated
is an important consideration of the entire formalism.
It was shown in Ref.~\cite{Haussling:1991ns} that
for a generic Lie superalgebra $su(N/M)$, demanding $d_M^2(X)=0$
with $d_M$ defined as above for all $X\in su(N/M)$ leads to the condition $N=M$.
Indeed, since $\eta$ is an odd element of $su(N/M)$ it has the form
\begin{equation}
\eta \;=\; 
\begin{bmatrix}
\mathbf{0}_{N\times N} & \zeta \\ \zeta^\dagger & \mathbf{0}_{M\times M} 
\end{bmatrix}
\label{etadef}
\end{equation}
where $\zeta$ is an $N\times M$ matrix.
To impose $(d_M^2 X)=0$ we must have
\begin{equation}
(d_M^2 X) 
\;=\; -[\,\eta,\,[\,\eta,\,X\,]_s\,]_s 
\;=\; -i[\,\eta^2,\,X\,]
\;=\; 0\;,
\label{dM2X}
\end{equation}
which means that 
\begin{equation}
\eta^2 \;=\;
\begin{bmatrix}
\zeta\zeta^\dagger & \mathbf{0}_{N\times M} \\
\mathbf{0}_{M\times N} & \zeta^\dagger\zeta
\end{bmatrix}
\end{equation}
must commute with all elements of $su(N/M)$.
This requires $\eta^2$ to be a multiple of a unit matrix, that is
\begin{equation}
\zeta\zeta^\dagger \;=\; v^2\mathbf{1}_{N\times N}\;,\qquad
\zeta^\dagger\zeta \;=\; v^2\mathbf{1}_{M\times M}\;,
\label{zetanorm}
\end{equation}
with $v^2$ a constant,
which is possible only when $N=M$.

Because of this, Coquereaux et al. in Ref.~\cite{Coquereaux:1990ev} work in 4 dimensions
by adding an extra row and column of zeroes to the $su(2/1)$ matrices 
to make them into $4\times 4$ $su(2/2)$ matrices.
The $\eta$-matrix for $su(2/2)$ will have the form of Eq.~(\ref{etadef}) 
with $\zeta$ a multiple of a $2\times 2$ unitary matrix.  
The supercommutator of $\eta$ and a generic $su(2/1)$ matrix embedded into $su(2/2)$ 
will have non-zero elements in the fourth row and fourth column, but these are dropped 
projecting the result back into $su(2/1)$.

Haussling et al. in Ref.~\cite{Haussling:1991ns} take a different approach and work in 3 dimensions throughout
by dropping the fourth row and fourth column from the $\eta$ matrix for $su(2/2)$.\footnote{%
In the representation we use for $su(2/2)$ in a later section, it is more precise to say 
that Ref.~\cite{Haussling:1991ns} drops the third row and third column
corresponding to the right-handed neutrino.
}
Writing the first column of $\zeta$ as $\xi$, 
the $\eta$-matrix used in Ref.~\cite{Haussling:1991ns} is
\begin{equation}
\eta \;=\; 
\begin{bmatrix}
\mathbf{0}_{2\times 2} & \xi \\
\xi^\dagger            & 0  
\end{bmatrix}
\;,
\label{3Deta}
\end{equation}
where $\xi^\dagger\xi=v^2$.
Since the condition $N=M$ is not met, $d_M$
defined with this $\eta$ is not nilpotent.

Thus, to define a matrix derivative for $su(2/1)$ 
one must either work in $su(2/2)$ and project back into $su(2/1)$, or forgo nilpotency.
However, it turns out that either way the resulting supercurvature and other
physical quantities will be the same,
so we will adopt the three dimensional version, Eq.~(\ref{3Deta}),
in our definition of $d_M$ on $su(2/1)$.

%%%%%%%%%%%%%%%%%%%%%%%%%%%%%%%%%%%%%%%%%%%%%%%%%%%%%%%%%%%%%%%%%%%%%%%%%%%%%%%%%%%%%%%%%%%%%
%\subsubsection{Extension to $\Omega(M)\otimes su(2/1)$}

We extend $d_M$ acting on $su(2/1)$ to 
$\mathbf{d}_M$ acting on $\Omega(M)\otimes su(2/1)$
by defining the operation of $\mathbf{d}_M$ on $a\otimes X \in \Omega(M)\otimes su(2/1)$ 
to be given by
\begin{equation}
\mathbf{d}_M(a\otimes X) \;=\; (-1)^{|a|}a\otimes (d_M X)\;,
\end{equation}
which can also be written as
\begin{equation}
\mathbf{d}_M(a\otimes X) \;=\; \bigl[\,\mathbf{d}_M,\,a\otimes X\,\bigr]_S\;,
\end{equation}
where the supercommutator here is that of Eq.~(\ref{supercommutator1}).
It is straightforward to show that $\mathbf{d}_M$ satisfies the super-Leibniz rule.

%%%%%%%%%%%%%%%%%%%%%%%%%%%%%%%%%%%%%%%%%%%%%%%%%%%%%%%%%%%%%%%%%%%%%%%%%%%%%%%%%%%%%%%%%%%%%
\subsubsection{Short note on nilpotency}

At this point, we would like to bring to the reader's attention the fact that
the statements $d_M^2=0$ and $(d_M^2 X)=0$ are not equivalent.
While the first guarantees the second, the converse is not true.
Indeed, using Eq.~(\ref{SuperLieConditions}) we can rewrite
Eq.~(\ref{dM2X}) as
\begin{equation}
(d_M^2 X)
\;=\; \bigl[d_M,\bigl[d_M,X\bigr]_s\bigr]_s
\;=\; \dfrac{1}{2}
      \bigl[\bigl[d_M,d_M\bigr]_s,X\bigr]_s
\;,
\end{equation}
and we can make the identification
\begin{equation}
\dfrac{1}{2}\bigl[d_M,d_M\bigr]_s
\;=\; d_M^2 \;=\; -i\eta^2\;,
\end{equation}
where $\eta^2$ is a non-zero even element of $su(N/M)$.
For the $N=M$ case, it becomes a multiple of the unit matrix which 
constitutes the center of the superalgebra
($\lambda^s_{15}$ in the case of $su(2/2)$ to be discussed later).
Thus, it is not clear whether $d_M$ as defined here
truly qualifies as a generalization of the `exterior derivative' operator.
Furthermore,
whether $d_M^2$, and consequently $\mathbf{d}_M^2$, 
can be considered to vanish or not is an important consideration when
calculating the supercurvature as we will see in the following.

%%%%%%%%%%%%%%%%%%%%%%%%%%%%%%%%%%%%%%%%%%%%%%%%%%%%%%%%%%%%%%%%%%%%%%%%%%%%%%%%%%%%%%%%%%%%%
%%%%%%%%%%%%%%%%%%%%%%%%%%%%%%%%%%%%%%%%%%%%%%%%%%%%%%%%%%%%%%%%%%%%%%%%%%%%%%%%%%%%%%%%%%%%%
%%%%%%%%%%%%%%%%%%%%%%%%%%%%%%%%%%%%%%%%%%%%%%%%%%%%%%%%%%%%%%%%%%%%%%%%%%%%%%%%%%%%%%%%%%%%%
\subsubsection{Derivation of $\mathcal{F}$}

Let us now look at the terms contributing to Eq.~(\ref{supercurvatureF}) one by one.
$(\mathbf{d}\mathcal{J})$ is simply\footnote{\label{sign-convention}%
If the superconnection $\mathcal{J}$ is considered an element of 
$su(2/1)\otimes\Omega(M)$ instead of $\Omega(M)\otimes su(2/1)$, then the
result of $\mathbf{d}$ acting on $\mathcal{J}$ will be 
\begin{equation}
(\mathbf{d}\mathcal{J})
\;=\; 
i\begin{bmatrix}
d\mathcal{W}-\frac{1}{\sqrt{3}}\,dB\cdot\mathbf{1}_{2\times 2} & -\sqrt{2}\,d\phi \\
-\sqrt{2}\,d\phi^\dagger & -\frac{2}{\sqrt{3}}\,dB
\end{bmatrix}
\;.
\end{equation}
Note the minus signs on the off diagonal terms which results when $\mathbf{d}$
commutes through the odd $su(2/1)$ matrix multiplying the zero-form fields.
This choice is a matter of convention and does not affect the final results.
}
\begin{eqnarray}
(\mathbf{d}\mathcal{J})
& = & 
i\begin{bmatrix}
d\mathcal{W}-\frac{1}{\sqrt{3}}\,dB\cdot\mathbf{1}_{2\times 2} & \sqrt{2}\,d\phi \\
\sqrt{2}\,d\phi^\dagger & -\frac{2}{\sqrt{3}}\,dB
\end{bmatrix}
\;.
\end{eqnarray}
while $(\mathbf{d}_M\mathcal{J})$ is given by
\begin{eqnarray}
\lefteqn{(\mathbf{d}_M\mathcal{J})
\;=\; i\bigl[\,\eta,\,\mathcal{J}\,\bigr]_S 
}
\cr
& = & i
\begin{bmatrix}
-\sqrt{2}\left(\xi\phi^\dagger + \phi\xi^\dagger\right) & 
i\!\left(\mathcal{W}\xi+\frac{1}{\sqrt{3}}B\xi\right) \\	%Chen_Correction
-i\!\left(\xi^\dagger\mathcal{W}+\frac{1}{\sqrt{3}}\xi^\dagger B\right) &  %Chen_Correction
-\sqrt{2}\left(\xi^\dagger\phi + \phi^\dagger\xi\right)
\end{bmatrix}
\;.
\cr
& &
\end{eqnarray}

To calculate supercommutator of $\mathcal{J}$ with itself,
we decompose $\mathcal{J}$ into two parts as
\begin{eqnarray}
\mathcal{J}
& = & 
\underbrace{i
\begin{bmatrix}
   \mathcal{W}-\frac{1}{\sqrt{3}}B\cdot \mathbf{1}_{2\times 2} & \mathbf{0}_{2\times 1} \\
   \mathbf{0}_{1\times 2} & -\frac{2}{\sqrt{3}}B 
\end{bmatrix}
}_{\displaystyle \mathcal{J}_{10}}
+
\underbrace{i
\begin{bmatrix}
   \mathbf{0}_{2\times 2} & \sqrt{2}\phi \vphantom{\big|} \\
   \sqrt{2}\phi^\dagger & 0\vphantom{\Big|}
\end{bmatrix}
}_{\displaystyle \mathcal{J}_{01}}
\;,
\cr
& &
\label{Jseparation}
\end{eqnarray}
where the two subscripts refer to the gradings in $\Omega(M)$ and $su(2/1)$, respectively,
in that order.
Keeping in mind the product rule given in Eq.~(\ref{OmegaLproduct}) for
$\Omega(M)\otimes su(2/1)$, we find
\begin{eqnarray}
[\,\mathcal{J}_{10},\,\mathcal{J}_{10}\,]_S
& = & -2i
\begin{bmatrix}
\;\varepsilon_{ijk}(W_i\wedge W_j)\tau_k & \mathbf{0}_{2\times 1}\; \\
\mathbf{0}_{1\times 2} & 0
\end{bmatrix}
\;,
\cr
%%%
[\,\mathcal{J}_{01},\,\mathcal{J}_{01}\,]_S
& = & -4i
\begin{bmatrix}
\;\phi\phi^\dagger & \mathbf{0}_{2\times 1}\; \\
\;\mathbf{0}_{1\times 2} & \phi^\dagger\phi\;
\end{bmatrix}
\;,
\end{eqnarray}
and\footnote{%
As mentioned in footnote~\ref{sign-convention},
we are assuming that the supercurvature is an element of $\Omega(M)\otimes su(2/1)$,
not $su(2/1)\otimes \Omega(M)$.  The latter choice would reverse the signs of
$[\,\mathcal{J}_{10},\,\mathcal{J}_{01}\,]_S$ and 
$[\,\mathcal{J}_{01},\,\mathcal{J}_{10}\,]_S$.
Again, this is a matter of convention and does not affect the final result
as long as the convention is consistently applied.
}
\begin{eqnarray}
\lefteqn{
[\,\mathcal{J}_{10},\,\mathcal{J}_{01}\,]_S \;=\; 
[\,\mathcal{J}_{01},\,\mathcal{J}_{10}\,]_S
}
\cr
& = & \sqrt{2}i
\begin{bmatrix}
\mathbf{0}_{2\times 2} & i\left(\mathcal{W}\phi+\frac{1}{\sqrt{3}}B\phi\right) \\	%Chen_Correction
-i\left(\phi^\dagger\mathcal{W}+\frac{1}{\sqrt{3}}\phi^\dagger B\right) & 0 		%Chen_Correction
\end{bmatrix}
\;.
\cr
& &
%%%
%[\,\mathcal{J}_1,\,\mathcal{J}_0\,]
%& = &
%\cr
%%%
\end{eqnarray}
Therefore,\footnote{%
\label{MSmultiplication}
Instead of calculating the supercommutator $\frac{1}{2}[\mathcal{J},\mathcal{J}]_S$
as we have done here, some papers treat the superconnection $\mathcal{J}$
as a super-endomorphism of a superspace and calculate the product $\mathcal{J}\odot\mathcal{J}$,
using the Ne'eman-Sternberg rule for supermatrix multiplication 
\cite{Hwang:1995wk,Ne'eman:1990nr,Sternberg:2012}:
\begin{eqnarray*}\label{rule}
\lefteqn{
\left[
  \begin{array}{cc}
    A & C \\
    D & B \\
  \end{array}
\right]\odot
\left[
  \begin{array}{cc}
    A' & C' \\
    D' & B' \\
  \end{array}
\right]}
\cr
& \!\!\!=\!\!\! &
\left[\!
  \begin{array}{cc}
    A\wedge A'+(-1)^{|D'|} C\wedge D' &\,  A\wedge C'+(-1)^{|B'|}C\wedge B' \\
    (-1)^{|A'|} D\wedge A'+B\wedge D' &\, (-1)^{|C'|} D\wedge C'+B\wedge B' \\
  \end{array}
\!\right]
\;.
%\cr
%& &
\end{eqnarray*}
The resulting supercurvature $\mathcal{F}$ is not an element of $\Omega(M)\otimes su(2/1)$,
and the definition of the inner product of $\mathcal{F}$ with itself must be 
changed accordingly in the calculation of the action.
However, the resulting action turns out to be the same.
The above multiplication rule is derived in Appendix~\ref{SuperMatrixMultiplication}.
}
\begin{eqnarray}
\lefteqn{[\,\mathcal{J},\,\mathcal{J}\,]_S} 
\cr
& = &
 [\,\mathcal{J}_{10},\,\mathcal{J}_{10}\,]_S
+[\,\mathcal{J}_{10},\,\mathcal{J}_{01}\,]_S
\cr
& & \hspace{1.9cm}
+[\,\mathcal{J}_{01},\,\mathcal{J}_{10}\,]_S
+[\,\mathcal{J}_{01},\,\mathcal{J}_{01}\,]_S
\cr
& = & 2i
\begin{bmatrix}
-\varepsilon_{ijk}(W_i\wedge W_j)\tau_k - 2\phi\phi^\dagger &
+\sqrt{2}i\left(\mathcal{W}\phi+\frac{1}{\sqrt{3}}B\phi\right) \\	%Chen_Correction
-\sqrt{2}i\left(\phi^\dagger\mathcal{W}+\frac{1}{\sqrt{3}}\phi^\dagger B\right) & 	%Chen_Correction
-2\phi^\dagger\phi
\end{bmatrix}
\;.
\cr
& &
\end{eqnarray}
%

%%%%%%%%%
%%%%%%%%%
%%%%%%%%%
\newpage
\begin{widetext}
Putting everything together, the supercurvature reads as
\begin{eqnarray}
\mathcal{F}
& = & 
\mathbf{d}\mathcal{J}
+\; \mathbf{d}_M\mathcal{J}
+\; \dfrac{1}{2}[\,\mathcal{J},\,\mathcal{J}\,]_S
%+\; (\mathbf{d}\mathbf{d}_M) + \mathbf{d}^2 
\cr
%%%
%& = & 
%i
%\begin{bmatrix}
%d\mathcal{W}-\frac{1}{\sqrt{3}}\,dB\cdot\mathbf{1}_{2\times 2} & \sqrt{2}\,d\phi \\
%\sqrt{2}\,d\phi^\dagger & -\frac{2}{\sqrt{3}}\,dB
%\end{bmatrix}
%+i
%\begin{bmatrix}
%-\sqrt{2}\left(\xi\phi^\dagger + \phi\xi^\dagger\right) & 
%-i\left(\mathcal{W}\xi+\frac{1}{\sqrt{3}}B\xi\right) \\
%i\left(\xi^\dagger\mathcal{W}+\frac{1}{\sqrt{3}}\xi^\dagger B\right) & 
%-\sqrt{2}\left(\xi^\dagger\phi + \phi^\dagger\xi\right)
%\end{bmatrix}
%\cr
%& & 
%+i
%\begin{bmatrix}
%-\varepsilon_{ijk}(W_i\wedge W_j)\tau_k - 2\phi\phi^\dagger &
%-\sqrt{2}i\left(\mathcal{W}\phi+\frac{1}{\sqrt{3}}B\phi\right) \\
%\sqrt{2}i\left(\phi^\dagger\mathcal{W}+\frac{1}{\sqrt{3}}\phi^\dagger B\right) & 
%-2\phi^\dagger\phi
%\end{bmatrix}
%+i
%\begin{bmatrix}
%\mathbf{0}_{2\times 2} & d\xi \\
%d\xi^\dagger           & 0
%\end{bmatrix}
%-i
%\begin{bmatrix}
%\xi\xi^\dagger & 0 \\ 0 & \xi^\dagger\xi
%\end{bmatrix}
%\cr
%%%
& = & i
\begin{bmatrix}
F_W-\frac{1}{\sqrt{3}}F_B\cdot\mathbf{1}_{2\times 2}-2\phi\phi^{\dagger}
-\sqrt{2}\left(\xi\phi^\dagger + \phi\xi^\dagger\right) %- \xi\xi^\dagger 
& 
\sqrt{2}\,D\phi %+ D\xi
+ \left(i\mathcal{W}\xi+\frac{i}{\sqrt{3}}B\xi\right) %Chen_Correction
\\
\sqrt{2}(D\phi)^{\dagger} 
- \left(i\xi^\dagger\mathcal{W}+\frac{i}{\sqrt{3}}\xi^\dagger B\right) 	%Chen_Correction
%+(D\xi)^{\dagger}
& 
-\frac{2}{\sqrt{3}}F_B-2\phi^{\dagger}\phi - \sqrt{2}\left(\xi^\dagger\phi + \phi^\dagger\xi\right) 
%-\xi^\dagger\xi\\
\end{bmatrix}
\cr
& = & i
\begin{bmatrix}
F_W-\frac{1}{\sqrt{3}}F_B\cdot\mathbf{1}_{2\times 2}
-2\hat{\phi}\hat{\phi}^{\dagger}
+ \xi\xi^\dagger 
& 
\sqrt{2}\,D\hat{\phi} \\
\sqrt{2}(D\hat{\phi})^{\dagger}  & 
-\frac{2}{\sqrt{3}}F_B
-2\hat{\phi}^\dagger\hat{\phi} 
+ v^2 
\\
\end{bmatrix}
\;,
\label{Fsu21}
\end{eqnarray}
\end{widetext}
%%%%%%%%%
%%%%%%%%%
%%%%%%%%%
%
where we have introduced the shifted 0-form field
\begin{equation}
\hat{\phi} \;=\; \phi + \dfrac{\xi}{\sqrt{2}}\;,
\label{phihatdef}
\end{equation}
and
\begin{eqnarray}\label{tensors}
D\phi & = & d\phi + \left( i\mathcal{W}\phi+\frac{i}{\sqrt{3}}B\phi \right) \;, \cr	%Chen_Correction
%D\xi & = & d\xi - \left( i\mathcal{W}\xi+\frac{i}{\sqrt{3}}B\xi \right) \;, \cr
D\hat{\phi} & = & d\hat{\phi} + \left( i\mathcal{W}\hat{\phi}+\frac{i}{\sqrt{3}}B\hat{\phi} \right) \;, \cr	%Chen_Correction
F_W & = & \left(F_W\right)_k\tau_k
\;=\; \left( d W_k - \epsilon_{ijk} W_i \wedge W_j \right)\tau_k 
\;, \cr
%& & \hspace{1.5cm} 
%\;=\; \left[ d W^k + ig \left( W\wedge W \right)^k \right]\tau^k.
%\cr
F_B & = & dB\;.\vphantom{\bigg|}
\end{eqnarray}
We have also used $\xi^\dagger\xi=v^2$.

%%%%%%%%%%%%%%%%%%%%%%%%%%%%%%%%%%%%%%%%%%%%%%%%%%%%%%%%%%%%%%%%%%%%%%%%%%%%%%%%%%%%%%%%%%%%%
%%%%%%%%%%%%%%%%%%%%%%%%%%%%%%%%%%%%%%%%%%%%%%%%%%%%%%%%%%%%%%%%%%%%%%%%%%%%%%%%%%%%%%%%%%%%%
\subsubsection{Gauge Transformation Properties}

Recall that in the case of QCD, the curvature $F_G=dG+\frac{1}{2}[G,G]$ transforms as
\begin{equation}
F_G \;\rightarrow U F_G U^\dagger
\label{FGtransform}
\end{equation}
under $SU(3)$ gauge transformations:
\begin{equation}
U \;=\; \exp\left[i\sum_{j=1}^8 \theta_j\lambda_j\right] \;.
\end{equation}
Let us see whether the supercurvature $\mathcal{F}$ derived above 
transforms in an analogous fashion under
$SU(2)_L\times U(1)_Y$ gauge transformations
generated by the even part of the $su(2/1)$ superalgebra:
\begin{equation}
U 
\;=\; \exp\left[i\sum_{j=1,2,3,8} \theta_j^{\phantom{s}}\lambda_j^s\right] 
\;=\;
\begin{bmatrix}
u\,e^{-i\theta/\sqrt{3}} & \mathbf{0}_{2\times 1} \\
\mathbf{0}_{1\times 2}   & e^{-2i\theta/\sqrt{3}}
\end{bmatrix}
\;,
\end{equation}
where
\begin{equation}
u \,=\, \exp\left[i\sum_{j=1,2,3}\theta_j\tau_j\right] \,\in\, SU(2)_L\;,\quad
\theta \;=\; \theta_8\;.
\end{equation}
The 1-form gauge fields transform as
\begin{eqnarray}
\mathcal{W} & \rightarrow & u\mathcal{W}u^\dagger + i\,du\,u^\dagger \;,\cr
          B & \rightarrow & B - d\theta \;.
\label{GT1}
\end{eqnarray}
For the 0-form field, we assume that it is the shifted field 
$\hat{\phi}=\phi+\xi/\sqrt{2}$ which
transforms as
\begin{equation}
\hat{\phi} \;\rightarrow\; u\,e^{i\theta/\sqrt{3}}\,\hat{\phi}\;.
\label{GT2}
\end{equation}
The interpretation is that $\xi/\sqrt{2}$ is the vacuum expectation value (VEV)
of $\hat{\phi}$ and $\phi$ is the fluctuation around that VEV.
Then, 
\begin{eqnarray}
F_W   & \rightarrow & u F_W u^\dagger \;,\cr
F_B   & \rightarrow & F_B \;,\cr
D\hat{\phi} & \rightarrow & u\,e^{i\theta/\sqrt{3}}\,D\hat{\phi} \;.
\end{eqnarray}
Unfortunately,
the $\xi\xi^\dagger$ term in the upper-left block of $\mathcal{F}$ is a constant
projection matrix which does not transform under gauge transformations.
This term prevents $\mathcal{F}$ from transforming analogously to Eq.~(\ref{FGtransform}) as
$
\mathcal{F} \rightarrow U\mathcal{F}U^\dagger.
$
Since this transformation law would guarantee the gauge invariance of the action, 
which we will derive in the next subsection, the lack of such a law is somewhat problematic
(though in fact, it is found that the problem cures itself in the sense that the action derived from 
this supercurvature is still manifestly gauge invariant).
In the following,
we trace this problem back to the non-nilpotency of the matrix derivative $d_M$
in $su(2/1)$.
However,
this can already be seen by noticing that the problem would not exist if we could
replace $\xi\xi^\dagger$ with $v^2\mathbf{1}_{2\times 2}$.

%%%%%%%%%%%%%%%%%%%%%%%%%%%%%%%%%%%%%%%%%%%%%%%%%%%%%%%%%%%%%%%%%%%%%%%%%%%%%%%%%%%%%%%%%%%%%
\subsubsection{Covariant Derivative}

Given $\mathbf{d}$, $\mathbf{d}_M$, and the superconnection $\mathcal{J}$,
we can construct a covariant derivative operator via
\begin{equation}
\mathcal{D} \;=\; \mathbf{d} + \mathbf{d}_M + \mathcal{J}\;.
\label{CoD}
\end{equation}
Let $\alpha\in\Omega(M)\otimes su(2/1)$ be an object which gauge transforms as
$\alpha\rightarrow \alpha' =U\alpha U^\dagger$.
Then, $(\mathbf{d}\alpha)$ transforms as
\begin{eqnarray}
\lefteqn{
(\mathbf{d}\alpha)
\; \rightarrow \;
(\mathbf{d}\alpha')
}
\cr
& = & \mathbf{d}(U\alpha U^\dagger)
\cr
& = &
(\mathbf{d}U)\alpha U^\dagger + U(\mathbf{d}\alpha)U^\dagger + (-1)^{|\alpha|}U\alpha(\mathbf{d}U^\dagger) 
\cr
& = & 
(\mathbf{d}U)U^\dagger \alpha' + U\mathbf{d}U^\dagger(\alpha') + (-1)^{|\alpha'|} \alpha' U(\mathbf{d}U^\dagger)
\cr
& = & 
(\mathbf{d}U)U^\dagger \alpha' + U\mathbf{d}U^\dagger(\alpha') - (-1)^{|\alpha'|} \alpha' (\mathbf{d}U)U^\dagger
\cr
& = & \Bigl[ U\mathbf{d}U^\dagger + (\mathbf{d}U)U^\dagger,\,\alpha'\Bigr]_S 
\;,
\end{eqnarray}
and we can see that $\mathbf{d}$ transforms as
\begin{equation}
\mathbf{d}\;\rightarrow\; \mathbf{d}'=\,U\mathbf{d}U^\dagger + (\mathbf{d}U)U^\dagger\;.
\end{equation}
On the other hand,
from the gauge transformation properties we introduced in Eqs.~(\ref{GT1}) and (\ref{GT2}),
we can infer that the combination 
\begin{eqnarray}
\mathbf{d}_M + \mathcal{J}
& = & i\eta + \mathcal{J}
\cr
& = & i
  \begin{bmatrix}
    \mathcal{W}-\frac{1}{\sqrt{3}}B\cdot \mathbf{1}_{2\times 2} & \sqrt{2}\,\hat{\phi} \\
    \sqrt{2}\,\hat{\phi}^\dagger & -\frac{2}{\sqrt{3}}B \\
  \end{bmatrix}
\;,
\end{eqnarray}
transforms as
\begin{equation}
\mathbf{d}_M + \mathcal{J}
\;\rightarrow\;
U(\mathbf{d}_M + \mathcal{J})U^\dagger - (\mathbf{d}U)U^\dagger \;.
\label{dMJtrans}
\end{equation}
Therefore, the covariant derivative 
$\mathcal{D} = \mathbf{d} + \mathbf{d}_M + \mathcal{J}$ transforms as
\begin{equation}
\mathcal{D} \;\rightarrow\; U\mathcal{D} U^\dagger\;,
\label{Dtrans}
\end{equation}
and consequently, if $\alpha\rightarrow U\alpha U^\dagger$ then
\begin{equation}
(\mathcal{D}\alpha) \;\rightarrow U(\mathcal{D}\alpha)U^\dagger\;.
\end{equation}

%%%%%%%%%%%%%%%%%%%%%%%%%%%%%%%%%%%%%%%%%%%%%%%%%%%%%%%%%%%%%%%%%%%%%%%%%%%%%%%%%%%%%%%%%%%%%
%%%%%%%%%%%%%%%%%%%%%%%%%%%%%%%%%%%%%%%%%%%%%%%%%%%%%%%%%%%%%%%%%%%%%%%%%%%%%%%%%%%%%%%%%%%%%
\subsubsection{Supercurvature from the Covariant Derivative}

The supercurvature can be defined as the supercommutator of the covariant derivative with itself:
\begin{eqnarray}
\label{DDsupercurvature}
\mathcal{F} 
& = & \dfrac{1}{2}\bigl[\,\mathcal{D},\,\mathcal{D}\,\bigr]_S \cr
& = & \dfrac{1}{2}\bigl[\,\mathbf{d}_S + \mathcal{J},\,\mathbf{d}_S + \mathcal{J}\,\bigr]_S \cr
%& = & \dfrac{1}{2}\bigl[\,\mathbf{d}_S,\,\mathbf{d}_S\,\bigr]_S
& = & \mathbf{d}_S^2
+(\mathbf{d}_S\,\mathcal{J})+\dfrac{1}{2}\bigl[\,\mathcal{J},\,\mathcal{J}\,\bigr]_S
\;.
\end{eqnarray}
From Eq.~(\ref{Dtrans}), we can infer that $\mathcal{F}$ gauge transforms as
$\mathcal{F} \rightarrow U\mathcal{F}U^\dagger$.

In the above expression for $\mathcal{F}$, it is usually assumed that 
\begin{equation}
\mathbf{d}_S^2 \;=\; (\mathbf{d}+\mathbf{d}_M)^2
\;=\; \mathbf{d}^2 + \mathbf{d}_M^2
\;=\; 0\;,
\end{equation}
and the $\mathbf{d}_S^2$ term is dropped recovering Eq.~(\ref{supercurvatureF}).
However, we have found that the supercurvature calculated without the first term
in the $su(2/1)$ case did not gauge transform as $\mathcal{F} \rightarrow U\mathcal{F}U^\dagger$.
This can be understood as due to the non-vanishing of 
$\mathbf{d}_M^2$, and the mixing of $\mathbf{d}_M$ and $\mathcal{J}$ 
under gauge transformations as shown in Eq.~(\ref{dMJtrans}).
Indeed, in the current case, $\mathbf{d}_M^2$ is represented by the matrix
\begin{equation}
\mathbf{d}_M^2
\;=\; -i\eta^2 
%\;=\; -i
%\begin{bmatrix}
%\xi\xi^\dagger &  0 \\
%0 & \xi^\dagger\xi
%\end{bmatrix}
\;=\; -i
\begin{bmatrix}
\xi\xi^\dagger & 0 \\
0 & v^2
\end{bmatrix}
\;,
\end{equation}
which precisely cancels the problematic terms if added to Eq.~(\ref{Fsu21}):
\begin{eqnarray}
\lefteqn{\mathcal{F}+\mathbf{d}_M^2}\cr
& = &
i
\begin{bmatrix}
F_W-\frac{1}{\sqrt{3}}F_B\cdot\mathbf{1}_{2\times 2}
-2\hat{\phi}\hat{\phi}^{\dagger}
%+ \xi\xi^\dagger 
& 
\sqrt{2}\,D\hat{\phi} \\
\sqrt{2}(D\hat{\phi})^{\dagger}  & 
-\frac{2}{\sqrt{3}}F_B
-2\hat{\phi}^\dagger\hat{\phi} 
%+ v^2 
\\
\end{bmatrix}
\;.
\cr
&&
\label{Fsu21prime}
\end{eqnarray}
Thus, keeping the $\mathbf{d}_M^2$ term will give us a supercurvature which
the desired gauge transformation property.
However, we nevertheless argue that the $\mathbf{d}_M^2$ term should be dropped.
In the following, we calculate the action for 
$\mathcal{F}$ of Eq.~(\ref{Fsu21}), without the addition of $\mathbf{d}_M^2$,
and find that adding $\mathbf{d}_M^2$ will lead to inconsistencies 
which we would like to avoid.

Before continuing, we note that 
the situation is somewhat different in the $su(2/2)$ case to be considered in section~\ref{su22}.  
There, the matrix derivative is nilpotent in the sense that $(\mathbf{d}_M^2 X)=0$
for all $X\in su(2/2)$, and $\mathbf{d}_M^2 \propto \lambda^s_{15} = -\frac{1}{\sqrt{2}} \mathbf{1}_{4\times 4}$ which belongs to the center of the superalgebra.
Despite the mixing between $\mathbf{d}_M$ and $\mathcal{J}$
as given in Eq.~(\ref{dMJtrans}), $\mathbf{d}_M^2\propto \mathbf{1}_{4\times 4}$ 
is invariant under gauge transformations by itself
and can be dropped without any ill effects.
%Conversely, any constant multiple of $\lambda^s_{15}$, belonging to the center of $su(2/2)$, can be added to $\mathcal{F}$ without changing its algebraic properties.
%In a sense, one would only be adding `zeroes' to $\mathcal{F}$.
%However, we will see that those `zeroes' will affect whether the gauge symmetry is broken spontaneously or not.
%Thus, the question of whether $\mathbf{d}_M^2$ should be dropped (since it is a nilpotent operator) or not (since it is actually non-zero), or whether something else should be added to $\mathcal{F}$ is a subtle question.

%%%%%%%%%%%%%%%%%%%%%%%%%%%%%%%%%%%%%%%%%%%%%%%%%%%%%%%%%%%%%%%%%%%%%%%%%%%%%%%%%%%%%%%%%%%%%
%%%%%%%%%%%%%%%%%%%%%%%%%%%%%%%%%%%%%%%%%%%%%%%%%%%%%%%%%%%%%%%%%%%%%%%%%%%%%%%%%%%%%%%%%%%%%
\subsection{Inner Product and the Action}

In QCD, the action is given by the inner product of the gauge connection $F_G=dG+\frac{1}{2}[G,G]$ with itself:
\begin{equation}
S_{\mathrm{QCD}}
\;=\; \dfrac{1}{4}\langle F_G,F_G\rangle\;.
\end{equation}
Note that $F_G$ is an element of $\Omega(M)\otimes su(3)$.
For $a,b\in\Omega(M)$ and $X,Y\in su(3)$, 
the inner product of the elements 
$a\otimes X$ and $b\otimes Y$ is given by
\begin{equation}
\langle a\otimes X, b\otimes Y\rangle%_{\Omega(M)\otimes su(3)}
\;=\; \langle a, b\rangle_{\Omega(M)}\langle X, Y\rangle_{su(3)}
\;.
\end{equation}
The inner product in $su(3)$ is simply
\begin{equation}
\langle X, Y\rangle_{su(3)}
\;=\; \Tr(XY)\;.
\end{equation}
For $\Omega(M)=\oplus_{i=0}^4\Omega^i(M)$, the inner product
in each of the subspaces $\Omega^i(M)$ is given by
\begin{equation}
\langle a_i, b_i\rangle_{\Omega^i(M)} \;=\; \int *a_i\wedge b_i\;,
\end{equation}
where $*$ indicates the Hodge dual,
that is:
\begin{eqnarray}
& &
*1 
\;=\; \dfrac{1}{24}\,\varepsilon^{\kappa\lambda\mu\nu}
dx_\kappa\wedge dx_\lambda\wedge dx_\mu\wedge dx_\nu
\;=\; d^4 x\;,
\cr
& & 
*dx^\nu 
\;=\; \dfrac{1}{6}\,\varepsilon^{\kappa\lambda\mu\nu}
dx_\kappa\wedge dx_\lambda\wedge dx_\mu
\;,
\cr
& & 
*(dx^\mu\wedge dx^\nu)
\;=\; \dfrac{1}{2}\,\varepsilon^{\kappa\lambda\mu\nu}
dx_\kappa\wedge dx_\lambda
\;,
\cr
& &
*(dx^\lambda\wedge dx^\mu\wedge dx^\nu)
\;=\; \varepsilon^{\kappa\lambda\mu\nu}
dx_\kappa
\;, \vphantom{\Big|}
\cr
& &
*(dx^\kappa\wedge dx^\lambda\wedge dx^\mu\wedge dx^\nu)
\;=\; \varepsilon^{\kappa\lambda\mu\nu}
\;, \vphantom{\Big|}
\end{eqnarray}
where we assume the metric $g^{\mu\nu}=\mathrm{diag}(1,-1,-1,-1)$
and $\varepsilon_{0123}=1$.
For $a,b\in\Omega(M)$, we decompose
$a=\sum_i a_i$, $b=\sum_i b_i$, where $a_i,b_i\in\Omega^i(M)$,
and define
\begin{equation}
\langle a, b\rangle_{\Omega(M)}
\;=\; \sum_i \langle a_i, b_i\rangle_{\Omega^i(M)}
\;=\; \int \sum_i *a_i\wedge b_i\;.
\end{equation}
In the case of $F_G$, which is an $su(3)$ valued 2-form, we can write
(with a slight abuse of notation) 
\begin{equation}
\langle{F_G},{F_G}\rangle
\;=\; \int \Tr\Bigl[*F_G\wedge F_G\Bigr] 
%\;=\; \int d^4 x\;G_{\mu\nu}^a G^{a\mu\nu}
\;.
\end{equation}
Note that $\Tr[\,*F_G\wedge F_G\,]$
is a real scalar-valued 4-form which is gauge invariant.
This guarantees the Lorentz and gauge invariance of the action $S_{\mathrm{QCD}}$.

Let us rewrite the above action in terms of the field strength tensor.
Since $G = iG^a\lambda_a = i(G^a_\mu\,dx^\mu)\lambda_a$, we have
\begin{eqnarray}
F_G 
& = & dG + \dfrac{1}{2}[\,G,\,G\,] 
\;=\; \dfrac{i}{2}\left(G_{\mu\nu}^a\,dx^\mu\wedge dx^\nu\right)\lambda_a 
\;,\cr
*F_G 
& = & \dfrac{i}{4}\left(G_{\rho\sigma}^b\,\varepsilon^{\rho\sigma}{}_{\kappa\lambda}
dx^\kappa\wedge dx^\lambda\right)\lambda_b
\;,
\end{eqnarray}
with
\begin{equation}
G_{\mu\nu}^a 
\;=\; \partial_\mu G_\nu^a - \partial_\nu G_\mu^a + if^{abc}G_\mu^b G_\nu^c
\;,
\end{equation}
from which we find
\begin{eqnarray}
%\lefteqn{
*F_G\wedge F_G
%}
%\cr
%& = & -\dfrac{1}{8}
%\left(
%G_{\rho\sigma}^a G_{\mu\nu}^b\,\varepsilon^{\rho\sigma}{}_{\kappa\lambda}
%\,dx^\kappa\wedge dx^\lambda\wedge dx^\mu\wedge dx^\nu
%\right)
%\lambda_a\lambda_b
%\cr
%& = & -\dfrac{1}{8}
%\left(
%G_{\rho\sigma}^a G_{\mu\nu}^b\,\varepsilon^{\rho\sigma}{}_{\kappa\lambda}
%\varepsilon^{\kappa\lambda\mu\nu}\,d^4x
%\right)
%\lambda_a\lambda_b
%\cr
%& = & -\dfrac{1}{4}
%\left\{
%G_{\rho\sigma}^a G_{\mu\nu}^b
%\left(
%g^{\rho\mu}g^{\sigma\nu}-g^{\rho\nu}g^{\sigma\mu}
%\right)d^4x
%\right\}
%\lambda_a\lambda_b
%\cr
& = & -\dfrac{1}{2}
\left(
G_{\mu\nu}^{a} G^{b\mu\nu}
d^4x
\right)
\lambda_a\lambda_b\;.
\end{eqnarray}
Recalling the normalization $\Tr(\lambda_a\lambda_b)=2\delta_{ab}$ for the
$su(3)$ Gell-Mann matrices,
we obtain
\begin{equation}
\Tr\Bigl[*F_G\wedge F_G\Bigr]
\;=\; -G_{\mu\nu}^{a} G^{a\mu\nu}\,d^4x
\;,
\end{equation}
and therefore
\begin{equation}
S_{\mathrm{QCD}}
\;=\; -\dfrac{1}{4}\int G_{\mu\nu}^{a} G^{a\mu\nu}\,d^4x
\;,
\end{equation}
which is the more familiar form.

%
%\begin{eqnarray}
%F_G 
%& = & dG + \dfrac{1}{2}[\,G,\,G\,] \cr
%& = & d(G_a^\mu dx_\mu \lambda_a^{\phantom{\mu}}) 
%\cr
%& &
%+ \dfrac{1}{2}
%\left(G_a^\mu dx_\mu \right) \wedge
%\left(G_b^\nu dx_\nu \right)
%[\,\lambda_a^{\phantom{\mu}},\,\lambda_b^{\phantom{\nu}}\,]
%\cr
%& = & \dfrac{1}{2}
%\Bigl\{
%\left(\partial^\mu G_a^\nu - \partial^\nu G_a^\mu \right) dx_\mu\wedge dx_\nu
%\Bigr\}\lambda_a
%\cr
%& & +
%\dfrac{1}{2}
%\Bigl\{
%i f_{abc}G_a^\mu G_b^\nu dx_\mu\wedge dx_\nu
%\Bigr\} \lambda_c
%\cr
%& = & \dfrac{1}{2}\lambda_a G_a^{\mu\nu}dx_\mu\wedge dx_\nu
%\;,
%\cr
%%%
%*F_G
%& = & \dfrac{1}{2}\lambda_a G_a^{\mu\nu} *(dx_\mu\wedge dx_\nu)
%\cr
%& = & ...
%\end{eqnarray}

In an analogous fashion, let us write
the action for $\mathcal{F}$, Eq.~(\ref{Fsu21}), 
as
\begin{equation}
\mathcal{S} \;=\; \dfrac{1}{4}\langle\mathcal{F},\mathcal{F}\rangle_S\;.
\end{equation}
Note that $\mathcal{F}$ is an element of $\Omega(M)\otimes su(2/1)$.
For $a,b\in\Omega(M)$ and $X,Y\in su(2/1)$, 
the inner product of the elements $a\otimes X$ and $b\otimes Y$ is given by
\begin{eqnarray}
\lefteqn{
\langle a\otimes X, b\otimes Y\rangle_S%_{\Omega(M)\otimes su(2/1)}
}
\cr
& = & (-1)^{|b||X|}
\langle a, b\rangle_{\Omega(M)}
\langle X, Y\rangle_{su(2/1)}
\;.\vphantom{\bigg|}
\end{eqnarray}
The inner product in $\Omega(M)$ is the same as before.
For the inner product on $su(2/1)$, we define it to be
\begin{equation}
\langle X, Y\rangle_{su(2/1)}
\;=\; \Tr(XY)\;,
\end{equation}
just as in $su(3)$.  
Note that our use of a trace instead of a supertrace here in this definition has phenomenological significance. 
First, it would break any internal $SU(2/1)$ symmetry that may exist, but maintain 
the diagonal $SU(2)_L\times U(1)_Y$ gauge invariance.
But more significantly, it provides all the gauge boson kinetic terms with the correct signs,
and also demand $\phi$ to be bosonic (commuting) instead of fermionic (anti-commuting) to
prevent the $\phi$-dependent terms in the action from vanishing.
Had a supertrace been used, $\phi$ would have been required to be fermionic.\footnote{%
We have been unable to find any mention in the literature
of the connection between the choice of trace or supertrace
in the inner product with the bosonic or fermionic nature of $\phi$.
Perhaps, this is a new observation.}

Unlike $F_G$ of QCD, $\mathcal{F}$ is a linear combination
of $su(2/1)$ valued 0-, 1-, and 2-forms.
Let us write
\begin{equation}
\mathcal{F} \;=\; \sum_{i=0}^2\mathcal{F}^i\;,
\end{equation}
where $\mathcal{F}^i\in\Omega^i(M)\otimes su(2/1)$.
Then,
\begin{eqnarray}
\langle\mathcal{F},\mathcal{F}\rangle_S
& = & \int\Tr\left[
\sum_{i=0}^2 *\mathcal{F}^i\wedge\mathcal{F}^i
\right]
\cr
& = & \sum_{i=0}^2 \langle\mathcal{F}^i,\mathcal{F}^i\rangle_S
\;.
\end{eqnarray}
Explicitly, we have
\begin{eqnarray}
\mathcal{F}^0
& = & i
\begin{bmatrix}
(-2\hat{\phi}\hat{\phi}^\dagger + \xi\xi^\dagger) & \mathbf{0}_{2\times 1} \\
\mathbf{0}_{1\times 2} & (-2\hat{\phi}^\dagger\hat{\phi} + v^2)
\end{bmatrix}
\;,
\cr
%%%
\mathcal{F}^1
& = & \sqrt{2}i
\begin{bmatrix}
\mathbf{0}_{2\times 2}  & (D\hat{\phi}) \\
(D\hat{\phi})^\dagger   & 0
\end{bmatrix}
\;,
\cr
%%%
\mathcal{F}^2
& = & i
\begin{bmatrix}
F_W - \frac{1}{\sqrt{3}}\,F_B\cdot\mathbf{1}_{2\times 2} & \mathbf{0}_{2\times 1} \\
\mathbf{0}_{1\times 2} & -\frac{2}{\sqrt{3}}\,F_B
\end{bmatrix}
\;,
\end{eqnarray}
and
%

%%%%%%%%%%%%%%%%%%%%
\newpage

%%%
%%%
%%%
\begin{widetext}
\begin{eqnarray}
\Tr\Bigl[*\mathcal{F}^0\wedge\mathcal{F}^0\Bigr]
& = & 
-\Tr
\begin{bmatrix}
*(-2\hat{\phi}\hat{\phi}^\dagger + \xi\xi^\dagger) & \mathbf{0}_{2\times 1} \\
\mathbf{0}_{1\times 2} & *(-2\hat{\phi}^\dagger\hat{\phi} + v^2)
\end{bmatrix}
\wedge
\begin{bmatrix}
(-2\hat{\phi}\hat{\phi}^\dagger + \xi\xi^\dagger) & \mathbf{0}_{2\times 1} \\
\mathbf{0}_{1\times 2} & (-2\hat{\phi}^\dagger\hat{\phi} + v^2)
\end{bmatrix}
\cr
& = & -\Tr
\begin{bmatrix}
*(-2\hat{\phi}\hat{\phi}^\dagger + \xi\xi^\dagger)\wedge(-2\hat{\phi}\hat{\phi}^\dagger + \xi\xi^\dagger) & \mathbf{0}_{2\times 1} \\
\mathbf{0}_{1\times 2} & *(-2\hat{\phi}^\dagger\hat{\phi} + v^2)\wedge(-2\hat{\phi}^\dagger\hat{\phi} + v^2)
\end{bmatrix}
\cr
& = & -8\left(\hat{\phi}^\dagger\hat{\phi}-\dfrac{v^2}{2}\right)^2\;d^4x\;,
\cr
%%%
\Tr\Bigl[*\mathcal{F}^1\wedge\mathcal{F}^1\Bigr]
& = & +2\,\Tr
\begin{bmatrix}
\mathbf{0}_{2\times 2}  & *(D\hat{\phi}) \\
*(D\hat{\phi})^\dagger         & 0
\end{bmatrix}
\wedge
\begin{bmatrix}
\mathbf{0}_{2\times 2}  & (D\hat{\phi}) \\
(D\hat{\phi})^\dagger         & 0
\end{bmatrix}
\cr
& = & +2\,\Tr
\begin{bmatrix}
*(D\hat{\phi})\wedge (D\hat{\phi})^\dagger  & \mathbf{0}_{2\times 1} \\
\mathbf{0}_{1\times 2}          & *(D\hat{\phi})^\dagger\wedge (D\hat{\phi}) 
\end{bmatrix}
\cr
& = & +4(D_\mu\hat{\phi})^\dagger(D^\mu\hat{\phi})\;d^4x\;,
\cr
%%%
\Tr\Bigl[*\mathcal{F}^2\wedge\mathcal{F}^2\Bigr]
& = & -\Tr
\begin{bmatrix}
*F_W - \frac{1}{\sqrt{3}}*\!F_B\cdot\mathbf{1}_{2\times 2} & \mathbf{0}_{2\times 1} \\
\mathbf{0}_{1\times 2} & -\frac{2}{\sqrt{3}}*\!F_B
\end{bmatrix}
\wedge
\begin{bmatrix}
F_W - \frac{1}{\sqrt{3}}\,F_B\cdot\mathbf{1}_{2\times 2} & \mathbf{0}_{2\times 1} \\
\mathbf{0}_{1\times 2} & -\frac{2}{\sqrt{3}}\,F_B
\end{bmatrix}
\cr
& = & -\Tr
\begin{bmatrix}
\left(*F_W - \frac{1}{\sqrt{3}}*\!F_B\cdot\mathbf{1}_{2\times 2}\right)\wedge
\left(F_W - \frac{1}{\sqrt{3}}\,F_B\cdot\mathbf{1}_{2\times 2}\right)
& \mathbf{0}_{2\times 1} \\
\mathbf{0}_{1\times 2} &
\frac{4}{3}*\!F_B\wedge F_B
\end{bmatrix}
\cr
& = & -\Tr
\begin{bmatrix}
*F_W\wedge F_W - \frac{1}{\sqrt{3}}
\left(*F_B\wedge F_W + *F_W\wedge F_B 
\right)
+ \frac{1}{3}*\!F_B\wedge F_B\cdot\mathbf{1}_{2\times 2} &
\mathbf{0}_{2\times 1} \\
\mathbf{0}_{1\times 2} &
\frac{4}{3}*\!F_B\wedge F_B
\end{bmatrix}
\cr
& = & -\left(
 F_{W\mu\nu}^i F_W^{i \mu\nu}
+F_{B\mu\nu}^{\phantom{\mu\nu}} F_B^{\mu\nu}
\right)d^4 x\;.
\end{eqnarray}
\end{widetext}
%%%
%%%
%%%

%
%\begin{eqnarray}
%\lefteqn{*(D\phi)^\dagger\wedge (D\phi)}
%\cr
%& = & 
%*\bigl(D_\nu\phi_i^\dagger\;dx^\nu\bigr)\wedge
% \bigl(D_\sigma\phi_i        \;dx^\sigma\bigr)
%\cr
%& = & 
%\dfrac{1}{6}
%\bigl(D_\nu\phi_i^\dagger \varepsilon^{\kappa\lambda\mu\nu}dx_\kappa\wedge dx_\lambda\wedge dx_\mu
%\bigr)\wedge
%\bigl(D_\sigma\phi_i\;dx^\sigma\bigr)
%\cr
%& = & \dfrac{1}{6}
%\bigl(D_\nu\phi_i^\dagger\bigr)
%\bigl(D_\sigma\phi_i        \bigr)
%\varepsilon^{\kappa\lambda\mu\nu}
%dx_\kappa\wedge dx_\lambda\wedge dx_\mu\wedge dx^\sigma
%\cr
%& = & -\dfrac{1}{6}
%\bigl(D_\nu\phi_i^\dagger\bigr)
%\bigl(D_\sigma\phi_i        \bigr)
%\varepsilon^{\kappa\lambda\mu\nu}
%\varepsilon_{\kappa\lambda\mu}{}^{\sigma}
%d^4x
%\cr
%& = & 
%\bigl(D_\nu\phi_i^\dagger\bigr)
%\bigl(D^\sigma\phi_i        \bigr)
%g^{\nu\sigma}
%d^4x
%\cr
%& = &
% \bigl(D_\nu\phi_i^\dagger\bigr)
% \bigl(D^\nu\phi_i\bigr)
%d^4x
%\;.
%\end{eqnarray}
%

\noindent
Notice that when calculating the traces of the 0- and 1- form contributions,
one respectively needs to commute $\hat{\phi}$ through $\hat{\phi}^\dagger\hat{\phi}\hat{\phi}^\dagger$, and
$(D\hat{\phi})$ through $(D\hat{\phi})^\dagger$, and $\hat{\phi}$ must be bosonic to prevent
the trace from vanishing.

Putting everything together, we find:
\begin{eqnarray}
\mathcal{S}
& = & \dfrac{1}{4}\langle\mathcal{F},\mathcal{F}\rangle_S
\cr
& = & 
\int d^4x\bigg[
-\frac{1}{4}
\left(
 F_{W\mu\nu}^i F_W^{i \mu\nu}
+F_{B\mu\nu}^{\phantom{\mu\nu}} F_B^{\mu\nu}
\right)
\cr
& & \hspace{1.2cm}
+ (D_\mu\hat{\phi})^\dagger (D^\mu\hat{\phi})
- V(\hat{\phi},\hat{\phi}^\dagger) \bigg]
\;,
%\cr
%& &
\end{eqnarray}
where
\begin{eqnarray}
F_{W\mu\nu}^i
& = & \partial_{\mu}W^{i}_{\nu}-\partial_{\nu}W^{i}_{\mu}+2i\varepsilon^{ijk} W^{j}_{\mu}W^{k}_{\nu}\;,
\vphantom{\Big|}
\cr
F_{B\mu\nu}
& = & \partial_{\mu}B_{\nu}-\partial_{\nu}B_{\mu}\;,
\cr
D_{\mu}\hat{\phi}
& = & \partial_{\mu}\hat{\phi} 
-i\left(\bm{\tau\cdot W}_{\mu}\right)\hat{\phi} 
-\frac{i}{\sqrt{3}}B_{\mu}\hat{\phi}
\;.
\label{FWFB}
\end{eqnarray}
and
\begin{equation}
V(\hat{\phi},\hat{\phi}^\dagger)
\;=\;
2\left(\hat{\phi}^\dagger\hat{\phi}-\dfrac{v^2}{2}\right)^2
\;.
\end{equation}

Note that the above action is manifestly $SU(2)_L\times U(1)_Y$ gauge invariant as required,
even though $\mathcal{F}$ did not have the desired gauge transformation property.
Furthermore, the Higgs potential is minimized when
$\phi^\dagger\phi = v^2/2$, consistent with our assumption that
\begin{equation}
\langle\hat{\phi}\rangle \;=\; \dfrac{\xi}{\sqrt{2}}
\;,
\end{equation}
and that $\phi$ is the fluctuation around it.
Had we used $\mathcal{F}+\mathbf{d}_M^2$ instead of $\mathcal{F}$, 
the Higgs potential would have been $2(\hat{\phi}^\dagger\hat{\phi})^2$
and $\hat{\phi}$ would not have developed a VEV.
So for the consistency of the formalism, we will drop the $\mathbf{d}_M^2$
term from our supercurvature.

The resulting model is quite interesting in that spontaneous symmetry breaking is
built into the model from the beginning.
The $\phi$ field appearing in the superconnection is already the fluctuation around
a symmetry breaking vacuum.
In other words,
as emphasized in Refs.~\cite{Coquereaux:1990ev,Haussling:1991ns}, the superconnection $\mathcal{J}$ already `knows' about the breaking of the symmetry.
Eq.~(\ref{dMJtrans}) suggests that the development of the Higgs VEV can be interpreted
as due to the separation of the matrix derivative $\mathbf{d}_M$ from the
superconnection $\mathcal{J}$, which would be the consequence of the two
branes separating from each other.
Thus, the spontaneous breaking of the gauge symmetry could be the result of
the brane dynamics at work.

%%%%%%%%%%%%%%%%%%%%%%%%%%%%%%%%%%%%%%%%%%%%%%%%%%%%%%%%%%%%%%%%%%%%%%%%%%%%%%%%%%%%%%%%%%%%%
%%%%%%%%%%%%%%%%%%%%%%%%%%%%%%%%%%%%%%%%%%%%%%%%%%%%%%%%%%%%%%%%%%%%%%%%%%%%%%%%%%%%%%%%%%%%%
\subsection{Symmetry Breaking}

Let us analyze the model further.
We take 
\begin{equation}
\langle\hat{\phi}\rangle \;=\; \dfrac{\xi}{\sqrt{2}} \;=\; 
\dfrac{v}{\sqrt{2}}\begin{bmatrix} 0 \\ 1 \end{bmatrix}\;,
\end{equation}
so that
\begin{equation}
\langle\hat{\phi}^\dagger\tau^1\hat{\phi}\rangle \;=\;
\langle\hat{\phi}^\dagger\tau^2\hat{\phi}\rangle \;=\; 0\;,\qquad
\langle\hat{\phi}^\dagger\tau^3\hat{\phi}\rangle \;=\; -\dfrac{v^2}{2}\;.
\end{equation}
Then, the linear combinations
\begin{equation}
W^{\pm} \;=\; \dfrac{W^1\mp W^2}{\sqrt{2}}\;,\qquad
Z \;=\; \dfrac{\sqrt{3}\,W^3 - B}{2}
\label{WZdef}
\end{equation}
obtain masses given by
\begin{equation}
M_W \;=\; v\;,\qquad
M_Z \;=\; \dfrac{2v}{\sqrt{3}}\;,
\label{WZmasses}
\end{equation}
while the linear combination
\begin{equation}
A \;=\; \dfrac{W^3 + \sqrt{3}\,B}{2}
\label{photondef}
\end{equation}
remains massless and couples to 
\begin{equation}
\dfrac{\lambda_3^s  + \sqrt{3}\lambda_8^s}{2}
\;=\;
\begin{bmatrix}
0 &  0 &  0 \\
0 & -1 &  0 \\
0 &  0 & -1
\end{bmatrix}
\;=\; Q
\;,
\label{Qdef1}
\end{equation}
which corresponds to the electromagnetic charge.
Comparison with the SM will be made after the introduction of the
coupling constant in the next subsection.

%%%%%%%%%%%%%%%%%%%%%%%%%%%%%%%%%%%%%%%%%%%%%%%%%%%%%%%%%%%%%%%%%%%%%%%%%%%%%%%%%%%%%%%%%%%%%
\subsection{The Coupling Constants and the value of $\mathbf{sin^2}\bm{\theta_W}$}

\subsubsection{The Higgs Quartic Coupling and the Higgs Mass}

We introduce the $SU(2)_L$ coupling constant $g$ by rescaling the superconnection $\mathcal{J}$, the action $\mathcal{S}$, and the matrix-derivative matrix $\eta$ as
\begin{equation}
\mathcal{J} \;\rightarrow\; \dfrac{g}{2}\mathcal{J}
\;,
\qquad
\mathcal{S} \;\rightarrow\; \dfrac{g^2}{4}\mathcal{S}
\;,
\qquad
\eta \;\rightarrow\; \dfrac{g}{2}\eta
\;.
\label{CouplingIntroduction}
\end{equation}
Extracting the Lagrangian from the action, we find
\begin{eqnarray}
\mathcal{L}
& = & -\frac{1}{4}F_{W\,\mu\nu}^i\,F^{i\,\mu\nu}_W-\frac{1}{4}F_{B\,\mu\nu}F_B^{\mu\nu}
\cr
& &
+\left(D_{\mu}\hat{\phi}\right)^{\dagger}\left(D^{\mu}\hat{\phi}\right)
-\dfrac{g^2}{2}\left(\hat{\phi}^{\dagger}\hat{\phi}-\dfrac{v^2}{2}\right)^2
\;,
\label{RescaledL}
\end{eqnarray}
with
\begin{eqnarray}
F_{W\,\mu\nu}^i
& = & \partial_{\mu}W^{i}_{\nu}-\partial_{\nu}W^{i}_{\mu}+ig\varepsilon^{ijk} W^{j}_{\mu}W^{k}_{\nu}
\;, \vphantom{\bigg|}\cr
F_{B\,\mu\nu}
& = & \partial_{\mu}B_{\nu}-\partial_{\nu}B_{\mu}\;,\cr
D_{\mu}\hat{\phi}
& = & \partial_{\mu}\hat{\phi}
-i\dfrac{g}{2}\left(\bm{\tau\cdot W}_{\mu}\right)\hat{\phi}
-i\dfrac{g}{2\sqrt{3}}B_{\mu}\hat{\phi}
\;.
\cr & &
\label{FWFB2}
\end{eqnarray}
The Higgs quartic coupling, which we normalize to
\begin{equation}
V(\hat{\phi}^\dagger,\hat{\phi}) 
\;=\; \lambda(\hat{\phi}^\dagger\hat{\phi})^2 + \cdots
\;,
\end{equation}
can be read off from Eq.~(\ref{RescaledL}) to be
\begin{equation}
\lambda\;=\;\dfrac{g^2}{2}\;.
\label{lambdavalue}
\end{equation}
Rewriting the Higgs field $\hat{\phi}$ as
\begin{equation}
\hat{\phi} \;=\; 
\begin{bmatrix}
\pi^+ \\
\dfrac{v+ h + i\pi^0}{\sqrt{2}}
\end{bmatrix}
\;,
\end{equation}
we find
\begin{equation}
V(\hat{\phi}^\dagger,\hat{\phi}) 
\;=\; \dfrac{1}{2}(2\lambda v^2) h^2 + \cdots
\end{equation}
so the Higgs mass (at tree level) is
\begin{equation}
M_h \;=\; \sqrt{2\lambda}\,v \;=\; gv\;.
\label{MHiggs}
\end{equation}

%
%%%%%%%%%%%%%%%%%%%%%%%%%%%%%%%%%%%%%%%%%%%%%%%%%%%%%%%%%%%%%%%%%%%%%%%%%%%%%%%%%%%%%%%%%%%%
%%%%%%%%%%%%%%%%%%%%%%%%%%%%%%%%%%%%%%%%%%%%%%%%%%%%%%%%%%%%%%%%%%%%%%%%%%%%%%%%%%%%%%%%%%%%
\subsubsection{$\sin^2\theta_W$ from the coupling to Higgs}
Since the Higgs doublet $\hat{\phi}$ has hypercharge $+1$, we can make the identification
\begin{equation}
g'\;=\; \dfrac{g}{\sqrt{3}}\;,
\label{gprimedef}
\end{equation}
so that
\begin{equation}
D_{\mu}\hat{\phi}
\;=\; \partial_{\mu}\hat{\phi}
-i\dfrac{g}{2}\left(\bm{\tau\cdot W}_{\mu}\right)\hat{\phi}
-i\dfrac{g'}{2}B_{\mu}\hat{\phi}
\;.
\end{equation}
Also, after symmetry breaking the photon field $A$ couples to $(g/2)Q$, where
$Q$ is given in Eq.~(\ref{Qdef1}).  Therefore,
\begin{equation}
e \;=\; \dfrac{g}{2} \;=\; g\sin\theta_W\;.
\end{equation}
This relation can also be confirmed from the matching condition of the gauge couplings:
\begin{equation}
\dfrac{1}{e^2} \;=\; \dfrac{1}{g^2} + \dfrac{1}{g^{\prime 2}} \;=\; \dfrac{4}{g^2}\;.
\end{equation}
Thus, this formalism predicts
\begin{equation}
\sin^2\theta_W 
\;=\; \dfrac{g^{\prime 2}}{g^2+g^{\prime 2}} 
\;=\; \dfrac{e^2}{g^2} 
\;=\; \dfrac{1}{4}
\;.
\end{equation}
Using $g$ and $g'$, the masses of the $W$ and $Z$ we listed earlier in 
Eq.~(\ref{WZmasses}) can be written
\begin{equation}
M_W \;=\; \dfrac{gv}{2}\;,\qquad
M_Z \;=\; \dfrac{\sqrt{g^2+g^{\prime 2}}\,v}{2}\;,
\end{equation}
while the linear combinations of $W^3$ and $B$ that constitute the $Z$ and the photon
listed in Eqs.~(\ref{WZdef}) and (\ref{photondef}) 
can be written
\begin{eqnarray}
Z & = & W^3\cos\theta_W - B\sin\theta_W \;,\cr
A & = & W^3\sin\theta_W + B\cos\theta_W \;,
\end{eqnarray}
just as in the SM.  
Note that together with Eq.~(\ref{MHiggs}), this model predicts
\begin{equation}
\dfrac{M_h}{M_W} \;=\; 2\;.
\end{equation}
This is clearly problematic, since it leads to the prediction $M_h \approx 160\,\mathrm{GeV}$.
However, we could argue that these tree level predictions are those that are valid
at the scale at which the SM emerges from the underlying NCG theory.
The question is whether renormalization group running from this
emergence scale to the electroweak scale will cure the Higgs mass.
This will be addressed in the next subsection.

%%%%%%%%%%%%%%%%%%%%%%%%%%%%%%%%%%%%%%%%%%%%%%%%%%%%%%%%%%%%%%%%%%%%%%%%%%%%%%%%%%%%%%%%%%%%
%%%%%%%%%%%%%%%%%%%%%%%%%%%%%%%%%%%%%%%%%%%%%%%%%%%%%%%%%%%%%%%%%%%%%%%%%%%%%%%%%%%%%%%%%%%%
\subsubsection{$\sin^2\theta_W$ from the coupling to fermions}

The value of $\sin^2\theta_W$ can also be checked by looking at the gauge couplings of the fermions \cite{Neeman:1979wp}.
The interaction of the leptons with the $SU(2)_L\times U(1)_Y$ gauge bosons in the SM 
is given by  
\begin{eqnarray}\label{leptonic1}
-\mathcal{L}_\ell^{\mathrm{SM}}
& = &
\frac{g}{2}\bigl(\,\overline{\ell}_L\gamma^{\mu}\tau_i\ell_L\,\bigr)W_{\mu}^i
\nonumber\\
& & -\frac{g'}{2}
\bigl(\,
\overline{\ell}_L\gamma^{\mu}\ell_L + 2\,\overline{\ell}_R\gamma^{\mu}\ell_R\,
\bigr) B_{\mu},
\end{eqnarray}
where
\begin{equation}
\ell_L \;=\; \begin{bmatrix} \nu_{L} \\ \ell_L^-  \end{bmatrix}\;,\qquad
\ell_R^{\phantom{0}} \;=\; \ell_R^{-}\;.
\end{equation}
In the $su(2/1)$ embedding, we demand invariance only under
transformations generated by the even part of the superalgebra, which leads to the
interaction
\begin{equation}
\mathcal{L}_\ell^{\mathrm{even}}
\;=\;
-\frac{g}{2}\sum_{i=1,2,3,8}\left(\,\overline{\psi}\gamma^{\mu}\lambda^s_i\psi\,\right) J_\mu^i
\end{equation}
where 
\begin{equation}
\psi \;=\; 
\begin{bmatrix}
\ell_L \\ \ell_R
\end{bmatrix}
\;=\; 
\begin{bmatrix}
\nu_L \\ \ell_L^- \\ \ell_R^-
\end{bmatrix}
\end{equation}
is a triplet under $su(2/1)$, and $J^i_\mu$ is the vector
field associated with the 1-form $J^i=J^i_\mu dx^\mu$.
Recalling that $J^{1,2,3}$ are identified with $W^{1,2,3}=W_\mu^{1,2,3}dx^\mu$,
while $J^8$ is identified with $B=B_\mu dx^\mu$, 
%the $SU(2)_L\times U(1)_Y$ part of this interaction is given by
this interaction can be written out as
\begin{eqnarray}\label{leptonic2}
-\mathcal{L}_\ell^{\mathrm{even}}
%& = &
%\frac{g}{2}\sum_{i=1,2,3,8}\left(\,\overline{\psi}\gamma^{\mu}\lambda^s_i\psi\,\right) J_\mu^i
%\cr
& = & \frac{g}{2}
\left(\,\overline{\ell}_L\gamma^{\mu}\tau_i\ell_L\,\right) W_\mu^i\nonumber\\
& & -\frac{g}{2\sqrt{3}}
\left(\,\overline{\ell}_L\gamma^{\mu}\ell_L+2\,\overline{\ell}_R\gamma^{\mu}\ell_R\,\right) B_{\mu}
\;.
%\cr
%& &
\end{eqnarray}
Comparing Eqs.~(\ref{leptonic1}) and (\ref{leptonic2}) we reproduce Eq.~(\ref{gprimedef}).

Here the requirement of $SU(2)_L\times U(1)_Y$ gauge invariance was used to determine the couplings between the even part of the $su(2/1)$ superconnection and the fermion fields.
The couplings between the odd part of the superconnection, namely the Higgs doublet $\phi$,
and the fermion fields must reproduce the SM Yukawa couplings.
How these can be accommodated within the superconnection formalism will be discussed
in section~\ref{Fermions}.

%%%%%%%%%%%%%%%%%%%%%%%%%%%%%%%%%%%%%%%%%%%%%%%%%%%%%%%%%%%%%%%%%%%%%%%%%%%%%%%%%%%%%%%%%%%%
%%%%%%%%%%%%%%%%%%%%%%%%%%%%%%%%%%%%%%%%%%%%%%%%%%%%%%%%%%%%%%%%%%%%%%%%%%%%%%%%%%%%%%%%%%%%
\subsection{The Emergence Scale and the Higgs boson mass from $\bm{su(2/1)}$}

Let us now address the prediction for the Higgs mass including radiative corrections.
In what follows we assume the relation
\begin{equation}
\dfrac{M_h^2}{M_W^2} \;=\; \dfrac{8\lambda}{g^2}
\end{equation}
to be invariant under renormalization group flow, and follow the evolutions of the
coupling constants $\lambda$ and $g$ assuming the SM particle content below the
scale at which the SM emerges from some underlying NCG theory.

The renormalization group equation (RGE) for $\lambda$
is coupled to those of the fermion Yukawa couplings, of which we only take  
that of the top quark to be relevant.
The RGE's for $\lambda$ and the top Yukawa coupling $h_t$ are \cite{Langacker:2010zza}
\begin{eqnarray}\label{RGLang}
\mu\frac{d h_t}{d\mu}
& = & \frac{h_t}{(4\pi)^2}
\left[\,
  \frac{9}{2}h_t^2-\left(\frac{17}{12}g'^2+\frac{9}{4}g^2+8g_s^2\right)
\right]\;,\cr
\mu\frac{d \lambda}{d\mu}
& = & \frac{1}{(4\pi)^2}
\bigg[
  \Bigl\{12 h_t^2 - \left(3 g'^2+9 g^2\right)
  \Bigr\} \lambda - 6 h_t^4
\cr
& & \qquad\quad + 24\lambda^2 + \frac{3}{8}\left(g'^4+2g'^2 g^2+3g^4\right)
\bigg]\;,
\cr
& &
\end{eqnarray}
where $g'$, $g$, and $g_s$ are the $U(1)_Y$, $SU(2)_L$ and $SU(3)_c$ coupling constants,  respectively,
and the top Yukawa coupling $h_t$ is normalized to
\begin{equation}
m_t \;=\; \dfrac{h_t v}{\sqrt{2}}\;.
\end{equation}
The most recent value of the top quark mass is
$m_t=173.21\pm 0.51\pm 0.71\;\mbox{GeV}$ \cite{Agashe:2014kda}.

\begin{figure}[ht] 
\rightline{
\includegraphics[width=0.50\textwidth]{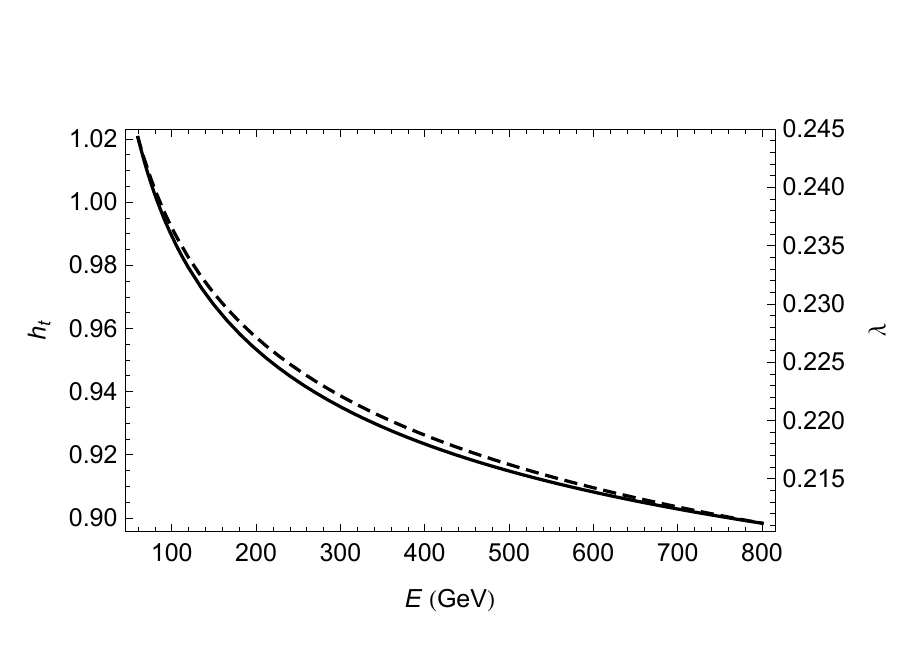}
}
\vspace{-2em}
\caption{The behavior of the top Yukawa coupling ($h_t$), which is represented as the dashed line, and the Higgs quartic coupling ($\lambda$).}
\label{couplings1}
\end{figure}

We will follow Ref.~\cite{Neeman:1979wp} to find the boundary condition on $\lambda$.
To fix the scale of emergence $\Lambda_s$ of the $su(2/1)$ structure, we look for the scale at which the relation
$g=\sqrt{3}g'$ (\textit{i.e.} $\sin^2{\theta_W}=1/4$) holds. 
We use the 1-loop relations
\begin{equation}
\frac{1}{[g_i(\Lambda_s)]^2}\,=\,\frac{1}{[g_i(\Lambda_0)]^2}-2 b_i\ln\frac{\Lambda_s}{\Lambda_0} 
\qquad (i=1,2,3)
\end{equation}
where  $g_1=g'$, $g_2=g$, $g_3=g_s$, and the respective constants $b_i$ read as \cite{Jones:1981we}:
\begin{eqnarray}
b_1 & = &  \frac{1}{16\pi^2}\left(\frac{20\,n_f}{9}+\frac{n_H}{6}\right)\;,\cr
b_2 & = & -\frac{1}{16\pi^2}\left(-\frac{4\,n_f}{3}-\frac{n_H}{6}+\frac{22}{3}\right)\;,\cr
b_3 & = & -\frac{1}{16\pi^2}\left(-\frac{4\,n_f}{3}+11\right)\;.
\end{eqnarray}
We will only need to look at the runnings of $g_1$ and $g_2$ to find $\Lambda_s$, but will
also need to look at the running of $g_3$ in the RGE's listed in Eq.~(\ref{RGLang}).
Setting the number of fermion families to $n_f=3$, and the number of Higgs doublets to $n_H=1$,
we have
\begin{equation}
b_1 = \dfrac{1}{16\pi^2}\left(\dfrac{41}{6}\right),\quad
b_2 = \dfrac{1}{16\pi^2}\left(-\dfrac{19}{6}\right),\quad
b_3 = -\dfrac{7}{16\pi^2}\,.
\label{b123}
\end{equation}
The numerical values ($\overline{\mathrm{MS}}$) of the coupling constants
at the scale $\Lambda_0=M_Z$ are given in Ref.~\cite{Agashe:2014kda} as
$\alpha_1^{-1}(M_Z)=98.36$,
$\alpha_2^{-1}(M_Z)=29.58$, and
$\alpha_3^{-1}(M_Z)=8.45$, 
where $\alpha_i^{-1}=4\pi/g_i^2$.
Note that $\alpha_1^{-1}(M_Z)/\alpha_2^{-1}(M_Z) = 3.325$.
Running this ratio up to where $\alpha_1^{-1}(\Lambda_s)/\alpha_2^{-1}(\Lambda_s) = 3$, 
we find the scale of emergence to be
\begin{eqnarray}
\Lambda_s \;\simeq\; 4\,\mathrm{TeV}\;.
\end{eqnarray}
Since this is the energy where the structure associated with $su(2/1)$ emerges, the constraint Eq.~(\ref{lambdavalue}), $\lambda=g_2^2/2$, is also expected to hold at this energy. This predicts the Higgs mass value as $M_h =2 M_W \simeq 160$ GeV, which should also be interpreted as the Higgs mass value at $\Lambda_s\simeq4$ TeV.
Using Eq.~(\ref{RGLang}) with the boundary conditions $\lambda=g_2^2/2$ at 4~TeV and
$h_t=\sqrt{2}m_t/v$ at $M_Z$, we find $\lambda(M_Z)\simeq0.24$ (Fig. \ref{couplings1}) and 
\begin{eqnarray} \label{higgsmass}
M_h (M_Z) \;\simeq\; 170\,\mathrm{GeV}\;.
\end{eqnarray}
Thus, 
the predicted Higgs mass is incorrect and it cannot be remedied within the $su(2/1)$ superconnection framework. 
However, as we will show in the next section, lowering it to $126$ GeV can be realized in the $su(2/2)$ extension which predicts the emergence of the left-right symmetric model (already broken to the SM) at the TeV scale.

%%%%%%%%%%%%%%%%%%%%%%%%%
%%%%%%%%%%%%%%%%%%%%%%%%%%%%%%

%%%%%%%%%%%%%%%%%%%%%%%%%%%%%%%%%%%%%%%%%%%%%%%%%%%%%%%%%%%%%%%%%%%%%%%%%%%%%%%%%%%%%%%%%%%%
%%%%%%%%%%%%%%%%%%%%%%%%%%%%%%%%%%%%%%%%%%%%%%%%%%%%%%%%%%%%%%%%%%%%%%%%%%%%%%%%%%%%%%%%%%%%
\subsection{$\bm{su(2/1)}$ summary}

In this section, we have reviewed the $su(2/1)$ superconnection formalism into which
the SM is embedded in some detail, including some sticking points,
and looked at its predictions. 
The embedding enforces the relations $\sin^2\theta_W = 1/4$ and $\lambda=g^2/2$.
The first is valid at the scale of $\Lambda_s \sim 4\,\mathrm{TeV}$, which we 
interpret as the scale at which the $su(2/1)$ formalism emerges from a
yet unknown underlying NCG theory. Assuming the second relation is also valid at that scale,
we obtain $M_h\sim 170\,\mathrm{GeV}$ as the prediction for the Higgs mass.

Though the Higgs mass prediction is clearly problematic, the formalism has a couple of
interesting and attractive features which deserve attention.
First and foremost, we note that the generalized exterior derivative in the discrete direction,
\textit{i.e.} the matrix derivative, carries in it information on the Higgs VEV
$\langle\hat{\phi}\rangle=\xi/\sqrt{2}$.  
The 0-form field $\phi$ which appears in the superconnection $\mathcal{J}$ is
the fluctuation of $\hat{\phi}$ around this VEV:
$\hat{\phi} = \langle\hat{\phi}\rangle + \phi$.
Thus, the $SU(2)_L\times U(1)_Y$ gauge symmetry is spontaneously broken to $U(1)_{em}$
from the beginning, and there is, in fact, no need to shift the field from $\phi$ 
to $\hat{\phi}$, only to shift it back again to account for the physical degrees of freedom
in the broken phase.
Since the matrix derivative is necessitated by the existence of the discrete extra dimension,
one could argue that it is the dynamical separation of the two branes itself that
broke the symmetry and shifted the Higgs field. 
In other words, it is the dynamics of the discrete geometry of the two branes that is responsible for
spontaneous symmetry breaking, and the Higgs field is just one manifestation of the
phenomenon.
This is in contrast to the usual SM point of view in which
the Higgs dynamics is independent of any dynamics of the background geometry.

Second, since the formalism fails to predict the correct Higgs mass, it begs an
extension to a formalism that would.
This can be viewed as an advantage instead of a drawback of the model
since it points us to new directions.
As we \cite{Aydemir:2013zua}, and other authors have pointed out 
\cite{Chamseddine:2012sw,Chamseddine:2013rta},
an additional singlet scalar degree of freedom in the
Higgs sector would mix with the Higgs boson to brings its mass down,
and a simple way to introduce such a degree of freedom would be to 
extend the SM electroweak gauge group
to $SU(2)_L\times SU(2)_R\times U(1)_{B-L}$.
In the superconnection formalism, this gauge group can be embedded into
$su(2/2)$.
Thus, despite the absence of our understanding on how these structures
arise from an underlying NCG theory,
the data already suggest an extension from $su(2/1)$ to $su(2/2)$.

%%%%%%%%%%%%%%%%%%%%%%%%%%%%%%%%%%%%%%%%%%%%%%%%%%%%%%%%%%%%%%%%%%%%%%%%%%%%%%%%%%%%%%%%%%%%
%%%%%%%%%%%%%%%%%%%%%%%%%%%%%%%%%%%%%%%%%%%%%%%%%%%%%%%%%%%%%%%%%%%%%%%%%%%%%%%%%%%%%%%%%%%%
%%%%%%%%%%%%%%%%%%%%%%%%%%%%%%%%%%%%%%%%%%%%%%%%%%%%%%%%%%%%%%%%%%%%%%%%%%%%%%%%%%%%%%%%%%%%
\section{Embedding of the Left-Right Symmetric Model into $\bm{su(2/2)}$}\label{su22}

Given the limitations of the $su(2/1)$ superconnection model outlined above, here
we explore the possibility of using an $su(2/2)$ superconnection to embed the SM electroweak gauge fields and the Higgs. 
The gauge group embedded will be $SU(2)_L\times SU(2)_R\times U(1)_{B-L}$ with the same
gauge couplings for $SU(2)_L$ and $SU(2)_R$.
Thus, we are working with the left-right symmetric model (LRSM) \cite{Mohapatra:1974hk,Mohapatra:1974gc,Senjanovic:1975rk,Mohapatra:1979ia,Mohapatra:1980yp,Chang:1983fu,Gunion:1989in,Deshpande:1990ip,Brahmachari:1991np,Czakon:1999ue,Duka:1999uc,Dev:2013oxa,Bambhaniya:2014cia,Bambhaniya:2013wza}.
%with seven gauge bosons: $W_L^\pm$, $W_R^\pm$, $\gamma$, $Z$, and $Z'$.
We will assume the breaking of $SU(2)_L\times SU(2)_R\times U(1)_{B-L}$ down to $U(1)_{em}$ so that the electromagnetic charge $Q$ will be
given by
\begin{equation}
Q \;=\; I_{L3} + I_{R3} + \dfrac{B-L}{2}\;,
\end{equation}
where $I_{L3}$ and $I_{R3}$ are respectively the third components of the left- and right-handed isospins.

We follow the same route as in the previous section:
we will first work out the superconnection $\mathcal{J}$ of the model and find that an
bi-doublet scalar field $\Phi$
appears in the odd part.  The supercurvature $\mathcal{F}$ and action $\mathcal{S}$ 
are derived from the superconnection $\mathcal{J}$,
and in the process it is discovered that the matrix derivative $\mathbf{d}_M$ 
in this case can be made nilpotent, and as a consequence, the supercurvature $\mathcal{F}$ has a simple gauge transformation property
which guarantees the gauge invariance of the action $\mathcal{S}$.
To achieve the breaking of $SU(2)_L\times SU(2)_R\times U(1)_{B-L}$ down to $U(1)_{em}$,
two complex triplet scalar fields $\Delta_{L}$ and $\Delta_{R}$ are introduced as matter fields living, respectively, on the left- and right-handed branes.
We find that $\Delta_{L,R}$ can naturally be placed in an $su(2/2)$ representation,
suggesting that their introduction is not entirely ad-hoc.
The formalism predicts the ratios of gauge coupling constants, and thus the value
of $\sin^2\theta_W$, and the self-couplings of the $\Phi$, but not the various couplings
involving $\Delta_{L,R}$ in the most generic Higgs potential \cite{Deshpande:1990ip}.
However, this is sufficient to fix the scale $\Lambda_s$ at which
the structure is expected to emerge from an underlying NCG theory, and also
suggest that the measured Higgs mass can be accommodated within the framework.

%%%%%%%%%%%%%%%%%%%%%%%%%%%%%%%%%%%%%%%%%%%%%%%%%%%%%%%%%%%%%%%%%%%%%%%%%%%%%%%%%%%%%%%%%%%%
\subsection{$\bm{su(2/2)}$ Superalgebra}

The superalgebra $su(2/2)$ consists of $4\times 4$ supertraceless Hermitian matrices, 
in which the even and odd parts are $2\times 2$ submatrices.
The basis for $su(2/2)$ can be chosen to be
%
%\begin{widetext}
\begin{eqnarray}\label{generators}
\lambda^s_1 & = &
\begin{bmatrix}
                0 & 1 & 0 & 0 \\
                1 & 0 & 0 & 0 \\
                0 & 0 & 0 & 0 \\
                0 & 0 & 0 & 0 \\
\end{bmatrix}
,\;
\lambda^s_2 \,=\, 
\begin{bmatrix}
                0 & -i & 0 & 0 \\
                i & 0 & 0 & 0 \\
                0 & 0 & 0 & 0 \\
                0 & 0 & 0 & 0 \\
\end{bmatrix}
,\;
\cr
\lambda^s_3 & = & 
\begin{bmatrix}
                1 & 0 & 0 & 0 \\
                0 & -1 & 0 & 0 \\
                0 & 0 & 0 & 0 \\
                0 & 0 & 0 & 0 \\
\end{bmatrix}
,\;
\lambda^s_4 \,=\, 
\begin{bmatrix}
                0 & 0 & 1 & 0 \\
                0 & 0 & 0 & 0 \\
                1 & 0 & 0 & 0 \\
                0 & 0 & 0 & 0 \\
\end{bmatrix}
,\;
\cr
\lambda^s_5 & = & 
\begin{bmatrix}
                0 & 0 & -i & 0 \\
                0 & 0 & 0 & 0 \\
                i & 0 & 0 & 0 \\
                0 & 0 & 0 & 0 \\
\end{bmatrix}
,\;
\lambda^s_6 \,=\, 
\begin{bmatrix}
                0 & 0 & 0 & 0 \\
                0 & 0 & 1 & 0 \\
                0 & 1 & 0 & 0 \\
                0 & 0 & 0 & 0 \\
\end{bmatrix}
,\;
\cr
\lambda^s_7 & = & 
\begin{bmatrix}
                0 & 0 & 0 & 0 \\
                0 & 0 & -i & 0 \\
                0 & i & 0 & 0 \\
                0 & 0 & 0 & 0 \\
\end{bmatrix}
,\;
\lambda^s_8 \,=\, 
\begin{bmatrix}
                0 & 0 & 0 & 0 \\
                0 & 0 & 0 & 0 \\
                0 & 0 & 1 & 0 \\
                0 & 0 & 0 & -1 \\
\end{bmatrix}
,\;
\cr
\lambda^s_9 & = & 
\begin{bmatrix}
                0 & 0 & 0 & 1 \\
                0 & 0 & 0 & 0 \\
                0 & 0 & 0 & 0 \\
                1 & 0 & 0 & 0 \\
\end{bmatrix}
,\;
\lambda^s_{10} \,=\, 
\begin{bmatrix}
                0 & 0 & 0 & -i \\
                0 & 0 & 0 & 0 \\
                0 & 0 & 0 & 0 \\
                i & 0 & 0 & 0 \\
\end{bmatrix}
,\;
\cr
\lambda^s_{11} & = & 
\begin{bmatrix}
                0 & 0 & 0 & 0 \\
                0 & 0 & 0 & 1 \\
                0 & 0 & 0 & 0 \\
                0 & 1 & 0 & 0 \\
\end{bmatrix}
,\;
\lambda^s_{12} \,=\, 
\begin{bmatrix}
                0 & 0 & 0 & 0 \\
                0 & 0 & 0 & -i \\
                0 & 0 & 0 & 0 \\
                0 & i & 0 & 0 \\
\end{bmatrix}
,\;
\cr
\lambda^s_{13} & = & 
\begin{bmatrix}
                0 & 0 & 0 & 0 \\
                0 & 0 & 0 & 0 \\
                0 & 0 & 0 & 1 \\
                0 & 0 & 1 & 0 \\
\end{bmatrix}
,\;
\lambda^s_{14} \,=\, 
\begin{bmatrix}
                0 & 0 & 0 & 0 \\
                0 & 0 & 0 & 0 \\
                0 & 0 & 0 & -i \\
                0 & 0 & i & 0 \\
\end{bmatrix}
,\;
\cr
\lambda^s_{15} & = &
-\frac{1}{\sqrt{2}}
\begin{bmatrix}
                1 & 0 & 0 & 0 \\
                0 & 1 & 0 & 0 \\
                0 & 0 & 1 & 0 \\
                0 & 0 & 0 & 1 \\
\end{bmatrix}
.\;
\end{eqnarray}
%\end{widetext}
%
These matrices are normalized to satisfy the orthogonality condition
\begin{eqnarray}\label{orthogonality}
\Tr(\lambda^s_a\lambda^s_b)\,=\,2\delta_{ab}\,,\quad\mbox{where}\;\;a,b\,=\,1\,,\,2\,,\cdots \,15\,.
\end{eqnarray}
The even elements of the superalgebra are spanned by
$\lambda^s_1$, $\lambda^s_2$, $\lambda^s_3$, $\lambda^s_8$, $\lambda^s_{13}$, $\lambda^s_{14}$,
$\lambda^s_{15}$, while the odd elements are spanned by
$\lambda^s_4$, $\lambda^s_5$, $\lambda^s_6$, $\lambda^s_7$, $\lambda^s_{9}$, $\lambda^s_{10}$,
$\lambda^s_{11}$, $\lambda^s_{12}$.
The only matrix different from its $su(4)$ counterpart is $\lambda_{15}^s$ due to the
supertracelessness condition.
These matrices close under commutation and anti-commutation relations as
\begin{eqnarray}
\dfrac{1}{i}[\lambda^s_i,\lambda^s_j] & = & 2\,f_{ijk}^{\phantom{s}}\lambda^s_k\;,\cr
\dfrac{1}{i}[\lambda^s_i,\lambda^s_m] & = & 2\,f_{iml}^{\phantom{s}}\lambda^s_l\;,\cr
\{\lambda^s_m,\lambda^s_n\} & = & 2\,d_{mnk}^{\phantom{s}}\lambda^s_k -\sqrt{2}\,\delta_{mn}^{\phantom{s}}\lambda^s_{15}\;,
\end{eqnarray}
where $i,j,k$ denote the even indices $1,2,3,8,13,14,15$, while $m,n,l$ denote the
odd indices $4,5,6,7,9,10,11,12$.
The $f$'s and $d$'s are the same as the structure constants for $su(4)$:
\begin{eqnarray}
\dfrac{1}{i}[\lambda_a,\lambda_b] & = & 2\,f_{abc}\lambda_c\;,\cr
\{\lambda_a,\lambda_b\} & = & 2\,d_{abc}\lambda_c + \delta_{ab}\;,
\end{eqnarray}
where $\lambda_a = \lambda^s_a$ for $a=1,2,\cdots 14$, and
\begin{equation}
\lambda_{15}
\;=\; \frac{1}{\sqrt{2}}
\begin{bmatrix}
                1 & 0 & 0 & 0 \\
                0 & 1 & 0 & 0 \\
                0 & 0 & -1 & 0 \\
                0 & 0 & 0 & -1 \\
\end{bmatrix}
\;.
\end{equation}

%%%%%%%%%%%%%%%%%%%%%%%%%%%%%%%%%%%%%%%%%%%%%%%%%%%%%%%%%%%%%%%%%%%%%%%%%%%
\subsection{$\bm{su(2/2)}$ Superconnection}

The superconnection $\mathcal{J}$ of this model is expressed as 
$\mathcal{J}=iJ_a^{\phantom{s}}\lambda_a^s$, where $a=1,2,\cdots 15$. We make the identifications\footnote{%
We switch from subscripts to superscripts for the $SU(2)$ indices to make room for the subscripts $L$ and $R$.
}
\begin{equation}
\begin{array}{lcllcl}
J_{1,2,3} & = & W_L^{1,2,3}\,,\qquad &
J_{13,14,8} & = & W_R^{1,2,3}\,,
\cr
J_4-iJ_5 & = & \sqrt{2}\phi_1^{0}\,,\qquad &
J_4+iJ_5 & = & \sqrt{2}\phi_1^{0\ast}\,,
\cr
J_6-iJ_7 & = & \sqrt{2}\phi_1^{-}\,,\qquad &
J_6+iJ_7 & = & \sqrt{2}\phi_1^{+}\,,
\cr
J_9-iJ_{10} & = & \sqrt{2}\phi_2^{+}\,,\qquad &
J_9+iJ_{10} & = & \sqrt{2}\phi_2^{-}\,,
\cr
J_{11}-iJ_{12} & = & \sqrt{2}\phi_2^{0}\,,\quad &
J_{11}+iJ_{12} & = & \sqrt{2}\phi_2^{0\ast}\,,
\cr
J_{15} & = & W_{BL}\,,& & 
\end{array}
\end{equation}
where $W_L^i = W_L^{i\mu}dx_\mu$, $W_R^i = W_R^{i\mu}dx_\mu$, and $W_{BL} = W_{BL}^{\mu}dx_\mu$ are 1-form fields while the $\phi$'s are 0-form fields, corresponding,
respectively, to the $SU(2)_L\times SU(2)_R\times U(1)_{B-L}$ gauge fields and 
the bi-doublet Higgs field:\footnote{%
Here, we use the subscripts 1 and 2 to label the two $SU(2)_L$ doublets
embedded in $\Phi$.
Some papers in the literature use subscripts to label
$SU(2)_R$ doublets, \textit{e.g.} Refs.~\cite{Gunion:1989in,Deshpande:1990ip},
so care is necessary when comparing formulae.
}
\begin{equation}\label{bidoublet}
\Phi 
\;=\;
\begin{bmatrix}
\phi_1^{0\,} & \phi_2^{+} \\
\phi_1^{-}   & \phi_2^{0\,}
\end{bmatrix}
\;.
\end{equation}
The resulting superconnection has the form
\begin{eqnarray}\label{superconnection2}
\mathcal{J}
& = & i
\begin{bmatrix}
                \;\mathcal{W}_L-\frac{1}{\sqrt{2}}W_{BL}\cdot \mathbf{1}_{2\times 2} & \sqrt{2}\,\Phi \\
                \sqrt{2}\,\Phi^{\dagger} & \mathcal{W}_R-\frac{1}{\sqrt{2}}W_{BL}\cdot \mathbf{1}_{2\times 2}\; \\
\end{bmatrix}
\;,
\cr
& &
\end{eqnarray}
where
\begin{equation}
\mathcal{W}_L^{\phantom{i}} \;=\; W^i_{L}\,\tau^i\;,\qquad 
\mathcal{W}_R^{\phantom{i}} \;=\; W^i_{R}\,\tau^i\;.
\end{equation}
In this assignment, we have assumed that the ordering of the rows of the $su(2/2)$ matrix,
from top to bottom, correspond to
left-handed isospin up, left-handed isospin down, right-handed isospin up, then right-handed isospin down.
So the leptons will be placed in a 4 dimensional representation of $su(2/2)$ in the order
\begin{equation}
\psi \;=\;
\begin{bmatrix}
\ell_L \\
%\hdotsfor[0.5]{1} \\
\ell_R 
\end{bmatrix}
\;=\;
\begin{bmatrix}
\nu_L \\ \ell_L^- \\ 
%\hdotsfor[0.5]{1} \\
\nu_R \\ \ell_R^-
\end{bmatrix}
\;.
\label{quartet}
\end{equation}

%%%%%%%%%%%%%%%%%%%%%%%%%%%%%%%%%%%%%%%%%%%%%%%%%%%%%%%%%%%%%%%%%%%%%%%%%%%
\subsection{$\bm{su(2/2)}$ Supercurvature}

As reviewed in the discussion of the $su(2/1)$ case,
the supercurvature is given by
\begin{equation}
\mathcal{F} \;=\; (\mathbf{d}\mathcal{J}) + (\mathbf{d}_M\mathcal{J}) + \dfrac{1}{2}[\,\mathcal{J},\,\mathcal{J}\,]_S\;,
\end{equation}
where the $\mathbf{d}_M^2$ term has been assumed not to contribute and has been dropped.

The $(\mathbf{d}\mathcal{J})$ term is
\begin{eqnarray}
\lefteqn{(\mathbf{d}\mathcal{J})}
\cr
& = & i
\begin{bmatrix}
  \;d\mathcal{W}_L-\frac{1}{\sqrt{2}}\,dW_{BL}\cdot \mathbf{1}_{2\times 2} & \sqrt{2}\,d\Phi \\
  \sqrt{2}\,d\Phi^{\dagger} & d\mathcal{W}_R-\frac{1}{\sqrt{2}}\,dW_{BL}\cdot \mathbf{1}_{2\times 2}\; \\
\end{bmatrix}
\;.
\cr & & 
\end{eqnarray}
The matrix derivative is defined with the $4\times 4$ $\eta$-matrix given by
\begin{equation}
\eta \;=\;
\begin{bmatrix}
\mathbf{0}_{2\times 2} & \zeta \\
\zeta^\dagger          & \mathbf{0}_{2\times 2}
\end{bmatrix}
\;,
\end{equation}
where $\zeta$ is a multiple of a $2\times 2$ unitary matrix, that is
$\zeta^\dagger\zeta = \zeta\zeta^\dagger = v^2\mathbf{1}_{2\times 2}$.
This time, the matrix derivative is nilpotent: $(\mathbf{d}_M^2 X)=0$ for all $X\in su(2/2)$.
We find
\begin{eqnarray}
\lefteqn{(\mathbf{d}_M\mathcal{J})} 
\cr
& = & i[\,\eta,\,\mathcal{J}\,]_S \cr
& = & i
\begin{bmatrix}
-\sqrt{2}(\zeta\Phi^\dagger + \Phi\zeta^\dagger) &
+i(\mathcal{W}_L \zeta-\zeta \mathcal{W}_R) \\		%Chen_Correction
-i(\zeta^\dagger \mathcal{W}_L-\mathcal{W}_R \zeta^\dagger) &	%Chen_Correction
-\sqrt{2}(\zeta^\dagger\Phi+\Phi^\dagger\zeta)
\end{bmatrix}
\;.
\cr
& &
\end{eqnarray}
To calculate the supercommutator of $\mathcal{J}$ with itself, we separate it into the 1-0 and 0-1 parts as before:

\begin{eqnarray}
\mathcal{J}
& = & \underbrace{i
\begin{bmatrix}
                \;\mathcal{W}_L-\frac{1}{\sqrt{2}}W_{BL}\cdot \mathbf{1}_{2\times 2} & \mathbf{0}_{2\times 2} \\
                \mathbf{0}_{2\times 2} & \mathcal{W}_R-\frac{1}{\sqrt{2}}W_{BL}\cdot \mathbf{1}_{2\times 2}\; \\
\end{bmatrix}}_{\displaystyle \mathcal{J}_{10}}
\cr
& &
+ \underbrace{i
\begin{bmatrix}
                \;\mathbf{0}_{2\times 2} & \sqrt{2}\,\Phi\;  \\
                \;\sqrt{2}\,\Phi^{\dagger} & \mathbf{0}_{2\times 2} \vphantom{\Big|}\; \\
\end{bmatrix}}_{\displaystyle \mathcal{J}_{01}}
\;.
\end{eqnarray}
We find:
\begin{eqnarray}
\lefteqn{[\,\mathcal{J}_{10},\,\mathcal{J}_{10}\,]_S \vphantom{\bigg|}} \cr
& = & -2i
\begin{bmatrix}
\;\varepsilon^{ijk}(W_L^i\wedge W_L^j)\tau^k & \mathbf{0}_{2\times 2} \\
\mathbf{0}_{2\times 2} & \varepsilon_{ijk}(W_R^i\wedge W_R^j)\tau^k\;
\end{bmatrix}
\;,
\cr
& & \vphantom{|}\cr
%%%
\lefteqn{[\,\mathcal{J}_{01},\,\mathcal{J}_{01}\,]_S
\;=\; -4i
\begin{bmatrix}
\;\Phi\Phi^\dagger       & \mathbf{0}_{2\times 2}\; \\
\;\mathbf{0}_{2\times 2} & \Phi^\dagger\Phi\;
\end{bmatrix}
\;,
}
\cr
%%%
\lefteqn{[\,\mathcal{J}_{10},\,\mathcal{J}_{01}\,]_S
\;=\; [\,\mathcal{J}_{01},\,\mathcal{J}_{10}\,]_S 
\vphantom{\Bigg|}}
\cr
& = & \sqrt{2}i
\begin{bmatrix}
\mathbf{0}_{2\times 2} & +i(\mathcal{W}_L\Phi-\Phi \mathcal{W}_R) \\	%Chen_Correction
-i(\Phi^\dagger \mathcal{W}_L - \mathcal{W}_R\Phi^\dagger) & \mathbf{0}_{2\times 2}	%Chen_Correction
\end{bmatrix}
\;.
\cr 
& &
\end{eqnarray}

\newpage
%%%
%%%
%%%
\begin{widetext}
%\begin{eqnarray}
%\mathcal{J}
%& = & \underbrace{i
%\begin{bmatrix}
%                \;\mathcal{W}_L-\frac{1}{\sqrt{2}}W_{BL}\cdot \mathbf{1}_{2\times 2} & \mathbf{0}_{2\times 2} \\
%                \mathbf{0}_{2\times 2} & \mathcal{W}_R-\frac{1}{\sqrt{2}}W_{BL}\cdot \mathbf{1}_{2\times 2}\; \\
%\end{bmatrix}}_{\displaystyle \mathcal{J}_{10}}
%\cr
%& &
%+ \underbrace{i
%\begin{bmatrix}
%                \;\mathbf{0}_{2\times 2} & \sqrt{2}\,\Phi\;  \\
%                \;\sqrt{2}\,\Phi^{\dagger} & \mathbf{0}_{2\times 2} \vphantom{\Big|}\; \\
%\end{bmatrix}}_{\displaystyle \mathcal{J}_{01}}
%\;.
%\end{eqnarray}
%
%We find:
%
%\begin{eqnarray}
%[\,\mathcal{J}_{10},\,\mathcal{J}_{10}\,]_S \vphantom{\bigg|}
%& = & -2i
%\begin{bmatrix}
%\;\varepsilon^{ijk}(W_L^i\wedge W_L^j)\tau^k & \mathbf{0}_{2\times 2} \\
%\mathbf{0}_{2\times 2} & \varepsilon_{ijk}(W_R^i\wedge W_R^j)\tau^k\;
%\end{bmatrix}
%\;,
%\cr
%%%
%[\,\mathcal{J}_{01},\,\mathcal{J}_{01}\,]_S
%& = & -4i
%\begin{bmatrix}
%\;\Phi\Phi^\dagger       & \mathbf{0}_{2\times 2}\; \\
%\;\mathbf{0}_{2\times 2} & \Phi^\dagger\Phi\;
%\end{bmatrix}
%\;,
%\cr
%%%
%[\,\mathcal{J}_{10},\,\mathcal{J}_{01}\,]_S
%& = & [\,\mathcal{J}_{01},\,\mathcal{J}_{10}\,]_S 
%\;=\; \sqrt{2}i
%\begin{bmatrix}
%\mathbf{0}_{2\times 2} & +i(\mathcal{W}_L\Phi-\Phi \mathcal{W}_R) \\	%Chen_Correction
%-i(\Phi^\dagger \mathcal{W}_L - \mathcal{W}_R\Phi^\dagger) & \mathbf{0}_{2\times 2}	%Chen_Correction
%\end{bmatrix}
%\;.
%\end{eqnarray}
%

\noindent
Putting everything together, we obtain
\begin{eqnarray}
\mathcal{F}
& = & i
\begin{bmatrix}
F_L -\frac{1}{\sqrt{2}}F_{BL}-2\Phi\Phi^{\dagger}-\sqrt{2}(\zeta\Phi^\dagger + \Phi\zeta^\dagger) & 
\sqrt{2}\,D\Phi + i(\mathcal{W}_L \zeta-\zeta \mathcal{W}_R) \\	%Chen_Correction
\sqrt{2}(D\Phi)^{\dagger} -i(\zeta^\dagger \mathcal{W}_L-\mathcal{W}_R \zeta^\dagger) & 	%Chen_Correction
F_R-\frac{1}{\sqrt{2}}F_{BL}-2\Phi^{\dagger}\Phi - \sqrt{2}(\zeta^\dagger\Phi+\Phi^\dagger\zeta) \\
\end{bmatrix}
\cr
& = & i
\begin{bmatrix}
F_L -\frac{1}{\sqrt{2}}F_{BL}-2\hat{\Phi}\hat{\Phi}^{\dagger} + v^2\mathbf{1}_{2\times 2} & 
\sqrt{2}\,D\hat{\Phi} \\
\sqrt{2}(D\hat{\Phi})^{\dagger} & 
F_R-\frac{1}{\sqrt{2}}F_{BL}-2\hat{\Phi}^{\dagger}\hat{\Phi} + v^2\mathbf{1}_{2\times 2} \\
\end{bmatrix}
\;,
\end{eqnarray}
\end{widetext}
%%%
%%%
%%%
%
where we have introduced the shifted Higgs field
\begin{equation}
\hat{\Phi} \;=\; \Phi + \dfrac{\zeta}{\sqrt{2}}\;,
\end{equation}
and
\begin{eqnarray}
F_{L,R}
& = & (F_{L,R})^a\tau^a
\;=\; (d W_{L,R}^i - (W_{L,R}\wedge W_{L,R})^i)\tau^i  
\cr
& & \hspace{1.4cm}\phantom{\Big|} 
\;=\;(dW_{L,R}^i - \varepsilon^{ijk} W_{L,R}^j\wedge W_{L,R}^k)\tau^i \;,
\cr
%W_{LR} & = & W_{LR}^i\tau^i \;,
%\cr
F_{BL}
& = & d W_{BL}\cdot \mathbf{1}_{2\times 2}
\;, \phantom{\bigg|}
\cr
D\Phi
& = & d\Phi+i\mathcal{W}_L\Phi-i\Phi \mathcal{W}_R		%Chen_Correction
\;,
\cr
D\hat{\Phi}
& = & d\hat{\Phi}+i\mathcal{W}_L\hat{\Phi}-i\hat{\Phi}\mathcal{W}_R		%Chen_Correction
\;.
\end{eqnarray}
We have also used $\zeta^\dagger\zeta=\zeta\zeta^\dagger=v^2\mathbf{1}_{2\times 2}$.

%%%%%%%%%%%%%%%%%%%%%%%%%%%%%%%%%%%%%%%%%%%%%%%%%%%%%%%%%%%%%%%%%%%%%%%%%%%
\subsection{Gauge Transformation Properties}

The even part of $su(2/2)$ generate the gauge transformations 
in $SU(2)_L\times SU(2)_R\times U(1)_{B-L}$:
\begin{eqnarray}
U 
& = &
\exp\left[i\sum_{i=1,2,3,8,13,14,15}\theta_i\lambda_i^s\right]
\cr
& = &
\begin{bmatrix}
u_L\,e^{-i\theta/\sqrt{2}} & \mathbf{0}_{2\times 2} \\
\mathbf{0}_{2\times 2}   & u_R\,e^{-i\theta/\sqrt{2}}
\end{bmatrix}
\;,
\label{Usu2Lsu2Ru1BL}
\end{eqnarray}
where
\begin{eqnarray}
u_L & = & \exp\Bigl[\,i(\theta_{1} \tau_1 + \theta_{2} \tau_2 + \theta_3\tau_3) \,\Bigr] \;\in\; SU(2)_L \;,\cr
u_R & = & \exp\Bigl[\,i(\theta_{13}\tau_1 + \theta_{14}\tau_2 + \theta_8\tau_3) \,\Bigr] \;\in\; SU(2)_R \;,\cr
\theta & = & \theta_{15}\;.\vphantom{\Big|}
\end{eqnarray}
The 1-form gauge fields transform as
\begin{eqnarray}
\mathcal{W}_L^{\phantom{\dagger}} & \rightarrow & u_L^{\phantom{\dagger}}\mathcal{W}_L^{\phantom{\dagger}} u_L^\dagger + i\,du_L^{\phantom{\dagger}} u_L^\dagger \;,\cr
\mathcal{W}_R^{\phantom{\dagger}} & \rightarrow & u_R^{\phantom{\dagger}}\mathcal{W}_R^{\phantom{\dagger}} u_R^\dagger + i\,du_R^{\phantom{\dagger}} u_R^\dagger \;,\cr
W_{BL} & \rightarrow & W_{BL}-d\theta \;.
\end{eqnarray}
For the 0-form field, we assume that it is the shifted field $\hat{\Phi}$ which transforms as
\begin{equation}
\hat{\Phi} \;\rightarrow\; u_L^{\phantom{\dagger}} \hat{\Phi}\, u_R^\dagger\;.
\end{equation}
$\zeta/\sqrt{2}$ is interpreted as the VEV of $\hat{\Phi}$, and $\Phi$ as the fluctuation around it.
Thus,
\begin{eqnarray}
F_{L}^{\phantom{\dagger}} & \rightarrow & u_L^{\phantom{\dagger}} F_{L}^{\phantom{\dagger}} u_L^\dagger \;,\cr
F_{R}^{\phantom{\dagger}} & \rightarrow & u_R^{\phantom{\dagger}} F_{R}^{\phantom{\dagger}} u_R^\dagger \;,\cr
F_{BL} & \rightarrow & F_{BL} \;,\cr
D\hat{\Phi} & \rightarrow & u_L^{\phantom{\dagger}}(D\hat{\Phi})u_R^\dagger\;.
\end{eqnarray}
This time, the terms coming from $\zeta$ causes no problems due to
$\zeta^\dagger\zeta=\zeta\zeta^\dagger=v^2\mathbf{1}_{2\times 2}$,
and $\mathcal{F}$ can be seen to transform as
\begin{equation}
\mathcal{F} \;\rightarrow\; U\mathcal{F} U^\dagger\;.
\end{equation}
%

%%%%%%%%%%%%%%%%%%%%%%%%%%%%%%%%%%%%%%%%%%%%%%%%%%%%%%%%%%%%%%%%%%%%%%%%%%%
\subsection{The Action}

Following the same procedure as the $su(2/1)$ case, we find that the action
in the $su(2/2)$ case is given by
%
%%%
%%%
%%%
%\begin{widetext}
%
\begin{eqnarray}
\mathcal{S}
& = & \dfrac{1}{4}\langle\mathcal{F},\mathcal{F}\rangle_S \cr
& = & \int d^4x 
\bigg[ 
- \dfrac{1}{4}
\left(
 F^{i\phantom{\mu\nu}}_{L\mu\nu}F^{i\mu\nu}_L 
+F^{i\phantom{\mu\nu}}_{R\mu\nu}F^{i\mu\nu}_R
+F_{BL\mu\nu}^{\phantom{\mu\nu}}F^{\mu\nu}_{BL}
\right)
\cr
& & \qquad\qquad
+\;\Tr\Bigl[
(D_\mu\hat{\Phi})^\dagger(D^\mu\hat{\Phi})
\Bigr]
-V(\hat{\Phi}^\dagger,\hat{\Phi})
\biggr]
\;,
\cr & &
\end{eqnarray}
%
%%%
%%%
%%%
%\end{widetext}
%
where 
\begin{eqnarray}
F^{i\mu\nu}_{L} & = & \partial^{\mu}W^{i\nu}_{L}-\partial^{\nu}W^{i\mu}_{L}+2i\varepsilon^{ijk} W^{j\mu}_{L}W^{k\nu}_{L}\;,
\cr
F^{i\mu\nu}_{R} & = & \partial^{\mu}W^{i\nu}_{R}-\partial^{\nu}W^{i\mu}_{R}+2i\varepsilon^{ijk} W^{j\mu}_{R}W^{k\nu}_{R}\;, \vphantom{\bigg|}
\cr
F^{\mu\nu}_{BL} & = & \partial^\mu W_{BL}^\nu-\partial^\nu W_{BL}^\mu \;,
\cr
D^\mu\hat{\Phi}
& = & \partial^\mu\hat{\Phi} 
- iW_{L}^{i\mu} \tau^i\hat{\Phi} + iW_{R}^{i\mu}\,\hat{\Phi}\,\tau^i
\;,
\cr
%%%
V(\hat{\Phi}^\dagger,\hat{\Phi})
& = & 2\,\Tr
\Biggr[
\left(
\hat{\Phi}^\dagger\hat{\Phi}^{\phantom{\dagger}}-\dfrac{v^2}{2}\mathbf{1}_{2\times 2}
\right)^2
\Biggr]
\;.
\label{DPhiandVPhi}
\end{eqnarray}
Thus, we obtain a manifestly gauge invariant action as required.
The Higgs potential is minimized when
\begin{equation}
\hat{\Phi}^\dagger\hat{\Phi} \;=\; \dfrac{v^2}{2}\mathbf{1}_{2\times 2}\;,
\end{equation}
which is consistent with our interpretation that the VEV of $\hat{\Phi}$ is given by
\begin{equation}
\langle\hat{\Phi}\rangle \;=\; \dfrac{\zeta}{\sqrt{2}}\;.
\end{equation}
%
%The unshifted field $\Phi = \hat{\Phi}-\zeta/\sqrt{2}$ which appears in the superconnection $\mathcal{J}$ is understood as the fluctuation away from this VEV:
%
%\begin{equation}
%\hat{\Phi} \;=\; \langle\hat{\Phi}\rangle + \Phi\;.
%\end{equation}
%

%%%%%%%%%%%%%%%%%%%%%%%%%%%%%%%%%%%%%%%%%%%%%%%%%%%%%%%%%%%%%%%%%%%%%%%%%%%%%%%%%%%%%%%%%%%%
%%%%%%%%%%%%%%%%%%%%%%%%%%%%%%%%%%%%%%%%%%%%%%%%%%%%%%%%%%%%%%%%%%%%%%%%%%%%%%%%%%%%%%%%%%%%
\subsection{Symmetry Breaking}\label{SSBinsu22}

%%%%%%%%%%%%%%%%%%%%%%%%%%%%%%%%%%%%%%%%%%%%%%%%%%%%%%%%%%%%%%%%%%%%%%%%%%%%%%%%%%%%%%%%%%%%
\subsubsection{Breaking with the Bi-doublet}

It is a well known fact that the bidoublet $\hat{\Phi}$ alone acquiring a vacuum expectation value (VEV) will
not break $SU(2)_L\times SU(2)_R\times U(1)_{B-L}$ all the way down to $U(1)_{em}$.
Indeed, if we assume non-zero VEV's for the (would be) neutral components of $\Phi$ as
\begin{equation}
\langle\hat{\Phi}\rangle
\;=\; \dfrac{\zeta}{\sqrt{2}}
\;=\; \dfrac{1}{\sqrt{2}}
\begin{bmatrix}
\kappa_1 & 0 \\ 0 & \kappa_2
\end{bmatrix}
\;,
\label{PhiVEV}
\end{equation}
where $\kappa_1$ and $\kappa_2$ are in general complex,
then the unitarity of $\zeta$ demands
\begin{equation}
|\kappa_1|\;=\;|\kappa_2|\;=\;v\;.
\end{equation}
So, up to a possible relative phase between $\kappa_1$ and $\kappa_2$, we have
\begin{equation}
\langle\hat{\Phi}\rangle
\;=\; \dfrac{v}{\sqrt{2}}
\begin{bmatrix}
1 & 0 \\ 0 & 1
\end{bmatrix}
\;.
\end{equation}
Since $\hat{\Phi}$ transforms as $\hat{\Phi}\rightarrow u_L\hat{\Phi}u_R^\dagger$ under
local gauge transformations, 
this VEV remains
invariant under $U(1)_{B-L}$, and under the vectorial combination of
$SU(2)_L$ and $SU(2)_R$, that is, when $u_L=u_R$.
Thus $\langle\hat{\Phi}\rangle$ only breaks $SU(2)_L\times SU(2)_R\times U(1)_{B-L}$
down to $SU(2)_V\times U(1)_{B-L}$, providing only three massive gauge bosons.

%%%%%%%%%%%%%%%%%%%%%%%%%%%%%%%%%%%%%%%%%%%%%%%%%%%%%%%%%%%%%%%%%%%%%%%%%%%%%%%%%%%%%%%%%%%%
\subsubsection{Introduction of Triplets}

To achieve the symmetry breaking we need, 
we follow Mohapatra and Senjanovic \cite{Mohapatra:1979ia} and introduce
the scalar fields $\Delta_{L}(3,1,2)$ and $\Delta_{R}(1,3,2)$,
where the first two numbers indicate the dimensions of the $SU(2)_L$ and $SU(2)_R$ representations these
fields belong to, and the third number is the $B-L$ charge.
We advocate the picture that 
$\Delta_L$ lives on the left-handed brane while
$\Delta_R$ lives on the right-handed brane
as matter fields and are not part of a generalized connection.

These triplet fields are often represented in the literature as $2\times 2$ complex traceless matrices:
\begin{eqnarray}
\Delta_{L,R}
& = & \dfrac{1}{\sqrt{2}}\left(
\delta^1_{L,R}\,\tau^1+\delta^2_{L,R}\,\tau^2+\delta^3_{L,R}\,\tau^3 
\right)
\cr
& = &
\begin{bmatrix}
\delta^+_{L,R}/\sqrt{2} &  \delta^{++}_{L,R}       \\
\delta^0_{L,R}          & -\delta^+_{L,R}/\sqrt{2} 
\end{bmatrix}
\;,
\label{Deltadef}
\end{eqnarray}
where
\begin{eqnarray}
\delta^{++}_{L,R} & = & \dfrac{1}{\sqrt{2}}\left(\delta^1_{L,R}-i\delta^2_{L,R}\right) \;,\cr
\delta^{+}_{L,R}  & = & \delta^3_{L,R} \;,\cr
\delta^{0}_{L,R}  & = & \dfrac{1}{\sqrt{2}}\left(\delta^1_{L,R}+i\delta^2_{L,R}\right) \;.
\end{eqnarray}
These fields transform as
\begin{eqnarray}
\Delta_L & \rightarrow & e^{+i\sqrt{2}\theta} u_L^{\phantom{\dagger}} \Delta_L^{\phantom{\mathrm{T}}} u_L^\dagger\;,\cr
\Delta_R & \rightarrow & e^{+i\sqrt{2}\theta} u_R^{\phantom{\dagger}} \Delta_R^{\phantom{\mathrm{T}}} u_R^\dagger\;,
\end{eqnarray}
where the $U(1)_{B-L}$ phase will be shown to correspond to $B-L=2$ later.

It is instructive to rewrite the $\Delta_{L,R}$ fields as complex symmetric matrices:
\begin{eqnarray}
\widetilde{\Delta}_{L,R}
& \equiv & i\tau^2\Delta_{L,R}^* 
\cr
& = &
\begin{bmatrix}
\delta^{0\ast}_{L,R}     & -\delta^{-}_{L,R}/\sqrt{2}        \\
-\delta^{-}_{L,R}/\sqrt{2} & -\delta^{--}_{L,R}
\end{bmatrix}
%\cr
%& \equiv &
%\begin{bmatrix}
%\widetilde{\Delta}^{0}_{L,R}          & \widetilde{\Delta}^{-}_{L,R}/\sqrt{2}        \\
%\widetilde{\Delta}^{-}_{L,R}/\sqrt{2} & \widetilde{\Delta}^{--}_{L,R}
%\end{bmatrix}
\;.
\end{eqnarray}
These fields transform as
\begin{eqnarray}
\widetilde{\Delta}_L & \rightarrow & e^{-i\sqrt{2}\theta} u_L^{\phantom{\mathrm{T}}} \widetilde{\Delta}_L^{\phantom{\mathrm{T}}} u_L^{\mathrm{T}}\;,\cr
\widetilde{\Delta}_R & \rightarrow & e^{-i\sqrt{2}\theta} u_R^{\phantom{\mathrm{T}}} \widetilde{\Delta}_R^{\phantom{\mathrm{T}}} u_R^{\mathrm{T}}\;.
\end{eqnarray}
Let us place these fields into a single $4\times 4$ matrix as
\begin{equation}
\widetilde{\Delta} \;=\;
\begin{bmatrix}
\widetilde{\Delta}_L & \mathbf{0}_{2\times 2} \\
\mathbf{0}_{2\times 2} & \widetilde{\Delta}_R
\end{bmatrix}
\;.
\end{equation}
Then under $SU(2)_L\times SU(2)_R\times U(1)_{B-L}$ gauge transformations, Eq.~(\ref{Usu2Lsu2Ru1BL}), 
it transforms as
\begin{equation}
\widetilde{\Delta}
\;\rightarrow\;
U
\widetilde{\Delta}
U^\mathrm{T}
\;,
\end{equation}
which shows that $\widetilde{\Delta}$ provides a module for an $su(2/2)$ representation.
The even elements of $su(2/2)$ correspond to the gauge transformations, while
the odd elements would interchange $\widetilde{\Delta}_L$ and $\widetilde{\Delta}_R$.
Since the leptons $\psi$ and its charge conjugate $\psi^c = C\bar{\psi}^{\mathrm{T}}$
transform as
\begin{equation}
\psi   \;\rightarrow\; U\psi\;,\qquad
\psi^c \;\rightarrow\; U^*\psi^c\;,
\end{equation}
we can construct the gauge invariant interaction
\begin{eqnarray}
\mathcal{L}_M 
& = & y_M
\left(
\bar{\psi}^c\,\widetilde{\Delta}^\dagger\,\psi +
\bar{\psi}  \,\widetilde{\Delta}        \,\psi^c 
\right)
\vphantom{\bigg|} 
\cr
& = & 
y_M \left[
\left(
\ell_{L}^{\mathrm{T}} \,C\, \Delta_L^{\phantom{\mathrm{T}}} i \tau^2  \ell_{L}^{\phantom{\mathrm{T}}}
+ 
\ell_{R}^{\mathrm{T}} \,C\, \Delta_R^{\phantom{\mathrm{T}}} i \tau^2 \ell_{R}^{\phantom{\mathrm{T}}}
\right)
+ h.c.
\right]
\;,
\cr & & 
\label{MajoranaInteraction}
\end{eqnarray}
which will lead to Majorana masses for the neutrinos after symmetry breaking.
Thus, the triplet scalars have a natural place in the $su(2/2)$ framework,
as do Majorana neutrinos.

%%%%%%%%%%%%%%%%%%%%%%%%%%%%%%%%%%%%%%%%%%%%%%%%%%%%%%%%%%%%%%%%%%%%%%%%%%%%%%%%%%%%%%%%%%%%
\subsubsection{The Higgs Potential}

Reverting to the original traceless matrix representation, the Lagrangian for the 
$\Delta_{L,R}$ is given by
\begin{eqnarray}
\mathcal{L}
& = & \Tr\Bigl[
 (D^\mu\Delta_L)^\dagger(D_\mu\Delta_L)
+(D^\mu\Delta_R)^\dagger(D_\mu\Delta_R)
\Bigr]
\cr
& & -V(\Delta_L^\dagger,\Delta_L^{\phantom{\dagger}},\Delta_R^\dagger,\Delta_R^{\phantom{\dagger}},\hat{\Phi}^\dagger,\hat{\Phi})
\;,\vphantom{\bigg|}
\end{eqnarray}
where the covariant derivatives are given by
\begin{eqnarray}
D^\mu\Delta_L & = & \partial^\mu\Delta_L - iW_{L}^{i\mu}\left[\tau^i,\Delta_L\right]-i\sqrt{2}W_{BL}^\mu\Delta_L \;,\cr
D^\mu\Delta_R & = & \partial^\mu\Delta_R - iW_{R}^{i\mu}\left[\tau^i,\Delta_R\right]-i\sqrt{2}W_{BL}^\mu\Delta_R \;.\cr
& &
\end{eqnarray}

The most general form of the Higgs potential 
$V(\Delta_L^\dagger,\Delta_L^{\phantom{\dagger}},\Delta_R^\dagger,\Delta_R^{\phantom{\dagger}},\hat{\Phi}^\dagger,\hat{\Phi})$
which respects the 
$SU(2)_L\times SU(2)_R\times U(1)_{B-L}$ gauge symmetry 
and the discrete left-right symmetry under the interchanges
\begin{equation}
\Delta_L\;\leftrightarrow\;\Delta_R\;,\qquad
\hat{\Phi}\;\leftrightarrow\;\hat{\Phi}^\dagger\;,
\end{equation}
and is at most quartic in the fields
is given in Ref.~\cite{Deshpande:1990ip} as%\footnote{%
%The parameters used in this potential are different from those used in Eq.~(2.9) of Ref.~\cite{Gunion:1989in}.  See Appendix~\ref{GunionComparison} for a translation chart.
%} 
%
%%%
%%%
%%%
\begin{widetext}
\begin{eqnarray}
V
& = & -\mu_1^2\Tr\Bigl[\Phi^\dagger\Phi\Bigr]
-\mu_2^2\left( \Tr\Bigl[\Phi^\dagger\widetilde{\Phi}\Bigr] + \Tr\Bigl[\widetilde{\Phi}^\dagger\Phi\Bigr] \right)
-\mu_3^2\left( \Tr\Bigl[\Delta_L^\dagger\Delta_L^{\phantom{\dagger}}\Bigr] + \Tr\Bigl[\Delta_R^\dagger\Delta_R^{\phantom{\dagger}}\Bigr]\right)
\cr
& & 
+ \lambda_1\left(\Tr\Bigl[\Phi^\dagger\Phi\Bigr]\right)^2
+ \lambda_2\left\{
  \left(\Tr\Bigl[\Phi^\dagger\widetilde{\Phi}\Bigr]\right)^2
+ \left(\Tr\Bigl[\widetilde{\Phi}^\dagger\Phi\Bigr]\right)^2
\right\}
\cr
& &
+ \lambda_3\,\Tr\Bigl[\Phi^\dagger\widetilde{\Phi}\Bigr]\Tr\Bigl[\widetilde{\Phi}^\dagger\Phi\Bigr]
+ \lambda_4\,\Tr\Bigl[\Phi^\dagger\Phi\Bigr]\left( \Tr\Bigl[\Phi^\dagger\widetilde{\Phi}\Bigr] + \Tr\Bigl[\widetilde{\Phi}^\dagger\Phi\Bigr]\right)
\cr
& & 
+ \rho_1\left\{
  \left(\Tr\Bigl[\Delta_L^\dagger\Delta_L^{\phantom{\dagger}}\Bigr]\right)^2
+ \left(\Tr\Bigl[\Delta_R^\dagger\Delta_R^{\phantom{\dagger}}\Bigr]\right)^2
\right\}
+ \rho_2\left( 
  \Tr\Bigl[\Delta_L\Delta_L\Bigr]\Tr\Bigl[\Delta_L^\dagger\Delta_L^\dagger\Bigr]
+ \Tr\Bigl[\Delta_R\Delta_R\Bigr]\Tr\Bigl[\Delta_R^\dagger\Delta_R^\dagger\Bigr]
\right)
\cr
& &
+ \rho_3\,\Tr\Bigl[\Delta_L^\dagger\Delta_L^{\phantom{\dagger}}\Bigr]\Tr\Bigl[\Delta_R^\dagger\Delta_R^{\phantom{\dagger}}\Bigr]
+ \rho_4\left(
  \Tr\Bigl[\Delta_L\Delta_L\Bigr]\Tr\Bigl[\Delta_R^\dagger\Delta_R^\dagger\Bigr]
+ \Tr\Bigl[\Delta_L^\dagger\Delta_L^\dagger\Bigr]\Tr\Bigl[\Delta_R\Delta_R\Bigr]
\right)
\cr
& &
+ \alpha_1\,\Tr\Bigl[\Phi^\dagger\Phi\Bigr]
\left(
  \Tr\Bigl[\Delta_L^\dagger\Delta_L^{\phantom{\dagger}}\Bigr]
+ \Tr\Bigl[\Delta_R^\dagger\Delta_R^{\phantom{\dagger}}\Bigr]
\right)
+ \alpha_2
\left(
  \Tr\Bigl[\Phi^\dagger\tilde{\Phi}\Bigr]\Tr\Bigl[\Delta_L^\dagger\Delta_L^{\phantom{\dagger}}\Bigr]
+ \Tr\Bigl[\tilde{\Phi}^\dagger\Phi\Bigr]\Tr\Bigl[\Delta_R^\dagger\Delta_R^{\phantom{\dagger}}\Bigr]
\right)
\cr
& &
+ \alpha_2^*
\left(
  \Tr\Bigl[\Phi^\dagger\tilde{\Phi}\Bigr]\Tr\Bigl[\Delta_R^\dagger\Delta_R^{\phantom{\dagger}}\Bigr]
+ \Tr\Bigl[\tilde{\Phi}^\dagger\Phi\Bigr]\Tr\Bigl[\Delta_L^\dagger\Delta_L^{\phantom{\dagger}}\Bigr]
\right)
+ \alpha_3
\left(
  \Tr\Bigl[\Phi\Phi^\dagger\Delta_L^{\phantom{\dagger}}\Delta_L^\dagger\Bigr]
+ \Tr\Bigl[\Phi^\dagger\Phi\Delta_R^{\phantom{\dagger}}\Delta_R^\dagger\Bigr]
\right)
\cr
& &
+ \beta_1
\left(
  \Tr\Bigl[\Phi\Delta_R^{\phantom{\dagger}}\Phi^\dagger\Delta_L^\dagger\Bigr]
+ \Tr\Bigl[\Phi^\dagger\Delta_L^{\phantom{\dagger}}\Phi\Delta_R^\dagger\Bigr]
\right)
+ \beta_2
\left(
  \Tr\Bigl[\widetilde{\Phi}\Delta_R^{\phantom{\dagger}}\Phi^\dagger\Delta_L^\dagger\Bigr]
+ \Tr\Bigl[\widetilde{\Phi}^\dagger\Delta_L^{\phantom{\dagger}}\Phi\Delta_R^\dagger\Bigr]
\right)
\cr
& &
+ \beta_3
\left(
  \Tr\Bigl[\Phi\Delta_R^{\phantom{\dagger}}\widetilde{\Phi}^\dagger\Delta_L^\dagger\Bigr]
+ \Tr\Bigl[\Phi^\dagger\Delta_L^{\phantom{\dagger}}\widetilde{\Phi}\Delta_R^\dagger\Bigr]
\right)\;,
\label{GeneralV}
\end{eqnarray}
\end{widetext}
%%%
%%%
%%%
%
where we have denoted $\hat{\Phi}$ simply as $\Phi$, and 
$\widetilde{\Phi} = \tau^2\Phi^*\tau^2$.
As we can see, it is fairly complicated with 18 free parameters:
3 negative mass-squared parameters $\mu_i^2$ ($i=1,2,3$), 4 parameters
$\lambda_i$ ($i=1,2,3,4$) for the quartic self-couplings of $\Phi$,  
4 parameters $\rho_i$ ($i=1,2,3,4$) for the quartic couplings of $\Delta_{L,R}$,
4 parameters $\alpha_i$ ($i=1,2,3$ with $\alpha_2$ complex) that couple
$\Phi$ to $\Delta_L$ or $\Delta_R$, and 3 parameters $\beta_i$ ($i=1,2,3$)
which couple all three.
The possible phase of $\alpha_2$ breaks CP explicitly.

Using the identity\footnote{%
See Appendix~\ref{GunionComparison} for a collection of useful identities of this type.
}
\begin{equation}
\Tr\Bigl[(\Phi^\dagger\Phi)^2\Bigr]
\;=\; \left(\Tr\Bigl[\Phi^\dagger\Phi\Bigr]\right)^2
- \dfrac{1}{2}\Tr\Bigl[\Phi^\dagger\widetilde{\Phi}\Bigr]\Tr\Bigl[\widetilde{\Phi}^\dagger\Phi\Bigr]
\;,
\end{equation}
we can see that the Higgs potential of Eq.~(\ref{DPhiandVPhi}) corresponds to
\begin{equation}
\lambda_1 \,=\,  2\,,\quad
\lambda_3 \,=\, -1\,,\quad
\lambda_2 \,=\, \lambda_4 \,=\, 0\,,
\label{lambdavalues}
\end{equation}
with
\begin{equation}
\mu_1^2 \;=\; 2v^2\;,\qquad
\mu_2^2 \;=\; 0\;. 
\end{equation}
We envision that the NCG theory from which the effective $su(2/2)$ model
emerges will determine all the other parameters in the potential as well.
For now, we follow Ref.~\cite{Mohapatra:1979ia} and simply
assume that the (would be) neutral components of the triplets acquire VEV's given by
\begin{equation}
\vev{\Delta_L} \;=\; \dfrac{1}{\sqrt{2}}\begin{bmatrix} 0 & 0\; \\ \; v_L & 0\; \end{bmatrix}\;,\quad
\vev{\Delta_R} \;=\; \dfrac{1}{\sqrt{2}}\begin{bmatrix} 0 & 0\; \\ \; v_R & 0\; \end{bmatrix}\;,
\end{equation}
where we set $v_L=0$ to avoid it from breaking $SU(2)_L$.
This breaks $SU(2)_R\times U(1)_{B-L}$ down to $U(1)_Y$.
The linear combinations
\begin{equation}
W_{R}^{\pm} \;=\; \dfrac{W_{R}^1\mp W_{R}^2}{\sqrt{2}}\;,\quad
Z'          \;=\; \dfrac{\sqrt{2}\,W_{R}^3 - W_{BL}^{\phantom{0}}}{\sqrt{3}}
\label{WRandZprime}
\end{equation}
obtain masses given by
\begin{equation}
M_{W_R} \;=\; \sqrt{2}v_R \;,\qquad
M_{Z'}  \;=\; \sqrt{6}v_R\;,
\end{equation}
while the linear combination
\begin{equation}
B \;=\; \dfrac{W_{R}^3 + \sqrt{2}\,W_{BL}^{\phantom{0}}}{\sqrt{3}}
\label{Bdef}
\end{equation}
remains massless and couples to 
\begin{equation}
\dfrac{\lambda_8^s + \sqrt{2}\,\lambda_{15}^s}{\sqrt{3}}
\;=\; \dfrac{1}{\sqrt{3}}
\begin{bmatrix}
-1 &  0 &  0 &  0 \\
 0 & -1 &  0 &  0 \\
 0 &  0 &  0 &  0 \\
 0 &  0 &  0 & -2
\end{bmatrix}
\;=\; \dfrac{Y}{\sqrt{3}}
\;,
\label{Ydef}
\end{equation}
which corresponds to the hypercharge $Y$ embedded in $su(2/1)$.

The presence of the VEV of $\hat{\Phi}$, Eq.~(\ref{PhiVEV}), breaks
the remaining $SU(2)_L\times U(1)_Y$ down to $U(1)_{em}$.
We have noted earlier that the nilpotency of the matrix derivative demands 
the unitarity of the $\zeta$ matrix, which in turn would demand $|\kappa_1|=|\kappa_2|$.
If the underlying NCG requires this condition to be maintained under the introduction of the triplet fields, it would constitute a robust prediction of the formalism
and provide an extra condition on the emergent LRSM. 
However, such a condition is phenomenologically problematic. If we
generate the quark masses via Yukawa couplings with the bi-doublet field $\Phi$,
that is, the interaction of the form
\begin{equation}
\mathcal{L}_Y \;=\;
\overline{q}_{Li}\left(y_{ij}\Phi+\tilde{y}_{ij}\widetilde{\Phi}\right)q_{Rj} + h.c.
\;,
\label{quarkYukawas}
\end{equation}
where $y_{ij}$ and $\tilde{y}_{ij}$ are the Yukawa coupling matrices, 
it would lead to mass matrices of the form
\begin{eqnarray}
\sqrt{2}M_u & = & \kappa_1^{\phantom{*}}\,y + \kappa_2^*\,\tilde{y}
\;=\; |\kappa_1|\,y + |\kappa_2|e^{-i\alpha}\tilde{y}
\;,\cr
\sqrt{2}M_d & = & \kappa_2^{\phantom{*}}\,y + \kappa_1^*\,\tilde{y}
\;=\; |\kappa_2|e^{i\alpha}y + |\kappa_1|\,\tilde{y}
\;,
\end{eqnarray}
where $\alpha$ is a possible CP violating phase difference between $\kappa_1$ and $\kappa_2$.
The condition $|\kappa_1|=|\kappa_2|$ would imply
\begin{equation}
M_u \;=\; e^{-i\alpha}M_d\;,
\end{equation}
leading to both $M_u$ and $M_d$ being diagonalized in the same basis with the same 
eigenvalues.
For this reason, it is normally assumed that $|\kappa_1|\gg|\kappa_2|$, which would
provide an explanation of $m_t\gg m_b$, and CKM mixing\footnote{The assumption $|\kappa_1|\gg|\kappa_2|$ is not mandatory in order to realize $m_t\gg m_b$. The smallness of CP violating parameter can be established by the interplay of $\sin\alpha$, $|\kappa_1|$, and $|\kappa_2|$ \cite{Senjanovic:2014pva}.}. Moreover, $|\kappa_1|$ and $|\kappa_2|$ being hierarchical is also required by the suppression of the flavor changing neutral-currents (FCNC) \cite{Deshpande:1990ip}. Therefore, we will allow for   
$|\kappa_1|\neq|\kappa_2|$ though the nilpotency of the matrix derivative
is destroyed.

We have seen in the $su(2/1)$ case that the lack of nilpotency of the matrix derivative
could lead to the gauge transformation property of the supercurvature $\mathcal{F}$
becoming non-standard, though otherwise it did not seem to have any ill effects.
However, it is somewhat worrisome that an exterior derivative is not nilpotent.
We conjecture a couple of reasons why this may be permissible.
First, the dynamics necessary for the breaking of the gauge symmetries
may be accompanied by some type of `phase' transition in the NCG from one in which 
the matrix derivative is nilpotent to one in which it is not.
Second, the $su(2/2)$ formalism is not a complete description of the
NCG (we are yet to include $SU(3)$ color or the generational structure) 
and needs to be extended to
a larger superalgebra in which the matrix derivative maintains nilpotency even after
symmetry breaking, just as the $su(2/1)$ formalism needed to be
extended to $su(2/2)$.
The two possibilities we have listed here could actually be compatible.
Recalling that the exterior derivative operator $d$ in standard differential geometry 
is the dual of the boundary operator $\partial$ \cite{Nakahara:2003nw}, $d^2\neq 0$ would imply
$\partial^2\neq 0$, that is, the boundary of a boundary does not vanish.
That could imply the appearance of some type of singularity in the geometry, 
which can be removed by going to higher dimensions.

The massive gauge bosons are now linear combinations of 
$W_L^\pm$, $Z$, $W_R^\pm$, and $Z'$ defined in Eqs.~(\ref{WZdef}) and (\ref{WRandZprime}).
Taking both $\kappa_1$ and $\kappa_2$ to be real for the moment, and 
setting $\kappa_+=\sqrt{\kappa_1^2+\kappa_2^2}$, they are
\begin{eqnarray}
W_2^\pm & = & W_R^\pm \cos\chi - W_L^\pm \sin\chi \;,\cr
W_1^\pm & = & W_R^\pm \sin\chi + W_L^\pm \cos\chi \;,
\end{eqnarray}
where
\begin{equation}
\tan 2\chi \;=\; \dfrac{2\kappa_1\kappa_2}{v_R^2}\;,
\end{equation}
with masses
\begin{equation}
M_{W_{1,2}}^2 \;=\; (v_R^2 + \kappa_+^2) \mp \sqrt{v_R^4 + 4\kappa_1^2\kappa_2^2}\;,
\end{equation}
and
\begin{eqnarray}
Z_2 & = & Z'\cos\varphi - Z\sin\varphi \;,\cr
Z_1 & = & Z'\sin\varphi + Z\cos\varphi \;,
\end{eqnarray}
where
\begin{equation}
\tan 2\varphi \;=\; \dfrac{4\sqrt{2}\kappa_+^2}{9v_R^2 - 2\kappa_+^2}\;,
\end{equation}
with masses
\begin{eqnarray}
M_{Z_1}^2 & = & \left(3v_R^2+\kappa_+^2\right)-\sqrt{9v_R^4-2 v_R^2 \kappa_+^2 + \kappa_+^4}\;,\cr
M_{Z_2}^2 & = & \left(3v_R^2+\kappa_+^2\right)+\sqrt{9v_R^4-2 v_R^2 \kappa_+^2 + \kappa_+^4}\;.\cr
&&
\end{eqnarray}
When $v_R\gg \kappa_+$, the masses are approximately
\begin{eqnarray}
M_{W_1}^2 & = & \kappa_+^2 - \dfrac{2\kappa_1^2\kappa_2^2}{v_R^2} + \cdots \;,\cr
M_{W_2}^2 & = & 2v_R^2 + \kappa_+^2 + \dfrac{2\kappa_1^2\kappa_2^2}{v_R^2} + \cdots \;,\cr
M_{Z_1}^2 & = & \dfrac{4\kappa_+^2}{3} - \dfrac{4\kappa_+^4}{27v_R^2} + \cdots \;,\cr
M_{Z_2}^2 & = & 6v_R^2 +\dfrac{2\kappa_+^2}{3} +\dfrac{4\kappa_+^4}{27v_R^2} + \cdots \;.
\end{eqnarray}
The remaining massless gauge boson is
\begin{equation}
A 
\;=\; \dfrac{W_{L}+\sqrt{3}B}{2}
\;=\; \dfrac{W_{L}^3+W_{R}^3+\sqrt{2}W_{BL}}{2}\;,
\label{Adef}
\end{equation}
coupled to
\begin{equation}
\dfrac{\lambda_{3}^{s} + \lambda_{8}^{s} + \sqrt{2}\lambda_{15}^{s}}{2}
\;=\;
\begin{bmatrix}
\;0 &  0 &  0 &  0 \\
\;0 & -1 &  0 &  0 \\
\;0 &  0 &  0 &  0 \\
\;0 &  0 &  0 & -1
\end{bmatrix}
\;=\; Q
\;.
\label{Qdef2}
\end{equation}
For the model to be phenomenologically viable, we need $v_R\gg \kappa_+ \approx 246\,\mathrm{GeV}$
to suppress the mixing between $W_L$ and $W_R$, and that between $Z$ and $Z'$.
The current experimental bounds on the LRSM parameters 
will be discussed in section~\ref{PHENO}.

%%%%%%%%%%%%%%%%%%%%%%%%%%%%%%%%%%%%%%%%%%%%%%%%%%%%%%%%%%%%%%%%%%%%%%%%%%%%%%%%%%%%%%%%%%%%
%%%%%%%%%%%%%%%%%%%%%%%%%%%%%%%%%%%%%%%%%%%%%%%%%%%%%%%%%%%%%%%%%%%%%%%%%%%%%%%%%%%%%%%%%%%%
\subsection{The Coupling Constants and the value of $\mathbf{sin^2}\bm{\theta_W}$}

%%%%%%%%%%%%%%%%%%%%%%%%%%%%%%%%%%%%%%%%%%%%%%%%%%%%%%%%%%%%%%%%%%%%%%%%%%%%%%%%%%%%%%%%%%%%
\subsubsection{Introduction of the Coupling Constant and $\sin^2\theta_W$ from the coupling to triplet Higgs}

We introduce the gauge coupling constant $g$ by rescaling the superconnection $\mathcal{J}$,
the action $\mathcal{S}$, and the matrix-derivative matrix $\eta$ as in Eq.~(\ref{CouplingIntroduction}).
The expression for the resulting action is the same as before except with the following
replacements:
\begin{eqnarray}
F^{i\mu\nu}_{L} & = & \partial^{\mu}W^{i\nu}_{L}-\partial^{\nu}W^{i\mu}_{L}+ig\varepsilon^{ijk} W^{j\mu}_{L}W^{k\nu}_{L}\;,
\cr
F^{i\mu\nu}_{R} & = & \partial^{\mu}W^{i\nu}_{R}-\partial^{\nu}W^{i\mu}_{R}+ig\varepsilon^{ijk} W^{j\mu}_{R}W^{k\nu}_{R}\;, \vphantom{\bigg|}
\cr
%F^{\mu\nu}_{BL} & = & \partial^\mu W_{BL}^\nu-\partial^\nu W_{BL}^\mu \;,
%%%
\cr
D^\mu\hat{\Phi}
& = & \partial^\mu\hat{\Phi} 
- i\dfrac{g}{2}W_{L}^{i\mu} \tau^i\hat{\Phi} 
+ i\dfrac{g}{2}W_{R}^{i\mu}\,\hat{\Phi}\,\tau^i
\;, \vphantom{\bigg|}
\cr
%%%
V(\hat{\Phi}^\dagger,\hat{\Phi})
& = & \dfrac{g^2}{2}\,\Tr
\Biggr[
\left(
\hat{\Phi}^\dagger\hat{\Phi}^{\phantom{\dagger}}-\dfrac{v^2}{2}
\right)^2
\Biggr]
\;.
%\cr & & 
\end{eqnarray}
Note that the gauge couplings for $SU(2)_L$ and $SU(2)_R$ are the same.
Thus, we have a left-right symmetric model (LRSM).
The ratio of the $U(1)_{B-L}$ coupling to the $SU(2)_{L,R}$ couplings cannot
be determined from the gauge couplings of the bi-doublet $\hat{\Phi}$ since it does not 
couple to $W_{BL}$.   
It can, however, be read off from the covariant derivatives of the triplet fields 
which are changed to:
\begin{eqnarray}
D^\mu\Delta_L & = & \partial^\mu\Delta_L - i\dfrac{g}{2}W_{L}^{i\mu}\left[\tau^i,\Delta_L\right]-i\dfrac{g}{\sqrt{2}}W_{BL}^\mu\Delta_L \;,\cr
D^\mu\Delta_R & = & \partial^\mu\Delta_R - i\dfrac{g}{2}W_{R}^{i\mu}\left[\tau^i,\Delta_R\right]-i\dfrac{g}{\sqrt{2}}W_{BL}^\mu\Delta_R \;.\cr
& &
\end{eqnarray}
Since the triplets are assigned the charge $B-L=2$, these needs to be compared with
the expressions
\begin{eqnarray}
D^\mu\Delta_L & = & \partial^\mu\Delta_L - i\dfrac{g}{2}W_{L}^{i\mu}\left[\tau^i,\Delta_L\right]-ig_{BL}W_{BL}^\mu\Delta_L \;,\cr
D^\mu\Delta_R & = & \partial^\mu\Delta_R - i\dfrac{g}{2}W_{R}^{i\mu}\left[\tau^i,\Delta_R\right]-ig_{BL}W_{BL}^\mu\Delta_R \;,\cr
& &
\end{eqnarray}
from which we conclude\footnote{%
Where this relation comes from can also seen by the requirement that the mass of $Z$,
\[
M_{Z_{1}}^2\;\simeq\;
\frac{g^2}{4}\kappa_+^2 \Bigl[1+4v_L^2/\kappa_+^2+O(\kappa_+^2/v_R^2)\Bigr]
\left(\frac{g^2+2g_{BL}^2}{g^2+g_{BL}^2}\right)
\]
should match the SM one $M_Z^2=(g^2+g'^2)v^2/4$ at the scale where $SU(2)_R$ is broken.  Basically, the Weinberg angle in the LRSM is defined as $\cos^2\theta_W=\frac{g^2+g_{BL}^2}{g^2+2g_{BL}^2}$, which is matched to the SM definition $\cos^2 \theta_W=\frac{g^2}{g^2+g'^2}$
\cite{Gunion:1989in}.
}
\begin{equation}
\dfrac{g_{BL}}{g} \;=\; \dfrac{1}{\sqrt{2}}\;.
\end{equation}
The $Z'$ and $B$ fields in Eqs.~(\ref{WRandZprime}) and (\ref{Bdef}) can then be written as
\begin{eqnarray}
Z' & = & \dfrac{g W_R^3 - g_{BL}^{\phantom{0}} W_{BL}^{\phantom{0}}}{\sqrt{g^2 + g_{BL}^2}} \;,\cr
B  & = & \dfrac{g_{BL}^{\phantom{0}} W_R^3 + g W_{BL}^{\phantom{0}}}{\sqrt{g^2 + g_{BL}^2}} \;.
\end{eqnarray}
The value of $g'$ can then be obtained from the matching condition associated with the
breaking $SU(2)_R\times U(1)_{B-L}\rightarrow U(1)_Y$:
\begin{equation}
\dfrac{1}{g^{\prime 2}} \,=\, \dfrac{1}{g^2} + \dfrac{1}{g_{BL}^2} \,=\, \dfrac{3}{g^2}
\quad\rightarrow\quad
\dfrac{g'}{g} \,=\, \dfrac{1}{\sqrt{3}}\;.
\end{equation}
Note also that the $B$ field couples to $(g/2)(Y/\sqrt{3})$, where $Y/\sqrt{3}$ is defined in Eq.~(\ref{Ydef}),
and the photon field of Eq.~(\ref{Adef}) couples to $(g/2)Q$, where $Q$ is defined in Eq.~(\ref{Qdef2}).
Therefore,
\begin{equation}
g'\;=\;\dfrac{g}{\sqrt{3}}\;,\qquad
e \;=\; \dfrac{g}{2}\;,
\end{equation}
and thus $\sin^2\theta_W=1/4$,
just as in the $su(2/1)$ case.  
Perhaps this is not surprising given that
our $su(2/1)$ embedding is a sub-embedding of the $su(2/2)$ embedding.

The masses of the massive gauge bosons are now expressed as
\begin{eqnarray}
M_{W_1}^2 & = & \dfrac{g^2\kappa_+^2}{4} - \dfrac{g^2\kappa_1^2\kappa_2^2}{2v_R^2} + \cdots \;,\cr
M_{W_2}^2 & = & \dfrac{g^2(2v_R^2+\kappa_+^2)}{4} + \dfrac{g^2\kappa_1^2\kappa_2^2}{2v_R^2} + \cdots \;,\cr
M_{Z_1}^2 & = & \dfrac{(g^2+g^{\prime 2})\kappa_+^2}{4} 
- \dfrac{g^2\kappa_+^4}{27 v_R^2} + \cdots \;,\cr
M_{Z_2}^2 & = & \dfrac{g^2 (9v_R^2 + \kappa_+^2)}{6}
+ \dfrac{g^2\kappa_+^4}{27 v_R^2} + \cdots \;,
\end{eqnarray}
and we can see that $\kappa_+$ plays the role of the SM Higgs VEV 
and thus $\kappa_+ \approx 246\,\mathrm{GeV}$.

%%%%%%%%%%%%%%%%%%%%%%%%%%%%%%%%%%%%%%%%%%%%%%%%%%%%%%%%%%%%%%%%%%%%%%%%%%%%%%%%%%%%%%%%%%%%
%%%%%%%%%%%%%%%%%%%%%%%%%%%%%%%%%%%%%%%%%%%%%%%%%%%%%%%%%%%%%%%%%%%%%%%%%%%%%%%%%%%%%%%%%%%%
\subsubsection{$\sin^2\theta_W$ from the coupling to fermions}

The value of $\sin^2\theta_W$ can also be determined from the gauge coupling of the fermions.
Collecting the lepton fields into an $su(2/2)$ quartet as shown in Eq.~(\ref{quartet}),
the requirement of $SU(2)_L\times SU(2)_R\times U(1)_{B-L}$ gauge invariance
leads to the interaction
\begin{eqnarray}
\label{leptonicSU(2/2)}
-\mathcal{L}_\ell^{\mathrm{even}}
& = &
\frac{g}{2}
\sum_{i=1,2,3,8,13,14,15}
\left(\,\overline{\psi}\gamma^{\mu}\lambda^s_i\psi\,\right)J_\mu^i
\cr
& = &
\frac{g}{2}
\bigg[\,
   \overline{\ell}_L \gamma_{\mu} \left( \bm{\tau\cdot W}^{\mu}_L \right) \ell_L
 + \overline{\ell}_R \gamma_{\mu} \left( \bm{\tau\cdot W}^{\mu}_R \right) \ell_R  \nonumber\\
& & \quad
 -\frac{1}{\sqrt{2}}\; \overline{\ell}_L \gamma_{\mu} W^{\mu}_{BL} \ell_L
 -\frac{1}{\sqrt{2}}\; \overline{\ell}_R \gamma_{\mu} W^{\mu}_{BL} \ell_R
\,\bigg]
\;.
\cr & &
\end{eqnarray}
This should be compared 
with the leptonic part of the $SU(2)_L\times SU(2)_R\times U(1)_{B-L}$ Lagrangian 
in which the left- and right-handed leptons are assigned the representations 
$\ell_L(2,1,-1)$ and $\ell_R(1,2,-1)$:
\begin{eqnarray}
-\mathcal{L}_\ell^{221} 
& = & 
\overline{\ell}_L\gamma_{\mu}
\biggl[\dfrac{g_L}{2}\bm{\tau\cdot W}^{\mu}_L-\frac{g_{BL}}{2}W^{\mu}_{BL}\biggr]\ell_L
\cr
& &
+\;
\overline{\ell}_R\gamma_{\mu}
\biggl[\dfrac{g_R}{2}\bm{\tau\cdot W}^{\mu}_R-\frac{g_{BL}}{2}W^{\mu}_{BL}\biggr]\ell_R
\;.
\cr & &
\end{eqnarray}
Identifying $g=g_L$, we find $g_R=g$ and
$\sqrt{2}g_{BL}=g$, leading to the same result as above.

%%%%%%%%%%%%%%%%%%%%%%%%%%%%%%%%%%%%%%%%%%%%%%%%%%%%%%%%%%%%%%%%%%%%%%%%%%%%%%%%%%%%%%%%%%%%
\subsubsection{The Higgs Quartic Couplings and the Higgs Mass}

With a bi-doublet $\hat{\Phi}$ and two complex triplets $\Delta_{L,R}$, the Higgs sector of the model has
20 degrees of freedom of which 6 are absorbed into the massive gauge bosons,
while 14 remain physical \cite{Gunion:1989in,Deshpande:1990ip}.
Of these 4 are doubly charged, 4 are singly charged, and 6 are neutral.
We identify the observed Higgs boson with the lightest neutral member.

The masses of the physical Higgs sector particles naturally depend on the
parameters in the Higgs potential, Eq.~(\ref{GeneralV}).
Of these, we are aware of the self-couplings of $\hat{\Phi}$, which,
due to the introduction of the coupling constant are rescaled from 
Eq.~(\ref{lambdavalues}) to
\begin{equation}
\lambda_1 \,=\,  \dfrac{g^2}{2}\,,\quad
\lambda_3 \,=\, -\dfrac{g^2}{4}\,,\quad
\lambda_2 \,=\, \lambda_4 \,=\, 0\,,
\end{equation}
as are the mass parameters to
\begin{equation}
\mu_1^2 \;=\; \dfrac{g^2v^2}{2}\;,\qquad
\mu_2^2 \;=\; 0\;.
\end{equation}
The other parameters are unknown except for requirement that 
they must lead to $v_R\gg \kappa_+ =\sqrt{\kappa_1^2+\kappa_2^2} \approx 246\,\mathrm{GeV} \gg v_L\approx 0$,
and $\kappa_1 \gg \kappa_2$.
Let us approximate $\kappa_2\approx 0$.
In this case, the mass of the lightest neutral scalar, which consists mostly of the
real part of $\phi_1^0$, is given
approximately by \cite{Gunion:1989in,Deshpande:1990ip}
\begin{equation}
M_{h}^2 \;\approx\; \kappa_1^2
\left(
2\lambda_1 - \dfrac{\alpha_1^2}{2\rho_1}
\right)
\;.
\label{MhApprox}
\end{equation}
The first term is what the Higgs mass would be if the couplings to the
triplets were non-existent: 
\begin{equation}
M_h \;=\; \sqrt{2\lambda_1}\kappa_1 \;=\; g\kappa_1\;.
\end{equation}
Given that the prediction for the left-handed $W$ mass would be $M_{W_L} = M_{W_1} = g\kappa_1/2$ in this case,
this would lead to the prediction $M_h/M_W = 2$ as in the $su(2/1)$ case.
The existence of the second term shows that mixing with the neutral
components of $\Delta_{L,R}$ could lower the mass to a more realistic value.
Renormalization group running could further lower $M_h$ toward $126$ GeV.
We will look at this possibility in the subsection after the next.

%%%%%%%%%%%%%%%%%%%%%%%%%%%%%%%%%%%%%%%%%%%%%%%%%%%%%%%%%%%%%%%%%%%%%%%%%%%%%%%%%%%%%%%%%%%%
%%%%%%%%%%%%%%%%%%%%%%%%%%%%%%%%%%%%%%%%%%%%%%%%%%%%%%%%%%%%%%%%%%%%%%%%%%%%%%%%%%%%%%%%%%%%
%\newpage
\subsection{The Emergence Scale and the Right-Handed Breaking Scale}

Let us now determine the scale $\Lambda_s$ at which we envision the
$su(2/2)\sim SU(2)_L\times SU(2)_R\times U(1)_{B-L}$ structure to emerge from
an underlying NCG theory.
This is the scale at which we expect the relation
$g_{BL}/g=1/\sqrt{2}$ to hold.

When one imagines the emergence of an LRSM from an underlying UV theory
at some scale $\Lambda_s$, one
usually thinks of the subsequent breaking of the gauge symmetry down to $U(1)_{em}$
to occur in steps at several scales, a schematic diagram of which would be:
\begin{eqnarray*}
\begin{array}{rl}
\mbox{NCG}& \mbox{theory?} \\
& \downarrow  \Lambda_s \vphantom{\bigg|} \\
su(2/2)\! & \sim \; SU(2)_L \times SU(2)_R \times U(1)_{B-L} \\
& \downarrow  \Lambda_R \sim g v_R \vphantom{\bigg|} \\
su(2/1)\! & \sim \; SU(2)_L \times U(1)_Y \\
& \downarrow M_W \sim gv \vphantom{\bigg|} \\
& \!\!U(1)_{em} 
\end{array}
\end{eqnarray*}
Thus, the theory would be effectively $SU(2)_L\times SU(2)_R\times U(1)_{B-L}$
between $\Lambda_s$ and $\Lambda_R$, and
$SU(2)_L\times U(1)_Y$ between $\Lambda_R$ and $M_W$,
that is, the gauge couplings would run with the LRSM particle content between $\Lambda_s$ and $\Lambda_R$, and with the SM particle content below $\Lambda_R$.
The boundary condition we would like to impose at $\Lambda_s$ is
\begin{equation}
\dfrac{g_{BL}(\Lambda_s)}{g(\Lambda_s)}\;=\;\dfrac{1}{\sqrt{2}}\;,
%\qquad
%\dfrac{g'(\Lambda_s)}{g(\Lambda_s)}\;=\;\dfrac{1}{\sqrt{3}}\;,
\label{relation}
\end{equation}
while the matching condition at $\Lambda_R$ requires
\begin{equation}
\dfrac{1}{g^{\prime 2}(\Lambda_R)}\;=\;
\dfrac{1}{g^2(\Lambda_R)} +
\dfrac{1}{g^2_{BL}(\Lambda_R)} \;.
\end{equation}
Therefore, we have, up to one-loop, the relations
\begin{eqnarray}
\dfrac{1}{g_L^2(\Lambda_s)}
& = & \dfrac{1}{g_2^2(\Lambda_R)} -2b_L\ln\frac{\Lambda_s}{\Lambda_R}
\cr
& = & \left(\dfrac{1}{g_2^2(M_W)} - 2b_2\ln\frac{\Lambda_R}{M_W}\right)-2b_L\ln\frac{\Lambda_s}{\Lambda_R}\;,\cr
%%%
\dfrac{1}{g_{BL}^2(\Lambda_s)}
& = & \frac{1}{g_{BL}^2(\Lambda_R)} - 2b_{BL}\ln\frac{\Lambda_s}{\Lambda_R}
\cr
& = & \left(\dfrac{1}{g^{\prime 2}(\Lambda_R)}-\dfrac{1}{g_2^2(\Lambda_R)}\right)
- 2b_{BL}\ln\frac{\Lambda_s}{\Lambda_R}
\;,
\cr
%%%
\dfrac{1}{g^{\prime 2}(\Lambda_R)}
& = & \dfrac{1}{g^{\prime 2}(M_W)} - 2b_1\ln\frac{\Lambda_R}{M_W}\;.
\label{su22RGE}
\end{eqnarray}
The $b_i$'s are given by \cite{Jones:1981we}
\begin{equation}
\label{1loopgeneral}
b_{i}\;=\;\frac{1}{16\pi^2}
\biggl[
-\frac{11}{3}C_{2}(G_i)+\frac{2}{3}{\underset{f}{\sum}}T_i(f)+\frac{1}{3}{\underset{s}{\sum}}T_i(s)
\biggr]
\;,
\end{equation}
where the summation is over Weyl fermions in the second term and over scalars in the third.
The index $i$ labels the gauge groups and we have $i=1,2$ below $\Lambda_R$, and
$i=L,R,BL$ above it, and $i=3$ for QCD.
$C_2(G_i)$ is the quadratic Casimir for the adjoint representation of the group $G_i$, 
and $T_i$ is the Dynkin index of each representation. 
For $SU(N)$, $C_2(G)=N$, $T=1/2$ for doublet representations and $T=2$ for triplets. 
For $U(1)$, $C_2(G)=0$ and
\begin{equation}
\sum_{f,s}T \;=\; \sum_{f,s}\left(\dfrac{Y}{2}\right)^2\;,
\label{U1Dynkin}
\end{equation}
where $Y/2$ is the $U(1)$ charge, the factor of $1/2$ coming from the traditional 
normalizations of the hypercharge $Y$ and $B-L$ charges.
In the LRSM, we have, for each generation, 6 left-handed and 6 right-handed quarks with 
$B-L=1/3$, 2 left-handed and 2 right-handed leptons with $B-L=-1$, 6 complex scalars (coming from the two triplets) with $B-L=2$, and a bi-doublet with $B-L=0$.

\begin{figure}[t]
\leftline{
\includegraphics[width=0.47\textwidth]{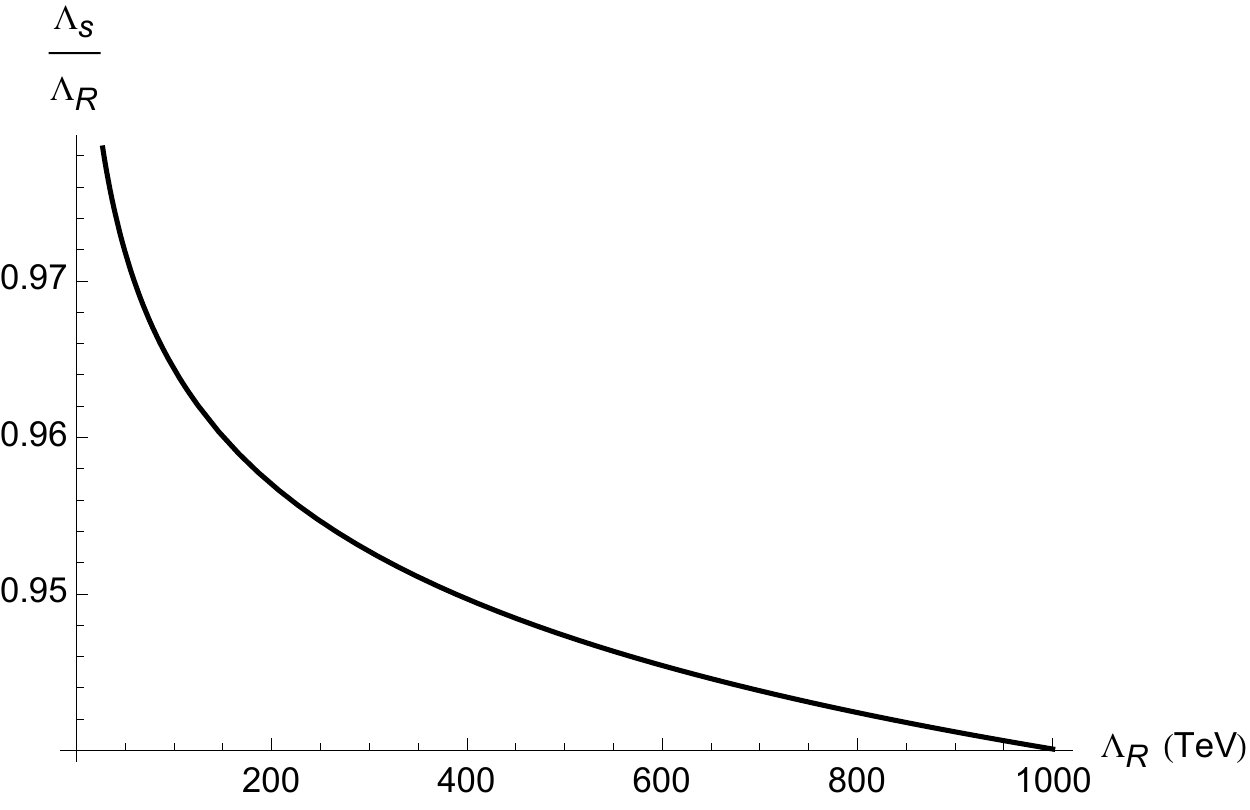}
}
\caption{The behavior of $\frac{\Lambda_s}{\Lambda_R}$ above 4 TeV, where it is exactly equal to unity.}
\label{graph}
\end{figure}

Therefore, we have
\begin{equation}
b_{L} \,=\, b_{R} \,=\, 
\frac{1}{16\pi^2}\left(-\frac{7}{3}\right) \;,\quad
b_{BL} \,=\, \frac{1}{16\pi^2}\left(\frac{14}{3}\right)\;.
\end{equation}
The values of $b_1$, $b_2$, and $b_3$ for the SM particle content have been listed earlier in
Eq.~(\ref{b123}).
Using the above RGE relations Eq.~(\ref{su22RGE}), we look for the scale $\Lambda_s$ at which the constraint Eq.~(\ref{relation}) is satisfied as a function $\Lambda_R$. 
The results are shown in Fig.~\ref{graph}. 

One possible solution is, of course, $\Lambda_s = \Lambda_R \approx 4\,\mathrm{TeV}$
since that was the scale at which $g'/g=1/\sqrt{3}$ when the couplings run with the SM particle content, and this automatically leads to $g_{BL}/g=1/\sqrt{2}$.
When $\Lambda_R$ is increased above 4~TeV, however,
as we can see from Fig.~\ref{graph}, we find that the scale $\Lambda_s$ at which
Eq.~(\ref{relation}) is satisfied is actually lower than $\Lambda_R$.
That is, we must run the couplings up to $\Lambda_R$ with the SM particle content,
and then run back down to a lower energy with the LRSM particle content to
satisfy the required boundary condition.
Obviously, this is an unphysical situation.
So we are led to conclude that our formalism demands
$\Lambda_s = \Lambda_R \approx 4\,\mathrm{TeV}$.
That is, the $SU(2)_L\times SU(2)_R\times U(1)_{B-L}$ gauge theory
emerges already broken to $SU(2)_L\times U(1)_Y$ of the SM at that scale.
In fact, it is already broken all the way down to $U(1)_{em}$ due to the
non-zero VEV of the bi-doublet field $\Phi$, though it may not necessarily be manifest
at $\Lambda_s$ due to $\Lambda_s = \Lambda_R \gg M_W$.

%%%%%%%%%%%%%%%%%%%%%%%%%
%%%%%%%%%%%%%%%%%%%%%%%%%%%%%%

This result is not inconsistent with the view we have been advocating based on our
$su(2/1)\sim SU(2)_L\times U(1)_Y$ model.
There, the theory emerged already broken to $U(1)_{em}$ at the emergence scale $\Lambda_s$,
with the Higgs field already shifted to fluctuations around its non-zero VEV.  
So schematically, we had:
\[
\begin{array}{c}
\mbox{NCG theory?}\hspace{0.5cm} \\
\downarrow  \Lambda_s\approx 4\,\mathrm{TeV}\vphantom{\bigg|} \\
\mbox{Broken $SU(2)_L \times U(1)_{Y} \xrightarrow{M_W} U(1)_{em}$} 
\end{array}
\]
For the $su(2/2)$ case, the breaking schematic is
\[
\begin{array}{c}
\mbox{NCG theory?}\hspace{1.2cm} \\
\downarrow  \Lambda_s = \Lambda_R \approx 4\,\mathrm{TeV} \vphantom{\bigg|} \\
\mbox{Broken $SU(2)_L \times SU(2)_R \times U(1)_{B-L} \xrightarrow{M_W} U(1)_{em}$} 
\end{array}
\]
These schematics suggest that 
the physics responsible for the emergence of the $su(2/1)$ or $su(2/2)$ patterns 
from the underlying theory
may also be responsible for the spontaneous breaking of the chiral gauge symmetries.
A possible and attractive scenario would be that this new physics is 
geometric in nature and is triggered by the separation of the two branes from each other, as evidenced in the fact that the matrix derivative encodes
information on the Higgs VEV.

Another attractive point about the scale $\Lambda_R$ being on the order of 4~TeV is
that it would place the masses of all the new particles associated with the LRSM at 
that scale, perhaps light enough to be discovered just beyond their current experimental limits.
We will return to this observation in section~\ref{PHENO}.

%%%%%%%%%%%%%%%%%%%%%%%%%%%%%%%%%%%%%%%%%%%%%%%%%%%%%%%%%%%%%%%%%%%%%%%%%%%%%%%%%%%%%%%%%%%%
%%%%%%%%%%%%%%%%%%%%%%%%%%%%%%%%%%%%%%%%%%%%%%%%%%%%%%%%%%%%%%%%%%%%%%%%%%%%%%%%%%%%%%%%%%%%
\subsection{The Higgs boson mass from $\bm{su(2/2)}$}

Let us now discuss how the observed Higgs mass can be accommodated in our $su(2/2)$
framework.
We have seen that both $su(2/1)$ and $su(2/2)$ embeddings 
place the emergence scale at $\sim 4$~TeV.
Moreover, they demand the same boundary condition on the Higgs quartic coupling 
($\lambda$ for $su(2/1)$, $\lambda_1$ for $su(2/2)$) at that scale.
In the $su(2/1)$ case, this led to a prediction of the Higgs mass of $M_h(M_Z)\sim 170\,\mathrm{GeV}$.
In the $su(2/2)$ case, however, the Higgs mass prediction can be lowered due to 
the mixing of the Higgs with other neutral scalars available in the model.

In this section, we will investigate the simplest option as an example. 
We will assume that only a SM singlet scalar $S$ survives dominantly at low energies ($\sim M_Z$) which is responsible for the mass of the right handed neutrino.  In our $su(2/2)$ embedding, this can be taken to be the real part of $\delta_R^0$: 
\begin{equation}
\delta_R^0 \;=\; \dfrac{v_R + S + iT}{\sqrt{2}} \;.
\end{equation}
We assume that $S$ couples to the SM Higgs field $H$ via
\begin{equation}\label{singletL}
\mathcal{L}_S
\;=\; \frac{1}{2}\partial_{\mu}S\partial^{\mu}S
-\frac{m^2}{2}S^2
-\frac{\lambda_S}{4}S^4
-\lambda_{HS}H^{\dagger}H S^2 
\;.
\end{equation}
This model, in which the SM is extended with a singlet scalar $S$, has been analyzed in detail 
previously in the contexts of the vacuum stability of the SM \cite{EliasMiro:2012ay,Lebedev:2012zw}, and dark matter \cite{Chen:2012faa,darkmatter}. 
In the $su(2/2)$ framework, terms in the above Lagrangian result from
the terms with coefficients $\rho_1$ and $\alpha_1$ in Eq. (\ref{GeneralV}), provided that
\begin{equation}
\lambda_S    \;=\;\rho_1\;,\qquad
\lambda_{HS} \;=\;\dfrac{\alpha_1}{2}\;.
\end{equation}
A similar singlet field is found in the Spectral SM \cite{Chamseddine:2012sw}.
See also Refs.~\cite{Devastato:2013oqa,Devastato:2014xga}.

%%%%%%%%%%%%%%%%%%%%%%%%%
\begin{figure}[t]
\leftline{
\includegraphics[width=0.47\textwidth]{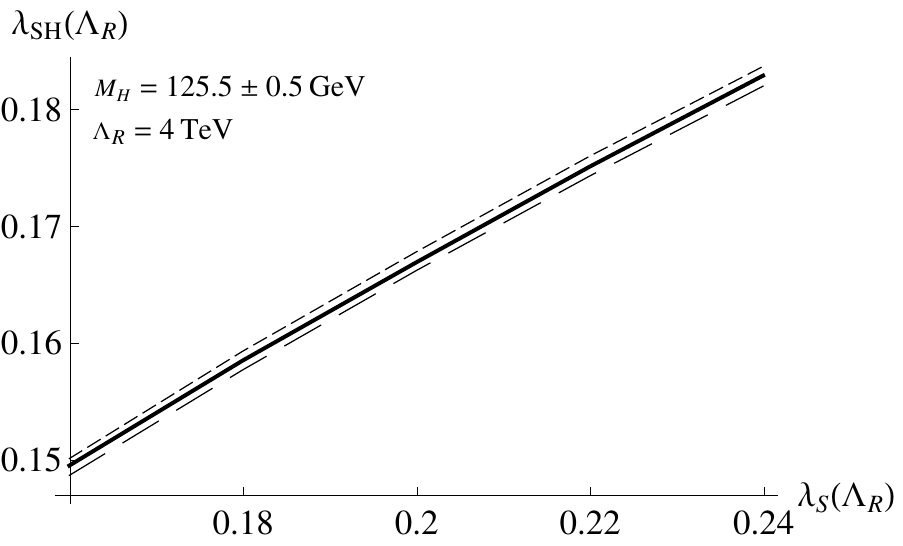}
}
\caption{A patch of the parameter space which gives the observed Higgs mass.}
\label{lmvalues}
\end{figure}
%%%%%%%%%%%%%%%%%%%%%%%%%%%%%%

Assuming that in addition to $S$, the right-handed neutrino survives a low energies,
we obtain the following renormalization group equations (RGE's)
for the evolution of the relevant parameters \cite{Chen:2012faa,Haba:2014zda}:
\begin{eqnarray}
\mu\frac{d h_t}{d\mu}
& = & \frac{h_t}{(4\pi)^2}
\left[
\frac{9}{2}h_t^2+ h_{\nu}^2-\left(\frac{17}{12}g'^2+\frac{9}{4}g^2+8g_s^2\right)
\right]
\;,\cr
\mu\frac{d h_{\nu}}{d\mu}
& = & \frac{h_{\nu}}{(4\pi)^2}
\left[
3h_t^2+\frac{5}{2}h_{\nu}^2-\left(\frac{3}{4}g'^2+\frac{9}{4}g^2\right)
\right]
\;,\cr
\mu\frac{d \lambda}{d\mu}
& = & \frac{1}{(4\pi)^2}
\biggl[
\Bigl\{12 h_t^2+4 h_{\nu}^2-\left(3 g'^2+9 g^2\right)\Bigr\}\lambda
\cr
& & \hspace{1.2cm}-2h_{\nu}^4 -6 h_t^4 +24\lambda^2 +2\lambda_{HS}^2
\cr
& & \hspace{1.2cm}+\frac{3}{8}\left(g'^4+2g'^2 g^2+3g^4\right)
\biggr]
\;,\cr
\mu\frac{d \lambda_{HS}}{d\mu}
& = & \frac{\lambda_{HS}}{(4\pi)^2}
\biggl[
6h_t^2+2h_{\nu}^2-\left(\frac{3}{2}g'^2+\frac{9}{2}g^2\right)
\cr
& & \hspace{1.3cm} +2\left(6\lambda+3\lambda_S+4\lambda_{HS}\right)
\biggr]
\;,\cr
\mu\frac{d \lambda_{S}}{d\mu}
& = & \frac{1}{(4\pi)^2}\left(8\lambda_{HS}^2+18\lambda_{S}^2\right)
\;,
\end{eqnarray}
where $h_t$ and $h_{\nu}$ are the top-quark and right-handed neutrino Yukawa couplings, $\lambda$ and $\lambda_S$ are the Higgs and the singlet quartic couplings, and $\lambda_{HS}$ is the Higgs-singlet coupling. The boundary conditions we use are $h_t(M_Z)=0.997$, obtained from $h_t(M_Z)=\sqrt{2}m_t/v$, and
$\lambda(\Lambda_R)=g^2(\Lambda_R)/2$,
where the latter is fixed by our $su(2/2)$ construction.
We also assume $h_{\nu}\sim 10^{-6}$, which is necessary to generate the correct light neutrino mass from the TeV scale seesaw, if the Dirac mass $M_D\approx M_e$. 
We also need the values of
$\lambda_S(\Lambda_R)$ and $\lambda_{HS}(\Lambda_R)$ as boundary conditions for our RGE's, 
but these are not fixed by our $su(2/2)$ framework (yet).
Thus, we allow these values to float to find the conditions that lead to the correct
Higgs mass.
The mass of the Higgs is determined from \cite{Chamseddine:2012sw}
\begin{eqnarray}
M_h^2
& = & \lambda v^2+\lambda_S v_R^2-\sqrt{\left(\lambda v^2-\lambda_S v_R^2\right)^2+4\lambda_{HS}^2v^2v_R^2} \cr
& \simeq & 2v^2\lambda\left(1 -\frac{\lambda_{HS}^2}{\lambda\lambda_S}\right)
\;,
\label{MhUfuk}
\end{eqnarray}
consistent with Eq.~(\ref{MhApprox}).
We have also set $v_R=\Lambda_R\simeq4$ TeV in this calculation.

The resulting range of values for $\lambda_S(\Lambda_R)$ and $\lambda_{HS}(\Lambda_R)$ 
which reproduce the correct Higgs mass is shown in Fig.~\ref{lmvalues}.
The plot shows the range of values ($0.15\sim 0.25$) in the perturbative region,
but larger values for these couplings are also possible as long as $(1-\lambda_{HS}^2/\lambda\,\lambda_S)\geq 0$, while $\lambda$ remains small.
Thus, the $su(2/2)$ structure can accommodate the correct Higgs mass, provided
the parameters in the Higgs potential are in the appropriate ranges.

%%%%%%%%%%%%%%%%%%%%%%%%%%%%%%%%%%%%%%%%%%%%%%%%%%%%%%%%%%%%%%%%%%%%%%%%%%%%%%%%%%%%%%%%%%%%
\bigskip
\subsection{$su(2/2)$ summary}

In this section, we have applied the formalism developed for the 
$su(2/1)$ embedding of the SM in the previous section
to an $su(2/2)$ superconnection into which the $SU(2)_L\times SU(2)_R\times U(1)_{B-L}$
gauge bosons of the LRSM were embedded in its diagonal even part, and a bi-doublet Higgs field $\Phi$ in its off-diagonal odd part.
Unlike the $su(2/1)$ case, the matrix derivative could be made nilpotent, and the
supercurvature $\mathcal{F}$ followed a simple transformation law under gauge
transformations.  
To the gauge invariant but spontaneously broken (to $SU(2)_V\times U(1)_{B-L}$)
action derived from $\mathcal{F}$, in which the bi-doublet $\Phi$ was already 
the shifted field fluctuating around the VEV $\zeta/\sqrt{2}$, we introduced two triplet fields 
$\Delta_{L,R}$ to achieve the breaking sequence
$SU(2)_L\times SU(2)_R\times U(1)_{B-L} \rightarrow SU(2)_L\times U(1)_Y \rightarrow U(1)_{em}$.

The predictions of the formalism were $g_{BL}/g=1/\sqrt{2}$ (or, equivalently, $g'/g=1/\sqrt{3}$) and $\lambda_1=-2\lambda_3=g^2/2$, $\lambda_2=\lambda_4=0$, where
the $\lambda_i$'s are quartic self couplings of the bi-doublet $\Phi$.
Of these, $\lambda_1$ corresponds to the quartic self coupling $\lambda$ of the SM Higgs.
Assuming the above symmetry breaking sequence, it was found that the
condition $g_{BL}/g=1/\sqrt{2}$ could only be imposed if the emergence scale $\Lambda_s$
of the $su(2/2)$ structure and the breaking scale $\Lambda_R$ of the LRSM down to the SM
were the same and at $\sim 4\,\mathrm{TeV}$.
Thus, the formalism demands that the LRSM emerge from the hypothetical underlying
NCG theory already fully broken.
Despite the formalism's predictions including the emergence scale
being essentially the same as in the $su(2/1)$ case,
the observed Higgs mass could still be accommodated due to the availability of 
other neutral scalar degrees of freedom in the model which mixed with the Higgs.

%%%%%%%%%%%%%%%%%%%%%%%%%%%%%%%%%%%%%%%%%%%%%%%%%%%%%%%%%%%%%%%%%%%%%%%%%%%%%%%%%%%%%%%%%%%%
%%%%%%%%%%%%%%%%%%%%%%%%%%%%%%%%%%%%%%%%%%%%%%%%%%%%%%%%%%%%%%%%%%%%%%%%%%%%%%%%%%%%%%%%%%%%
%%%%%%%%%%%%%%%%%%%%%%%%%%%%%%%%%%%%%%%%%%%%%%%%%%%%%%%%%%%%%%%%%%%%%%%%%%%%%%%%%%%%%%%%%%%%
\section{Fermions}\label{Fermions}

For the superconnection formalism presented here to be taken seriously, we must
have the freedom to couple fermions to the Higgs field with arbitrary Yukawa couplings,
or the formalism must be able to predict what these Yukawa couplings should be.
This is a difficult problem given that the superconnection formalism is essentially
a gauge-Higgs unified theory, and as with any such scenario, if one naively couples the
superconnection $\mathcal{J}$ to the fermions, the Yukawa couplings are forced to be equal to the gauge couplings \cite{Hosotani:1983xw,Hosotani:1983vn,Hosotani:1988bm,Hatanaka:1998yp,Hall:2001zb,Burdman:2002se,Haba:2002py}.

To see how this comes about in the $su(2/1)$ case,
let us assume that the Lagrangian of the lepton $su(2/1)$ triplet $\psi$ coupled to the
superconnection $\mathcal{J}$ is given, schematically, by
\begin{equation}
\mathcal{L}_\ell 
\;=\; i\bar{\psi}\mathcal{D}\psi
\;=\; 
i\bar{\psi}\left[
\mathbf{d}+\dfrac{g}{2}\left(\mathbf{d}_M + \mathcal{J}\right)
\right]\psi
\;,
\end{equation}
where all the operators within the parenthesis must be placed
in the appropriate representations.
In the spinorial representation, the 1-form $dx^\mu$ is represented by the
Dirac matrix $\gamma^\mu$ \cite{Kahler:1960zz,Graf:1978kr,Banks:1982iq,Benn:1982sr}.  Thus, we have the replacements
\begin{eqnarray}
d 
& = & dx^\mu\wedge\partial_\mu 
\;\rightarrow\; \gamma^\mu\partial_\mu \;=\; \slashed{\partial} \;, 
\cr
W^i
& = & W^i_\mu dx^\mu
\;\rightarrow\; W^i_\mu \gamma^\mu \;=\; \slashed{W}^i \;, 
\cr
B
& = & B_\mu dx^\mu
\;\rightarrow\; B_\mu \gamma^\mu \;=\; \slashed{B} \;, 
\end{eqnarray}
and consequently,
\begin{eqnarray}
\mathbf{d}
& \rightarrow & \slashed{\partial}\cdot\mathbf{1}_{3\times 3}\;,\cr
\mathcal{J}
& \rightarrow & i
\begin{bmatrix}
\bm{\tau\cdot\slashed{W}}-\frac{1}{\sqrt{3}}\slashed{B}\cdot\mathbf{1}_{2\times 2}
& \sqrt{2}\phi \\
\sqrt{2}\phi^\dagger 
& -\frac{2}{\sqrt{3}}\slashed{B}
\end{bmatrix}
\;\equiv\; J
\;.
\cr & &
\end{eqnarray}
We define the Dirac operator $D$ as
\begin{equation}
D \;=\; \slashed{\partial}\cdot\mathbf{1}_{3\times 3} + i\dfrac{g}{2}\eta\;,
\end{equation}
to represent the generalized exterior derivative $\mathbf{d}_S = \mathbf{d}+\frac{g}{2}\mathbf{d}_M$.
Then, the Lagrangian is found to be
\begin{eqnarray}
\lefteqn{\mathcal{L}_\ell
\;\rightarrow\;
i\bar{\psi}\left(D+\dfrac{g}{2}J\right)\psi
}
\cr
& = &
 \overline{\ell}_L i\slashed\partial\,\ell_L
+\overline{\ell}_R i\slashed\partial\,\ell_R
\vphantom{\bigg|}
\cr
& & - \frac{g}{2}\left(
      \overline{\ell}_L\xi          \,\ell_R
     +\overline{\ell}_R\xi^{\dagger}\,\ell_L
      \right) 
     -\frac{g}{\sqrt{2}}\left(
      \overline{\ell}_L\phi          \,\ell_R
     +\overline{\ell}_R\phi^{\dagger}\,\ell_L
      \right) \cr
& & - \frac{g}{2}\,\left[
(\overline{\ell}_L\gamma^{\mu}\tau_i\,\ell_L) W^i_{\mu}
-\frac{1}{\sqrt{3}}\left(\,
    \overline{\ell}_L\gamma^{\mu}\ell_L
+2\,\overline{\ell}_R\gamma^{\mu}\ell_R
\,\right)B_{\mu}
\right]
\cr
& = &
 \overline{\ell}_L i\slashed\partial\,\ell_L
     +\overline{\ell}_R i\slashed\partial\,\ell_R
     -\frac{g}{\sqrt{2}}\left(
      \overline{\ell}_L\hat{\phi}          \,\ell_R
     +\overline{\ell}_R\hat{\phi}^{\dagger}\,\ell_L
      \right) \cr
& & - \frac{g}{2}\,\left[
(\overline{\ell}_L\gamma^{\mu}\tau_i\,\ell_L) W^i_{\mu}
-\frac{1}{\sqrt{3}}\left(\,
    \overline{\ell}_L\gamma^{\mu}\ell_L
+2\,\overline{\ell}_R\gamma^{\mu}\ell_R
\,\right)B_{\mu}
\right]
\;,
\cr & &
\end{eqnarray}
where we have defined $\hat{\phi} = \phi+\xi/\sqrt{2}$ as before.
Thus, in addition to $g'/g=1/\sqrt{3}$, we find that
the lepton Yukawa coupling is fixed to $g/\sqrt{2}$,
and the matrix derivative terms couple the leptons to the
Higgs VEV and yield the charged lepton mass.
Note also that this Lagrangian is manifestly $SU(2)\times U(1)_Y$
gauge invariant when written in terms of the shifted field $\hat{\phi}$,
but the invariance is already spontaneously broken with fermion mass terms 
when written in terms of the $\phi$ field appearing in the superconnection.

In order to be able to change the Yukawa coupling to an
arbitrary value, one must 
have the freedom to multiply $\xi$ in the matrix derivative,
which determines $\langle\hat{\phi}\rangle$, 
and $\phi$ in the superconnection $\mathcal{J}$ by the same
constant for each fermion flavor to
maintain the gauge invariance of the Lagrangian when written in terms of 
$\hat{\phi}$.
In the Spectral SM approach \cite{Connes:1990qp,Chamseddine:1991qh,Connes:1994yd,Chamseddine:1996zu,Connes:2006qv,Chamseddine:2007hz,Chamseddine:2010ud,Chamseddine:2012sw,Chamseddine:2013rta}, this is accomplished by
writing the Dirac operator $D$ in full fermion flavor space,
including all three generations, and
inserting the complete fermion mass-mixing matrix into the
off-diagonal matrix derivative part.  
This is in accordance with the idea that the matrix derivative encodes
information on symmetry breaking.
The superconnection $J$ is also defined via the generalized
exterior derivative using $D$, which passes on the information 
included in the matrix derivative to the couplings of the $\phi$.  
The information also feeds into the supercurvature, from which one
determines the gauge-Higgs action and the Higgs VEV.
This procedure allows for the introduction of arbitrary masses and mixings
into the fermion sector.

Thus, the Spectral SM shows that it is possible to embed the required masses and mixings of the fermions to reproduce the SM into the `geometry' of the NCG discrete direction.
The interesting point is that in the Spectral SM approach, it is the fermion masses
that are the input and the Yukawa couplings the output, and not the other way around
as in the standard approach.  
The breaking of the gauge symmetry is encoded in the geometry, which is
given in terms of the fermion masses and from which one extracts the Higgs VEV,
and one could say that the Yukawa interactions themselves are consequences rather than the reason for fermion mass.

It must be said, though, that this is actually a highly unsatisfactory state of affairs.
One wishes the NCG to determine the fermion masses and mixings,
and not the other way around.
But for that a full theory of NCG dynamics would be necessary.
So for the time being, we leave the prediction of the fermion masses and mixings to
a possible future theory, and simply deal with the problem by 
assuming that when the $su(2/1)$ (or $su(2/2)$) structure
emerges from the underlying NCG theory at the emergence scale, the geometry,
a full description of which could be fairly complicated,
in addition to breaking the gauge symmetries also
fixes the fermion masses and mixings to the observed values.

%%%%%%%%%%%%%%%%%%%%%%%%%%%%%%%%%%%%%%%%%%%%%%%%%%%%%%%%%%%%%%%%%%%%%%%%%%%%%%%%%%%%%%%%%%%%
%%%%%%%%%%%%%%%%%%%%%%%%%%%%%%%%%%%%%%%%%%%%%%%%%%%%%%%%%%%%%%%%%%%%%%%%%%%%%%%%%%%%%%%%%%%%
%%%%%%%%%%%%%%%%%%%%%%%%%%%%%%%%%%%%%%%%%%%%%%%%%%%%%%%%%%%%%%%%%%%%%%%%%%%%%%%%%%%%%%%%%%%%
\section{Phenomenology}\label{PHENO}

We have been led to the possibility that a LRSM emerges
from an underlying NCG theory at the scale of $\Lambda_s\approx 4\,\mathrm{TeV}$,
which also breaks to $SU(2)_L\times U(1)_Y$ with a triplet VEV of $v_R=O(\Lambda_R/g)$ with
$\Lambda_R = \Lambda_s$.
An additional constraint that the model predicts is that the Higgs quartic coupling
$\lambda_1$ and the $SU(2)$ coupling $g$ are related by $\lambda_1 = g^2/2$ at that scale.
In this section, let us look at what the phenomenological consequences are of such a scenario.

Since we envision that the UV theory above the emergence scale of $\Lambda_s=4\,\mathrm{TeV}$ is a NCG theory with a discrete extra dimension, the
existence of the extra dimension at such a low scale should have 
observable consequences beyond predicting a LRSM with a particular boundary condition.
However, since the extra dimension is also discrete, 
it is not at all clear what it means to have a `scale' associated with it.
Having zero measure, it cannot be populated by extra degrees of freedom, which,
for a continuum extra dimension model, would lead to Kaluza-Klein states.
Lacking an understanding of the hypothetical UV NCG theory, 
it is difficult to state what to expect, so we will, for now,
concentrate on the more conventional phenomenology of the effective LRSM the formalism predicts.
Considerations of more exotic `smoking gun' signatures will be left to future works.

%%%%%%%%%%%%%%%%%%%%%%%%%%%%%%%%%%%%%%%%%%%%%%%%%%%%%%%%%%%%%%%%%%%%%%%%%%%%%%%%%%%%%%%%%%%%
\subsection{New Particles}

First and foremost, the LRSM we have been considering predicts a plethora of new and heavy particles: $W_2^\pm$ (which are mostly $W_R^\pm$), $Z_2$ (which is mostly $Z'$), and a variety of neutral, singly charged, and doubly charged scalars originating in the Higgs sector
that are denoted $H^0_1$, $H^0_2$, $H^0_3$, $A^0_1$, $A^0_2$,
$H^\pm_1$, $H^\pm_2$, $\delta_{L}^{\pm\pm}$ and $\delta_R^{\pm\pm}$ 
in Ref.~\cite{Zhang:2007da}.\footnote{%
The lightest neutral Higgs particle, denoted $h^0$ in Ref.~\cite{Zhang:2007da}, is
identified with the SM Higgs and is not included in this list.
}
The coupling of the triplet to the leptons, Eq.~(\ref{MajoranaInteraction}), 
will also generate massive Majorana neutrinos, which we will denote $N$ 
or $N_R$ (since they are mostly $\nu_R$) with a possible flavor index.

All these new particles will receive masses from the same triplet VEV, $v_R=O(\Lambda_R/g)$, so we can expect them all to have masses in the multi-TeV range, placing some of them, hopefully, within LHC reach.  
The actual masses will depend on the many parameters of the model, e.g. those
appearing in the Higgs potential. 
One concrete prediction we can make is that
$M_{Z_2}/M_{W_2}\approx\sqrt{3}$,
so, for instance, if $M_{W_2}=4\,\mathrm{TeV}$ then
$M_{Z_2} \approx \sqrt{3}\,M_{W_2} \approx 7\,\mathrm{TeV}$.
Thus, the actual particle masses need not all be concentrated at $4$~TeV,
and one expects a spread out spectrum.
The TeV scale Majorana masses of the $N_R$'s will also allow us to invoke 
a TeV scale see-saw mechanism to suppress regular neutrino masses.

Bounds on these new particle masses exist from various low energy observables, 
and from direct searches at the LHC.  
Let us take a look at what they are.

%%%%%%%%%%%%%%%%%%%%%%%%%%%%%%%%%%%%%%%%%%%%%%%%%%%%%%%%%%%%%%%%%%%%%%%%%%%%%%%%%%%%%%%%%%%%
%%%%%%%%%%%%%%%%%%%%%%%%%%%%%%%%%%%%%%%%%%%%%%%%%%%%%%%%%%%%%%%%%%%%%%%%%%%%%%%%%%%%%%%%%%%%
\subsection{Bounds from low energy processes}

%%%%%%%%%%%%%%%%%%%%%%%%%%%%%%
\begin{figure}[t]
\includegraphics[width=8.7cm]{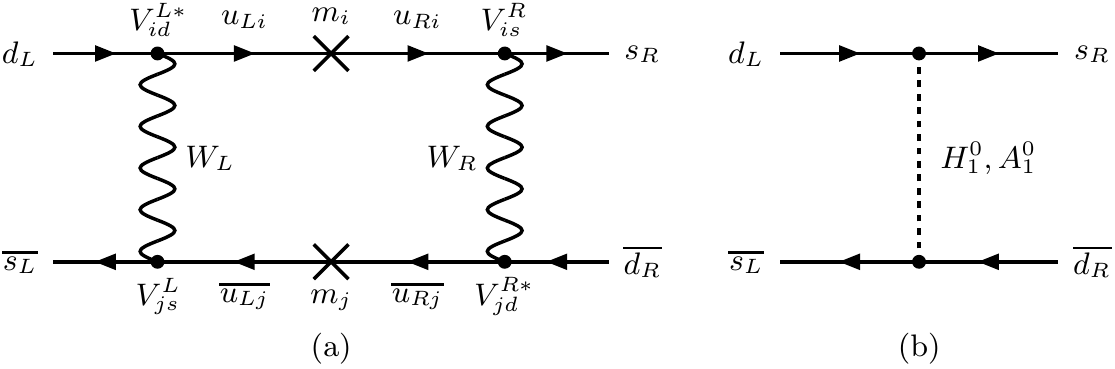}
\caption{Processes that contribute to $K^0$-$\overline{K^0}$ mixing in the LRSM.
In (a), the indices $i$ and $j$ run over the three generations of up-type quarks $u$, $c$, and $t$.
Diagrams rotated by $90^\circ$, $180^\circ$, and $270^\circ$ also contribute.
}
\label{KKbarMixing}
\end{figure}
%%%%%%%%%%%%%%%%%%%%%%%%%%%%%%

Constraints on the LRSM from low energy processes
have been heavily studied in the literature
for both the non-supersymmetric LRSM \cite{Beall:1981zq,Beall:1981ze,Mohapatra:1983ae,Gilman:1983bh,Ecker:1983uh,Ecker:1983dj,Ecker:1985ei,Ecker:1985vv,Leurer:1985rh,Herczeg:1985cx,He:1988th,Langacker:1989xa,London:1989cf,Chang:1990sfa,Aquino:1991mn,Barenboim:1996wz,Ball:1999yi,Kiers:2002cz,Raidal:2002ph,Czakon:2002wm,Zhang:2007da,Chen:2008se,Chen:2008kt,Xu:2009nt,Maiezza:2010ic,Bayes:2011zza,Chakrabortty:2012pp,Bertolini:2014sua,Babu:1993hx,Rizzo:1994aj,Rizzo:1997bs,Kim:1999wb},
and its supersymmetric extension \cite{Couture:1997eq,Frank:2003yi,Zhang:2007qma,Dutta:2007ue}.
Processes and observables that have been considered include
muon decay $\mu\rightarrow e \overline{\nu}_e\nu_{\mu}$,
neutron beta decay $n\rightarrow p\,e^- \overline{\nu}_e$,
the neutron electric dipole moment ($n$EDM),
$K^0\overline{K^0}$ mixing (\textit{i.e.} the $K_L^0$-$K_S^0$ mass difference
$\Delta M_K = M_{K_L^0}-M_{K_S^0}$ and the $K$-decay CP violation parameters $\epsilon$ and $\epsilon'$,
\textit{c.f.} Fig.~\ref{KKbarMixing}),
$D^0\overline{D^0}$ mixing, 
$B^0\overline{B^0}$ mixing ($\Delta M_{B_d}$, $\Delta M_{B_s}$, and CP violation in 
hadronic $B$ decays), $b$ semileptonic decay, and
$b\rightarrow s\gamma$.

Of the constraints thus obtained, those on the mass of $W_2$, which is mostly $W_R$,
are fairly robust and independent of the detailed form of the Higgs potential.
This is due to the $SU(2)_R$ gauge coupling being well known in the LRSM,
and for the quark sector, if one assumes the Yukawa interaction with the
bi-doublet $\Phi$, Eq.~(\ref{quarkYukawas}), to be solely responsible for 
the quark masses and mixings, then the right-handed CKM matrix $V^R$
can be fairly well constrained from quark masses and the usual left-handed CKM matrix $V^L$.
According to the analysis of Ref.~\cite{Zhang:2007da},
$\Delta M_K$ yields 2.5~TeV, while the combination of $\epsilon$ and $n$EDM yields 4~TeV, respectively, 
as the lower bound of $M_{W_2}$. 
The 4 TeV bound matches precisely our right-handed scale $\Lambda_R$ where we expect
typical new particle masses to be.

Bounds on scalar masses are more model dependent, and can be much stronger than that on $M_{W_R}$.
For instance, $H^0_1$ and $A^0_1$ exchange can contribute to $K^0\overline{K^0}$ mixing at tree-level
(see Fig.~\ref{KKbarMixing}(b)), and Ref.~\cite{Zhang:2007da} uses $\Delta M_K$ to place a lower bound of 15 TeV on their masses.

%%%%%%%%%%%%%%%%%%%%%%%%%%%%%%%%%%%%%%%%%%%%%%%%%%%%%%%%%%%%%%%%%%%%%%%%%%%%%%%%%%%%%%%%%%%%
%%%%%%%%%%%%%%%%%%%%%%%%%%%%%%%%%%%%%%%%%%%%%%%%%%%%%%%%%%%%%%%%%%%%%%%%%%%%%%%%%%%%%%%%%%%%
\subsection{LHC Signatures}

CMS and ATLAS have looked for BSM signals including $W_R$, $Z'$, and $\delta_{L,R}^{\pm\pm}$'s of the LRSM 
in proton-proton collisions of energies up to $\sqrt{s}=8\,\mathrm{TeV}$, 
and will continue their searches when the LHC resumes operation in 2015
at the center of mass energy of $\sqrt{s}=13\,\mathrm{TeV}$. 
Here, we cite some of their current bounds.

%%%%%%%%%%%%%%%%%%%%%%%%%%%%%%%%%%%%%%%%%%%%%%%%%%%%%%%%%%%%%%%%%%%%%%%%%%%%%%%%%%%%%%%%%%%%
\subsubsection{$W_R$ and $N_R$}

\begin{figure}[t]
\includegraphics[width=8.7cm]{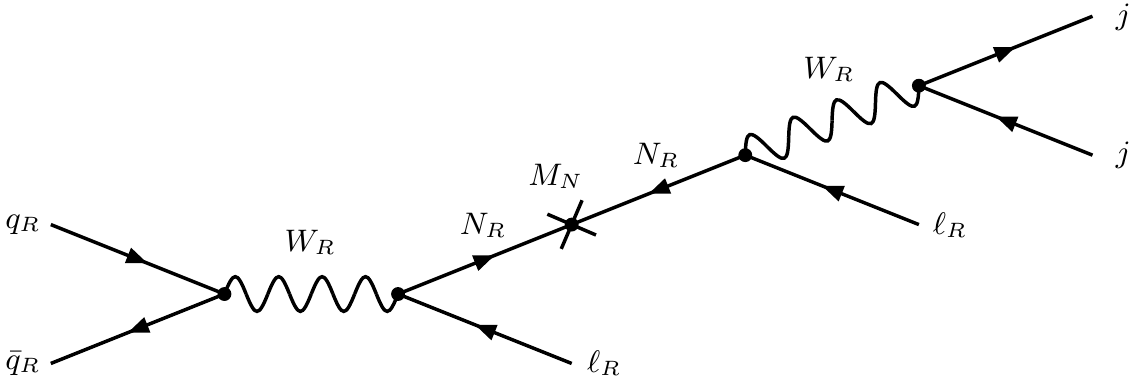}
\caption{The Feynman diagram for the production of a heavy right-handed neutrino and its decay to a dilepton and a dijet through $W_R$ exchange.
$M_N$ is the Majorana mass of $N_R$.}
\label{DrellYan}
\end{figure}

Since the heavy neutrino $N_R$ has a large Majorana mass in our construction, 
a distinctive signature for the model at the LHC is a same-sign dilepton + dijet final state with no missing energy via  $W_R^{\pm}\rightarrow N_R \ell^{\pm}\rightarrow \ell^{\pm}\ell^{\pm}jj$ \cite{Keung:1983uu}. 
The Drell-Yan diagram for this lepton number violating process is shown in Fig.~\ref{DrellYan},
the violation due to the Majorana mass insertion on the $N_R$ line.
Since a Dirac mass does not violate lepton number, a Dirac neutrino would only allow
oppositely-charged dileptons in the final state.
Therefore, LHC searches for the same-sign channel are important in determining whether
the heavy neutrino $N_R$ is Majorana or Dirac.

%%%%%%%%%%%%%%%%%%%%%%%%%%%%%%
%%%%%%%%%%%%%%%%%%%%%%%%%%%%%%

The cross section of the process naturally depends on both $W_R$ and $N_R$ masses, 
so the searches exclude correlated regions in $M_{W_R}$-$M_N$ space.
In particular, if $M_{W_R}<M_N$ then the process will be highly suppressed and
be undetectable, so certain regions cannot be probed.
If the intermediate state is $N_e$ ($N_\mu$), then the dilepton in the final state will be $ee$ ($\mu\mu$).  However, if the $N_e$ and $N_\mu$ mix, then one can also have $e\mu$ final states.
Thus, the bounds could also depend on which channels are included in the analysis.

In Ref.~\cite{ATLAS:2012ak},
ATLAS reports that for both no-mixing and maximal-mixing scenarios (between $N_e$ and $N_{\mu}$), they have excluded $W_R$ of mass up to 1.8 TeV (2.3 TeV) at $95\%$ C.L.,
assuming a mass difference between $W_R$ and $N_\ell$ larger than 0.3 TeV (0.9 TeV). 

In Ref.~\cite{Khachatryan:2014dka},
CMS reports that $W_R$ of mass up to 3.0 TeV have been excluded at $95\%$ C.L.
for the $ee$ and $\mu\mu$ channels separately, and also with the two channels 
combined assuming degenerate $N_e$ and $N_\mu$ masses.

%%%%%%%%%%%%%%%%%%%%%%%%%%%%%%%%%%%%%%%%%%%%%%%%%%%%%%%%%%%%%%%%%%%%%%%%%%%%%%%%%%%%%%%%%%%%
\subsubsection{Doubly Charged Higgs}

Another signature of the LRSM which could be observed at the
LHC comes from the triplet Higgs channel. 
In particular, the doubly-charged Higgs of the triplet decaying to two same-sign leptons
\begin{equation}
\delta_{L,R}^{\pm\pm} \;\rightarrow\; e^{\pm}e^{\pm},\,\mu^{\pm}\mu^{\pm},e^{\pm}\mu^{\pm}
\end{equation}
via the Majorana interaction, Eq.~(\ref{MajoranaInteraction}),
will present a very clear and distinctive signal. 

In Ref.~\cite{ATLAS:2012hi},
ATLAS reports that $\delta_L^{\pm\pm}$ of masses up to 409 GeV, 398 GeV, and 375 GeV, and $\delta_R^{\pm\pm}$ of masses up to 322 GeV, 306 GeV, and 310 GeV 
have been excluded at $95\%$ C.L. for the $ee$, $\mu\mu$, and $e\mu$ final states, respectively. 
These results assume a branching fraction of $100\%$ for each final state. 
For smaller branching fractions, the bounds will be weaker.

In Ref.~\cite{Chatrchyan:2012ya},
CMS reports the mass bounds on the doubly charged Higgs in the left-handed triplet
extension of the SM (type II seesaw model). 
At the $95\%$ C.L., the bounds are 
444 GeV, 453 GeV, 373 GeV, 459 GeV, 375 GeV, and 204 GeV respectively when
100\% decay to $ee$, $e\mu$, $e\tau$, $\mu\mu$, $\mu\tau$, and $\tau\tau$ final
states are assumed.

%%%%%%%%%%%%%%%%%%%%%%%%%%%%%%%%%%%%%%%%%%%%%%%%%%%%%%%%%%%%%%%%%%%%%%%%%%%%%%%%%%%%%%%%%%%%
%%%%%%%%%%%%%%%%%%%%%%%%%%%%%%%%%%%%%%%%%%%%%%%%%%%%%%%%%%%%%%%%%%%%%%%%%%%%%%%%%%%%%%%%%%%%
\subsection{Neutrinoless double-$\beta$ decay}

%%%%%%%%%%%%%%%%%%%%%%%%%%%%%%
\begin{figure}[t]
\includegraphics[width=8.5cm]{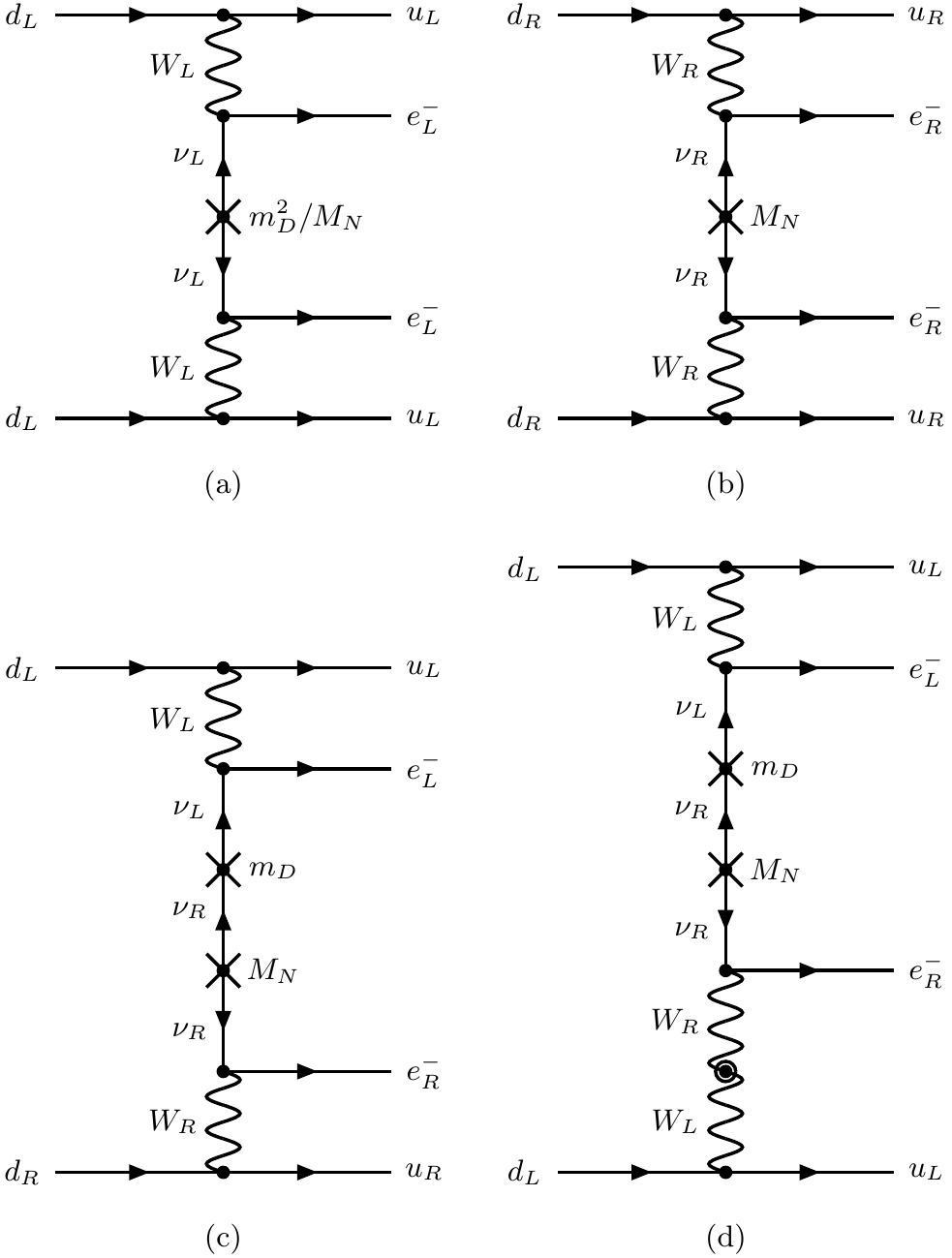}
\caption{Neutrinoless double-beta decay in the LRSM via the exchange of $W_L$, $W_R$ with 
$\nu_L$ and $\nu_R$ intermediate states. $m_D$ and $M_N$ are respectively the Dirac and Majorana masses of the neutrino.
Mass eigenstates are linear combinations of $\nu_L$ and $\nu_R$, with the light
state $\nu$ consisting mostly of $\nu_L$, and the heavy state $N$ consisting mostly of $\nu_R$.
The double-circle on the $W$ propagator in (d) indicates $W_L$-$W_R$ mixing.}
\label{doublebetafig}
\end{figure}
%%%%%%%%%%%%%%%%%%%%%%%%%%%%%%

%%%%%%%%%%%%%%%%%%%%%%%%%%%%%%
\begin{figure}[ht]
\centering
\includegraphics[width=6cm]{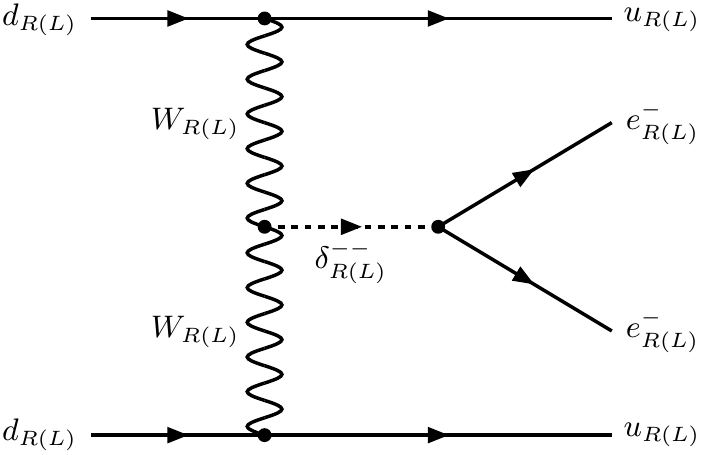}
\caption{Doubly-charged Higgs contributions to $0\nu\beta\beta$ in LRSM. }
\label{doublebetaHiggsfig}
\end{figure}
%%%%%%%%%%%%%%%%%%%%%%%%%%%%%%

Since the Majorana masses of the neutrinos violate lepton number by two units, 
their presence will lead to lepton number violating processes such as neutrinoless double-$\beta$ decay ($0\nu\beta\beta$):
\begin{equation}
(A,Z) \;\;\rightarrow\;\; (A,Z+2) \,+\, e^{-} +\, e^{-}\;.
\end{equation}
The rate of $0\nu\beta\beta$ can be written generically as
\begin{equation} 
\frac{\Gamma_{0\nu\beta\beta}}{\ln 2}
\;=\; G\,\frac{|\mathcal{M}|^2}{m_e^2}
\left|m_{\nu}^{\beta\beta}\right|^2
\;,
\label{doublebeta1}
\end{equation}
where $G$ denotes the kinematic factor, $\mathcal{M}$ is the nuclear matrix element, $m_e$ is the electron mass, and $|m_{\nu}^{\beta\beta}|$ is the effective neutrino mass: 
\begin{equation}
m_{\nu}^{\beta\beta}\;=\;\sum_i U^2_{ei}\,m_i^{\hphantom{2}}
.
\label{doublebeta2}
\end{equation}
Here, $U$ is the PMNS mixing matrix, and $m_i$ the mass of $i$-th mass eigenstate.

In the LRSM, $0\nu\beta\beta$ would occur via the $W_L$ and $W_R$ exchange processes 
with $\nu_L$ and $\nu_R$ intermediate states as shown in Fig.~\ref{doublebetafig}.
The dominant contributions are more likely to come from pure left-handed currents with light neutrino intermediate states, Fig.~\ref{doublebetafig}(a), and pure right-handed current with heavy neutrino intermediate states, Fig.~\ref{doublebetafig}(b). 
Contributions from left-right mixed currents, Figs.~\ref{doublebetafig}(c) and \ref{doublebetafig}(d), involve a suppressing factor due to the small left-right mixing ($m_D/M_N \sim 10^{-6}$ for the TeV scale LRSM with generic Yukawa couplings)
\cite{Hirsch:1996qw,Tello:2010am,Rodejohann:2011mu,Cirigliano:2004tc,Dev:2013vxa,Das:2012ii}. However, there is still room in the parameter space which allows significant contributions from the mixed diagrams, as discussed in \cite{Barry:2013xxa}.

In addition, there exist contributions to $0\nu\beta\beta$ 
from the doubly-charged Higgs mediated diagrams shown in
Fig.~\ref{doublebetaHiggsfig}. 
While those involving the left-handed currents and $\delta_L$ are suppressed by a factor of 
$p^2/M_{\delta}^2$, those involving the right-handed currents and $\delta_R$  are proportional to $M_{N}/M_{\delta}^2$, which may give significant contributions, depending on the masses in question.

The present limits on $0\nu\beta\beta$ are not in contradiction with the TeV scale LRSM. 
As stated in Ref.~\cite{Maiezza:2010ic}, the current signal on $0\nu\beta\beta$ can be accounted by, for example, $M_{W_2}=3$ TeV and $M_N=10$ GeV, where some fine-tuning is required but not in an unacceptable amount.

%%%%%%%%%%%%%%%%%%%%%%%%%%%%%%%%%%%%%%%%%%%%%%%%%%%%%%%%%%%%%%%%%%%%%%%%%%%%%%%%%%%%%%%%%%%%
%%%%%%%%%%%%%%%%%%%%%%%%%%%%%%%%%%%%%%%%%%%%%%%%%%%%%%%%%%%%%%%%%%%%%%%%%%%%%%%%%%%%%%%%%%%%
\subsection{Lepton Flavor Violating Processes}

%%%%%%%%%%%%%%%%%%%%%%%%%%%%%%
\begin{figure}[ht]
\centering
\includegraphics[width=5cm]{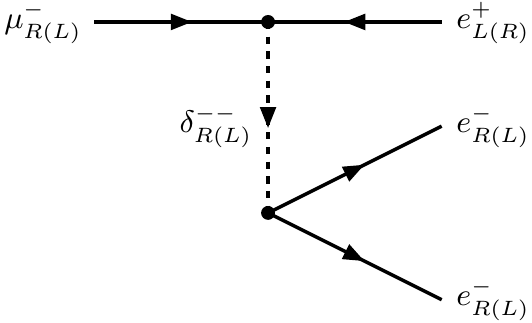}
\caption{Doubly-charged Higgs contribution to $\mu^-\rightarrow e^- e^- e^+$ in LRSM. }
\label{mu2eee}
\end{figure}
%%%%%%%%%%%%%%%%%%%%%%%%%%%%%%

The LRSM also allows for the lepton flavor violating (LFV) processes 
$\mu\rightarrow e\gamma$ \cite{Minkowski:1977sc,Adam:2013mnn,Dinh:2012bp}, $\mu\rightarrow 3e$ \cite{Dinh:2012bp,Bellgardt:1987du}, and $\mu\rightarrow e$ conversion in nuclei \cite{Dohmen:1993mp,Honecker:1996zf,Kaulard:1998rb,Bertl:2006up,Dinh:2012bp} 
though the predictions for these processes are dependent on how
much LFV is built into the Majorana interaction between the leptons and the
triplet scalars \cite{Cirigliano:2004tc,Cirigliano:2004mv,Tello:2010am,Das:2012ii,Dinh:2012bp}.
Current experimental bounds on the branching fractions are \cite{Agashe:2014kda}
\begin{eqnarray}
B(\mu\rightarrow e\gamma) & < &  5.7\times 10^{-13}\;,\cr
B(\mu\rightarrow eee) & < & 1.0\times 10^{-12}\;,\cr
\dfrac{\sigma(\mu^-\mathrm{Au}\rightarrow e^-\mathrm{Au})}{\sigma(\mu^-\mathrm{Au}\rightarrow\mathrm{capture})} & < & 7\times 10^{-13}\;,
\end{eqnarray}
all at 90\% C.L.

Among these processes, $\mu\rightarrow 3e$ is mediated by 
the doubly charged Higgses ($\delta^{\pm\pm}_{L, R}$) at tree level, Fig.~\ref{mu2eee}, and 
can be expected to have a relatively large branching fraction.
Since the Majorana couplings at the vertices of Fig.~\ref{mu2eee}
are proportional to the scale of the Majorana masses of the $N$'s, the diagram is roughly
proportional to $M_N^2/M_{\delta}^2$, and the branching fraction to $M_N^4/M_\delta^4$.
So the process provides a bound on $M_N/M_\delta$.
In the analysis of Ref.~\cite{Tello:2010am} 
the left-right symmetry operator of the model is taken to be charge conjugation $C$ \cite{Maiezza:2010ic}
which restricts the form of the Majorana interaction and allows one to constrain the 
right-handed PMNS matrix from the left-handed PMNS matrix, just as in the case of the
Yukawa interaction for quarks, providing some predictability.
Assuming $M_{W_2}=3.5\,\mathrm{TeV}$ and $M_N^{\max} = 0.5\,\mathrm{TeV}$, 
and taking all of the above LFV processes into account, Ref.~\cite{Tello:2010am}
reports $M_N^{\max}/M_\delta < 0.1$, which places a lower bound on $M_\delta$ of 5 TeV,
right in the ballpark of our $\Lambda_R=4\,\mathrm{TeV}$.

Belle, Babar, and LHCb have also searched for the LFV decays $\tau\rightarrow \ell\gamma$ \cite{Hayasaka:2007vc,Aubert:2009ag} and
$\tau\rightarrow\ell\ell\ell$ \cite{Miyazaki:2007zw,Lees:2010ez,Aaij:2013fia}, where $\ell=e$ or $\mu$and have placed bounds on the branching fractions to these decays at around a few times $10^{-8}$
at 90\% C.L..
According to the analysis of Ref.~\cite{Akeroyd:2006bb}, the LRSM 
with $W_2$, $H_1^{\pm}$, and $\delta_{L,R}^{\pm\pm}$ masses all set to 3 TeV
can accommodate a branching fraction of $\tau\rightarrow \ell\ell\ell$
as large as $10^{-9}$ which would be accessible at the
next generation of super $B$ factories \cite{Akeroyd:2004mj,Bona:2007qt}.

%%%%%%%%%%%%%%%%%%%%%%%%%%%%%%%%%%%%%%%%%%%%%%%%%%%%%%%%%%%%%%%%%%%%%%%%%%%%%%%%%%%%%%%%%%%%
\subsection{Phenomenological Outlook}

Based on these conventional phenomenological analyses, we conclude that the TeV scale LRSM 
predicted by the $su(2/2)$ superconnection formalism, possibly with an underlying NCG, 
provides a wealth of new particles and predictions within reach of LHC and other experiments.
The fact that the current experimental bounds on the LRSM and the corresponding predictions of the superconnection formalism are suspiciously close may be a sign that LHC is on the brink of discovering something new and exciting.

With the center of mass energy of $\sqrt{s}=13\,\mathrm{TeV}$ for its second run, the LHC is well capable of observing the new particles of the model among which the most important are the right-handed gauge bosons ($W_R^{\pm}$, $Z'$) whose masses are fixed by the formalism and range within the TeV scale. 
With the scale of 4~TeV, selected by the formalism itself, these masses will be within reach of the LHC, provided that the right handed neutrinos ($N_R$) are light enough to make the corresponding channels accessible.\footnote{There is nothing in the model which constraints the right-handed neutrinos $N_R$ to be light.  With $N_R$ heavier than $W_R^{\pm}$, the Drell-Yan interactions will be highly suppressed and thus, although theoretically the TeV scale LRSM could still be viable, it will be  very difficult for the LHC to detect its signature through the $W_R^{\pm}$ channels.} 

A number of relevant and important observations could be delivered in the LFV branch as well, especially in $\mu\rightarrow e$ conversion in nuclei, which we briefly discussed in an earlier section. 
With the next generation of machines, COMET \cite{Cui:2009zz} and Mu2e \cite{Abrams:2012er} collaborations target to increase their sensitivity for this process from $10^{-13}$ to $10^{-17}$, which will significantly improve the limits on new physics including LRSM.  
Moreover, the next generation of super $B$ factories aim to increase the limit on LFV $\tau$ decays to a level of $10^{-9}$ \cite{Akeroyd:2004mj,Bona:2007qt}, which will also provide useful information on the nature of new physics.

Thus, the early disappointments of the LHC (lack of other discoveries except for the discovery of the Higgs)
could have been nothing but the calm before the storm.

%%%%%%%%%%%%%%%%%%%%%%%%%%%%%%%%%%%%%%%%%%%%%%%%%%%%%%%%%%%%%%%%%%%%%%%%%%%%%%%%%%%%%%%%%%%%
%%%%%%%%%%%%%%%%%%%%%%%%%%%%%%%%%%%%%%%%%%%%%%%%%%%%%%%%%%%%%%%%%%%%%%%%%%%%%%%%%%%%%%%%%%%%
%%%%%%%%%%%%%%%%%%%%%%%%%%%%%%%%%%%%%%%%%%%%%%%%%%%%%%%%%%%%%%%%%%%%%%%%%%%%%%%%%%%%%%%%%%%%
\section{Summary, Questions, and Speculations}\label{Summary}

%%%%%%%%%%%%%%%%%%%%%%%%%%%%%%%%%%%%%%%%%%%%%%%%%%%%%%%%%%%%%%%%%%%%%%%%%%%%%%%%%%%%%%%%%%%%
\subsection{Strengths and Weaknesses of the NCG-Superconnection Approach}

In this paper, we have reviewed the $su(2/1)$ superconnection approach to the SM of Ne'eman et al. \cite{Neeman:1979wp,Ne'eman:1990nr,Hwang:1995wk}, supplemented by later developments by Coquereaux et al. \cite{Coquereaux:1990ev} and Haussling et al. \cite{Haussling:1991ns}.
The superconnection describes the connection in a model space in which our $3+1$ dimensional spacetime is extended by a discrete extra dimension consisting of only two points, \textit{i.e.}
the model space consists of two $3+1$ dimensional branes separated by a gap.
The left-handed fermions are assumed to inhabit one brane, and the right-handed fermions the other.
The even part of the superconnection describes the usual gauge connection within each $3+1$ dimensional brane, while the odd part of the superconnection, identified with the Higgs doublet, describes the connection in the discrete direction bridging the gap between the two branes.
Contrary to early misconceptions about the approach, the Higgs doublet enters as a bosonic scalar, and does not violate the spin-statistics theorem.

The $su(2/1)$ superconnection model predicts $\sin^2\theta_W=1/4$, a condition which can only be imposed on the $SU(2)_L\times U(1)_Y$ gauge couplings at $\sim 4$ TeV.
We interpret this to mean that the SM emerges from the underlying discrete extra dimension theory at that scale.  The model also predicts the Higgs mass, which, including RGE running down from 4~TeV, is $\sim 170$~GeV.

To remedy this problem, we extended the model to $su(2/2)$, in which the $SU(2)_L\times SU(2)_R\times U(1)_{B-L}$ gauge bosons and a bi-doublet field $\Phi$ were embedded into the superconnection.
This extended the SM to the LRSM, for which the emergence scale also turned out to be 4~TeV.
In this case, additional gauge triplet scalars were introduced to break $SU(2)_R\times U(1)_{B-L}$
down to $U(1)_Y$, and the scale of this breaking was also required to be at 4~TeV.
The bi-doublet field then broke the $SU(2)_L\times U(1)_Y$ symmetry down to $U(1)_{em}$.
It was shown that the lightest neutral scalar in the Higgs sector could have mass as light as $\sim 126$ GeV.
The model also predicts a plethora of new particles with masses in the multi-TeV range, within reach of the
LHC run 2.

Of the several salient features of the approach, the most interesting is that the
generalized exterior derivative in the discrete direction, \textit{i.e.} the matrix derivative,
shifts the VEV of the off-diagonal 0-forms to non-zero values, effectively breaking the gauge
symmetries.
Thus, gauge symmetry breaking is intimately connected to the geometry of the model spacetime,
in particular, to the separation of the two branes.
We envision a scenario in which the two branes, originally overlapping, separate from each other
dynamically and trigger gauge symmetry breaking.
In other words, the Higgs mechanism is not due to the Higgs dynamics which is independent
of any underlying geometry, but an integral part of the geometry itself, and is quantum 
gravitational in character.

Many problems still remain for the formalism to mature into a full fledged model building paradigm.
First, the Lie superalgebra structure is assumed to emerge from some underlying NCG theory, but we
have not clarified how the geometry enforces the structure yet.
We would also like to incorporate QCD, quarks, and fermion generations into the structure.
The Spectral SM of Connes et al. supposedly has already done this, but as commented on earlier,
the Spectral SM approach does not have much predictive power.
Also, after the incorporation of QCD into the model, we would like to unify it with the LRSM
via the Pati-Salam group.  The $U(1)_{B-L}$ gauge boson being part of the $su(2/2)$ superconnection,
this suggests that QCD cannot be simply tacked on to the model.

There is also the subtle problem of how the nilpotency of the matrix derivative should be treated.
In the $su(2/1)$ case, the term $\mathbf{d}_M^2$ in the definition of the supercurvature could not be ignored 
for gauge invariance, but including it led to internal inconsistencies.
In the $su(2/2)$ case, $\mathbf{d}_M^2$ belongs to the center of the superalgebra, so
it can be added or subtracted from the super curvature without changing its algebraic properties.
This suggests that one can decide to ignore $\mathbf{d}_M^2$ based on consistency requirements,
but one cannot shake the impression that the treatment is ad hoc.
Furthermore, in the $su(2/2)$ case, phenomenological requirements
demanded that the nilpotency of the matrix derivative be broken.
Whether this is another indication of a deep connection between
spontaneous symmetry breaking and the geometry of the underlying theory
remains to be seen.

These, and other questions will be addressed in future publications.

%%%%%%%%%%%%%%%%%%%%%%%%%%%%%%%%%%%%%%%%%%%%%%%%%%%%%%%%%%%%%%%%%%%%%%%%%%%%%%%%%%%%%%%%%%%%
\subsection{Comment on the Hierarchy Problem and the Unification of Couplings}

The major appeal of the
more traditional approaches to BSM model building such as supersymmetry (as well as
technicolor, extra dimensions, etc.)  
is that they address the hierarchy problem, 
and that they shed light on the apparent unification of couplings, 
both within the context of local effective field theory (EFT).

However, this apparent theoretical appeal of supersymmetry 
does not exclude approaches that do not necessarily follow the local EFT paradigm. 
For example, in the Spectral SM 
approach of Connes et al. \cite{Connes:1990qp,Chamseddine:1991qh,Connes:1994yd,Chamseddine:1996zu,Connes:2006qv,Chamseddine:2007hz,Chamseddine:2010ud,Chamseddine:2012sw,Chamseddine:2013rta}
the hierarchy problem can be addressed in a completely different fashion \cite{Chamseddine:2010ud}. 
The crucial NCG (and thus in some sense non-local) aspect of the SM is found in the Higgs sector, which in principle comes with an extra (second) scale, to be distinguished from the usual UV scale of local EFT. 
The hierarchy between the Higgs and the UV (Planck) scale can be associated (as shown by Chamseddine and Connes in Ref.~\cite{Chamseddine:2010ud}) with the natural exponential factor that comes from the dynamical discrete geometry of the Higgs sector. 
Similarly, the apparent gauge unification (in the guise of an effective $SO(10)$ relation between the gauge couplings) is also incorporated into the Spectral SM. 
These aspects of the NCG approach to the SM are almost completely unknown in the particle physics community, and at the moment, almost completely undeveloped from a phenomenological viewpoint.

One of our aims in our upcoming review of the Spectral SM \cite{AMST2} is to clarify these interesting features of the NCG approach to the SM and make them palatable to the wider phenomenological community. We are also motivated by a deeper need to understand the limitations of the local EFT paradigm from the point of view of the physics of quantum gravity, which is usually, rather naively, ignored at the currently interesting particle physics scales, by invoking the concept of decoupling, which represents another central feature of the local EFT and which is also challenged by the NCG approach to the SM. Finally, as we discuss in the next concluding subsection of this paper, the usual RG analysis of the local EFT should be re-examined in the new light of the non-commutative/non-local structure of the SM, and the apparent existence of two natural (and naturally related) physics scales.

%%%%%%%%%%%%%%%%%%%%%%%%%%%%%%%%%%%%%%%%%%%%%%%%%%%%%%%%%%%%%%%%%%%%%%%%%%%%%%%%%%%%%%%%%%%%
\subsection{The violation of decoupling and the possibility for UV/IR mixing}

In this concluding subsection we would like to comment on the observation made in Ref.~\cite{nondec} regarding the violation of decoupling in the Higgs sector, and how this 
violation %of decoupling in the Higgs sector 
may point to the more fundamental possibility of mixing of UV and IR degrees of freedom, 
given our view that a NCG underlies the Higgs sector. 
Such UV/IR mixing is known in the simpler context of non-commutative field theory \cite{uvir}, which we review below.

First, let us briefly recall the argument made in Ref.~\cite{nondec}:
Essentially Senjanovic and Sokorac found within the LRSM
that the Higgs scalars do not decouple at low energy due to the essential relation between the gauge couplings
and the Higgs mass. Note that this violation of decoupling will affect the scales of the electroweak breaking (taken as
the low energy scale) and the TeV scale (taken as the high energy scale of new physics).

Such a violation of decoupling might point to a more fundamental phenomenon of the UV/IR mixing of the short and long distance physics.
Here we briefly recall in slightly more detail the UV/IR mixing found in non-commutative field theory \cite{uvir}.
In this particular toy-model case (to be distinguished from NCG of Connes relevant for
our discussion) the non-commutative spatial coordinates are assumed to satisfy
\begin{equation}
[\,x_a,\, x_b\,] \;=\; i \theta_{ab}\;,
\end{equation}
where $\theta_{ab}$ is real and antisymmetric.
Note that when this relation is taken together with the fundamental commutation relation
\begin{equation}
[\,x_a,\, p_b\,] \;=\; i \delta_{ab}\;,
\end{equation}
they imply the possibility of a fundamentally new effect: UV/IR mixing, i.e.
\begin{equation}
\delta x_a \;\sim\; \theta_{ab}\, \delta p_b \;.
\end{equation}
This would mean, contrary to the usual intuition from local effective field theory,
that high energy processes are related to low energy distances. 
We have argued elsewhere that the UV/IR mixing
should be a fundamental feature of quantum gravity and string theory \cite{Freidel:2013zga,Freidel:2014qna}.

At the moment we are not aware of an explicit UV/IR relation in the context of the NCG of Connes that underlies the superconnection formalism and the new viewpoint on the
SM and the physics beyond it, as advocated in this paper.
However, there exists a very specific toy model of non-commutative field theory in which
such UV/IR mixing has been explicitly demonstrated. The nice feature of this toy model is that
it can be realized in a fundamental short distance theory, such as string theory \cite{uvir}.

The non-commutative field theory is defined by the
effective action
\begin{equation}
S_{nc} \;=\; \int d^4 x\; L[\phi]
\end{equation}
where the product of the fields $\phi$ is given by the Moyal (or star) product
\begin{equation}
(\phi_1 \star \phi_2)(x) \;\equiv\; 
\exp\left[\frac{i}{2} \theta^{ab} \partial_a^{y} \partial_b^z\right] \phi_1(y) \phi_2(z)
\Big|_{y=z=x}
\;.
\end{equation}
In what follows, motivated by the form of the Higgs Lagrangian, we take $L[\phi]$ to describe the massive
$\lambda \phi^4$ non-commutative field theory.

The main point made in Ref.~\cite{uvir} is that in the simplest case of the $\phi^{4}$ theory the
1PI two point function has the following non-trivial leading form (up to an overall coefficient) 
\begin{equation}
\Lambda_{\mathrm{eff}}^2 - m^2 \log\left(\frac{\Lambda_{\mathrm{eff}}^2}{m^2}\right)
\;,
\end{equation}
where $m$ is the mass of the $\phi$ field, and the effective cut-off $\Lambda_{\mathrm{eff}}$ is given by the following expression:
\begin{equation}
\Lambda_{\mathrm{eff}}^2 \;=\; \left(\frac{1}{\Lambda^2} - p_a \theta^2_{ab} p_b\right)^{-1}\;.
\end{equation}
Here, $\Lambda$ is the usual UV cut-off.
Note that the non-commutativity scale $\theta$ plays the role of the natural IR cut-off.

The UV/IR mixing, characteristic of this type of non-commutative field theory leads to the question of the existence of the proper continuum limit for non-commutative field theory. 
This question can be examined from the point of view of non-perturbative Renormalization Group (RG). 
The proper Wilsonian analysis of this type of non-commutative theory has been done in Ref.~\cite{Grosse:2004yu}. 
The UV/IR mixing leads to a new kind of the RG flow: a double RG flow, in
which one flows from the UV to IR and the IR to the UV and ends up, generically, at a self-dual fixed point. 
It would be tantalizing if the NCG set-up associated with the SM, and in particular, the LRSM generalization discussed in this paper, would lead to the phenomenon of the UV/IR mixing and the double RG flow with a self-dual fixed point. 
Finally, we remark that it has been argued in a recent work on quantum gravity and string theory that such UV/IR mixing and the double RG might be a generic feature of quantum gravity coupled to matter \cite{Freidel:2013zga,Freidel:2014qna}.

Even though the NCG in our case is different from this toy example, the
lesson in the essential physics of the UV/IR mixing is present in our situation as well:
the Higgs field can be associated with the natural scale of non-commutativity and thus natural IR scale, and therefore
even in our situation we might reasonably expect that the the Higgs scale is mixed with the
UV cut-off defined by some more fundamental theory.
Needless to say, at the moment this is only an exciting conjecture.

If this conjecture is true, given the results presented in this paper one could expect that the appearance of the LRSM degrees of freedom (as well as the embedded SM
degrees of freedom) at low energy is essentially a direct manifestation of some effective UV/IR mixing, and thus that on one hand the remnants of the UV physics can
be expected at a low energy scale of 4 TeV, and conversely that the LRSM structure point to some unique features of the high energy physics of quantum gravity.
In this context we recall the observations made in Ref.~\cite{dienes} about the special nature of the Pati-Salam model, which unifies the LRSM with QCD, in certain constructions of
string vacua. Even though this observation is mainly based on ``groupology'' and it is not deeply understood, this observation might
be indicative that the Pati-Salam model is the natural completion of the SM, as suggested in this paper, in which the infrared physics associated with the Higgs sector is mixed with the ultraviolet physics of some more fundamental physics, such as string theory.

%%%%%%%%%%%%%%%%%%%%%%%%%%%%%%%%%%%%%%%%%%%%%%%%%%%%%%%%%%%%%%%%%%%%%%%%%%%%%%%%%%%%%%%%%%%%
%%%%%%%%%%%%%%%%%%%%%%%%%%%%%%%%%%%%%%%%%%%%%%%%%%%%%%%%%%%%%%%%%%%%%%%%%%%%%%%%%%%%%%%%%%%%
\acknowledgments

We would like to thank Marvin Blecher, Lay Nam Chang, David Fairlie, Laurant Freidel, Murat Gunaydin, Rob Leigh, and Al Shapere 
for stimulating conversations.
We also thank the Miami Winter conference for providing a stimulating environment for work.
DM was supported in part by the U.S. Department of Energy, grant DE-FG02-13ER41917, task A. 
TT is grateful for the hospitality of the Kavli-IPMU during his
sabbatical year from fall 2012 to summer 2013 
where the initial stages of this work was performed, and 
where he was supported by the World Premier International
Research Center Initiative (WPI Initiative), MEXT, Japan.

%%%%%%%%%%%%%%%%%%%%%%%%%%%%%%%%%%%%%%%%%%%%%%%%%%%%%%%%%%%%%%%%%%%%%%%%%%%%%%%%%%%%%%%%%%%%
%%%%%%%%%%%%%%%%%%%%%%%%%%%%%%%%%%%%%%%%%%%%%%%%%%%%%%%%%%%%%%%%%%%%%%%%%%%%%%%%%%%%%%%%%%%%
%%%%%%%%%%%%%%%%%%%%%%%%%%%%%%%%%%%%%%%%%%%%%%%%%%%%%%%%%%%%%%%%%%%%%%%%%%%%%%%%%%%%%%%%%%%%
%%%%%%%%%%%%%%%%%%%%%%%%%%%%%%%%%%%%%%%%%%%%%%%%%%%%%%%%%%%%%%%%%%%%%%%%%%%%%%%%%%%%%%%%%%%%
%%%%%%%%%%%%%%%%%%%%%%%%%%%%%%%%%%%%%%%%%%%%%%%%%%%%%%%%%%%%%%%%%%%%%%%%%%%%%%%%%%%%%%%%%%%%
%%%%%%%%%%%%%%%%%%%%%%%%%%%%%%%%%%%%%%%%%%%%%%%%%%%%%%%%%%%%%%%%%%%%%%%%%%%%%%%%%%%%%%%%%%%%
\appendix

%%%%%%%%%%%%%%%%%%%%%%%%%%%%%%%%%%%%%%%%%%%%%%%%%%%%%%%%%%%%%%%%%%%%%%%%%%%%%%%%%%%%%%%%%%%%
%%%%%%%%%%%%%%%%%%%%%%%%%%%%%%%%%%%%%%%%%%%%%%%%%%%%%%%%%%%%%%%%%%%%%%%%%%%%%%%%%%%%%%%%%%%%
\section{The Ne'eman-Sternberg Rule for Supermatrix Multiplication}
\label{SuperMatrixMultiplication}

As stated in footnote~\ref{MSmultiplication}, 
some papers in the literature treat the superconnection $\mathcal{J}$
as a super-endomorphism of a superspace and calculate the supercurvature $\mathcal{F}$
differently.  
In this appendix, we derive the multiplication rule for super-endomorphisms 
with differential forms as elements
(or super-endomorphism valued differential forms).
We will follow the notation of Sternberg \cite{Sternberg:2012} in which
the $\mathbb{Z}_2$ gradings with be denoted with $\pm$ superscripts instead of $0$, $1$ subscripts.

A superspace $E$ is simply a vector space with a $\mathbb{Z}_2$ grading:
\begin{equation}
E \;=\; E^+ \oplus E^- \;.
\end{equation}
We denote the set of all endomorphisms, \textit{i.e.} linear transformations, on $E$ 
with $\mathrm{End}(E)$, which is already an associative algebra.
In matrix representation, the product of $\mathrm{End}(E)$ is just standard matrix multiplication.

When $E$ is provided with a $\mathbb{Z}_2$ grading as above, 
a $\mathbb{Z}_2$ grading can also be introduced into $\mathrm{End}(E)$ 
by simply letting $\mathrm{End}(E)^+$ consist of all endomorphisms that
map $E^{\pm}$ to $E^{\pm}$, and $\mathrm{End}(E)^-$ consist of all endomorphisms that
map $E^{\pm}$ to $E^{\mp}$. That is:
\begin{eqnarray}
\mathrm{End}(E)^+ & : & E^+\rightarrow\,E^+,\;\; E^-\rightarrow\, E^-\;, \cr
\mathrm{End}(E)^- & : & E^+\rightarrow\,E^-,\;\; E^-\rightarrow\, E^+\;.
\end{eqnarray}
In matrix representation, elements of $\mathrm{End}(E)^+$ will consist of matrices of the
form
\begin{equation}
\begin{bmatrix} A & 0 \\ 0 & B \end{bmatrix}
,\quad
A\,:\, E^+\rightarrow\,E^+,\quad
B\,:\, E^-\rightarrow\,E^-,
\end{equation}
while elements of $\mathrm{End}(E)^-$ will consist of those of the form
\begin{equation}
\begin{bmatrix} 0 & C \\ D & 0 \end{bmatrix}
,\quad
C\,:\, E^-\rightarrow\,E^+,\quad
D\,:\, E^+\rightarrow\,E^-.
\end{equation}
Then, clearly
\begin{equation}
\mathrm{End}(E) \;=\; \mathrm{End}(E)^+ \oplus \mathrm{End}(E)^- \;,
\end{equation}
and $\mathrm{End}(E)$ can be viewed as a superalgebra, its product satisfying
Eq.~(\ref{SAproductpm}).
Note that the product of the superalgebra here is just the product of the associative
algebra from which it was derived, \textit{i.e.} standard matrix multiplication.
So far, nothing has changed by viewing $\mathrm{End}(E)$ as a superalgebra.

The multiplication rule changes when the superalgebra 
$\mathrm{End}(E)=\mathrm{End}(E)^+\oplus\mathrm{End}(E)^-$ is tensored with the
commutative superalgebra of differential forms $\Omega(M)=\Omega^+(M)\oplus \Omega^-(M)$,
yielding a superalgebra of super-endomorphism valued differential forms, or
super-endormorphisms with differential forms as elements.
The rule depends slightly on whether we view super-endomorphism valued differential forms
as elements of the tensor product $\mathrm{End}(E)\otimes \Omega(M)$, or the
tensor product $\Omega(M)\otimes \mathrm{End}(E)$, since this
affects the sign in the definition of the product
for tensored superalgebras, Eq.~(\ref{TensoredSAproduct}).

For elements of $\mathrm{End}(E)\otimes \Omega(M)$, 
we have the Ne'eman-Sternberg multiplication rule
given in Refs.~\cite{Hwang:1995wk,Ne'eman:1990nr,Sternberg:2012,Coquereaux:1990ev} as:
%
%%%
%%%
%%%
\begin{widetext}
\begin{align}
	\begin{bmatrix}
	A& C \\ 
	D& B
	\end{bmatrix} 
\odot
	\begin{bmatrix}
	A' & C'\\
	D' & B'
	\end{bmatrix}
:= 
	\begin{bmatrix}
	A \wedge A' +(-1)^{| C|} C\wedge D' &
	C \wedge B' + (-1)^{| A|} A \wedge C'\\
	D\wedge A' +(-1)^{| B|} B\wedge D'&
	B \wedge B'+(-1)^{| D|} D \wedge C'
	\end{bmatrix},
\end{align}
where $A$, $B$, $C$, $D$, $A'$, $B'$, $C'$, and $D'$ are all 
matrices themselves whose elements are differential forms of a definite grading.
For the elements of $\Omega(M)\otimes \mathrm{End}(E)$, we have
\begin{align}
	\begin{bmatrix}
	A& C \\ 
	D& B
	\end{bmatrix} 
\odot
	\begin{bmatrix}
	A' & C'\\
	D' & B'
	\end{bmatrix}
:= 
	\begin{bmatrix}
	A \wedge A' +(-1)^{| D'|} C\wedge D' &
	A \wedge C' + (-1)^{| B'|} C \wedge B'  \\
	(-1)^{|A'|} D\wedge A' + B\wedge D'& 
	(-1)^{| C'|} D \wedge C' + B \wedge B'
	\end{bmatrix}.
\label{eq:neeman_sternberg_mult_rule}
\end{align}
\end{widetext}
%%%
%%%
%%%
%

These relations are simple consequences of Eq.~(\ref{TensoredSAproduct}).
First, rewrite each matrix in tensor product form, schematically, as
\begin{eqnarray}
\lefteqn{
	\begin{bmatrix}
	A & C \\
	D & B
	\end{bmatrix}
}\cr
& = & 
A \otimes 
	\begin{bmatrix}
	1&0\\
	0&0
	\end{bmatrix}
+
B \otimes
	\begin{bmatrix}
	0&0\\
	0&1
	\end{bmatrix}
+
C \otimes
	\begin{bmatrix}
	0&1\\
	0&0
	\end{bmatrix}
+
D \otimes
	\begin{bmatrix}
	0&0\\
	1&0
	\end{bmatrix}
,
%%%
\cr
\lefteqn{
	\begin{bmatrix}
	A' & C' \\
	D' & B'
	\end{bmatrix}
}\cr
& = & 
\!
A' \otimes 
	\begin{bmatrix}
	1&0\\
	0&0
	\end{bmatrix}
\!+
B' \otimes
	\begin{bmatrix}
	0&0\\
	0&1
	\end{bmatrix}
\!+
C' \otimes
	\begin{bmatrix}
	0&1\\
	0&0
	\end{bmatrix}
\!+
D' \otimes
	\begin{bmatrix}
	0&0\\
	1&0
	\end{bmatrix}
.
\cr & &
\end{eqnarray}
Note that we are using tensor products in $\Omega(M)\otimes \mathrm{End}(E)$
with the differential form on the left and the supermatrix on the right.
Then, we multiply the tensor products together, term by term.
For instance,
\begin{eqnarray}
\lefteqn{\left(
A\otimes 
	\begin{bmatrix}
	1& 0\\
	0& 0
	\end{bmatrix}
\right)
\odot 
\left(
A' \otimes
	\begin{bmatrix}
	1& 0\\
	0& 0
	\end{bmatrix}
\right)
} 
\cr
& = &
(-1)^{0 \times |A'|}
\left(A\wedge A'\right)
\otimes
\left(
	\begin{bmatrix}
	1 & 0\\
	0 & 0
	\end{bmatrix}
	\cdot
	\begin{bmatrix}
	1 & 0\\
	0 & 0
	\end{bmatrix}
\right)
\cr
& = & \left(A\wedge A'\right)
\otimes
	\begin{bmatrix}
	1 & 0\\
	0 & 0
	\end{bmatrix}
\cr
& = &
\begin{bmatrix}
A\wedge A' & 0 \\ 0 & 0 
\end{bmatrix}
,	
\end{eqnarray}
and
\begin{eqnarray}
\lefteqn{
\left(
C\otimes 
	\begin{bmatrix}
	0& 1\\
	0& 0
	\end{bmatrix}
\right)
\odot 
\left(
D' \otimes
	\begin{bmatrix}
	0& 0\\
	1& 0
	\end{bmatrix}
\right)
}
\cr
& = &
(-1)^{1 \times |D'|}
\left(C\wedge D'\right)
\otimes
\left(
	\begin{bmatrix}
	0 & 1\\
	0 & 0
	\end{bmatrix}
	\cdot
	\begin{bmatrix}
	0 & 0\\
	1 & 0
	\end{bmatrix}
\right)
\cr
& = &
(-1)^{|D'|} 
\left( C\wedge D' \right)
\otimes
	\begin{bmatrix}
	1 & 0\\
	0 & 0
	\end{bmatrix}
\cr
& = &
\begin{bmatrix}
(-1)^{|D'|}C\wedge D' & 0 \\ 0 & 0 
\end{bmatrix}
.
\end{eqnarray}
Summing, we obtain the 1-1 element of Eq.~(\ref{eq:neeman_sternberg_mult_rule}).
All other elements can be derived in a similar fashion.

%%%%%%%%%%%%%%%%%%%%%%%%%%%%%%%%%%%%%%%%%%%%%%%%%%%%%%%%%%%%%%%%%%%%%%%%%%%%%%%%%%%%%%%%%%%%
%%%%%%%%%%%%%%%%%%%%%%%%%%%%%%%%%%%%%%%%%%%%%%%%%%%%%%%%%%%%%%%%%%%%%%%%%%%%%%%%%%%%%%%%%%%%
\section{Useful Identities}\label{GunionComparison}

%
%%%
%%%
%%%
\begin{widetext}
To compare Eq.~(2.9) of Ref.~\cite{Gunion:1989in} and Eq.~(A2) of Ref.~\cite{Deshpande:1990ip}
we need the following identities:
\begin{eqnarray}
\Tr\Bigl[\bigl(\Phi^\dagger\Phi\bigr)^2\Bigr]
& = & \left(\Tr\Bigl[\Phi^\dagger\Phi\Bigr]\right)^2
- \dfrac{1}{2}\Tr\Bigl[\Phi^\dagger\widetilde{\Phi}\Bigr]\Tr\Bigl[\widetilde{\Phi}^\dagger\Phi\Bigr]
\;,\cr
%%%
\dfrac{1}{2}
\left(
\Tr\Bigl[\Phi^\dagger\widetilde{\Phi}\Bigr]+\Tr\Bigl[\widetilde{\Phi}^\dagger\Phi\Bigr]
\right)^2
& = & 
\dfrac{1}{2}\left\{
  \left(\Tr\Bigl[\Phi^\dagger\widetilde{\Phi}\Bigr]\right)^2
+ \left(\Tr\Bigl[\widetilde{\Phi}^\dagger\Phi\Bigr]\right)^2
\right\} 
+ \Tr\Bigl[\Phi^\dagger\widetilde{\Phi}\Bigr]\Tr\Bigl[\widetilde{\Phi}^\dagger\Phi\Bigr]
\;,\cr
%%%
\dfrac{1}{2}
\left(
\Tr\Bigl[\Phi^\dagger\widetilde{\Phi}\Bigr]-\Tr\Bigl[\widetilde{\Phi}^\dagger\Phi\Bigr]
\right)^2
& = & 
\dfrac{1}{2}\left\{
  \left(\Tr\Bigl[\Phi^\dagger\widetilde{\Phi}\Bigr]\right)^2
+ \left(\Tr\Bigl[\widetilde{\Phi}^\dagger\Phi\Bigr]\right)^2
\right\} 
- \Tr\Bigl[\Phi^\dagger\widetilde{\Phi}\Bigr]\Tr\Bigl[\widetilde{\Phi}^\dagger\Phi\Bigr]
\;,\cr
%%%
\Tr\Bigl[\Phi^\dagger\Phi\widetilde{\Phi}^\dagger\widetilde{\Phi}\Bigr]
& = & 
\dfrac{1}{2}\,
\Tr\Bigl[\Phi^\dagger\widetilde{\Phi}\Bigr]\Tr\Bigl[\widetilde{\Phi}^\dagger\Phi\Bigr]
\;,\cr
%%%
\Tr\Bigl[
\Phi^\dagger\widetilde{\Phi}\Phi^\dagger\widetilde{\Phi}
\Bigr]
+
\Tr\Bigl[
\widetilde{\Phi}^\dagger\Phi\widetilde{\Phi}^\dagger\Phi
\Bigr]
& = & \dfrac{1}{2}
\left\{
  \left(\Tr\Bigl[\Phi^\dagger\widetilde{\Phi}\Bigr]\right)^2
+ \left(\Tr\Bigl[\widetilde{\Phi}^\dagger\Phi\Bigr]\right)^2
\right\}
\;,
\cr
%%%
\Tr\Bigl[
\bigl(\Delta_L^\dagger\Delta_L^{\phantom{\dagger}}\bigr)^2
\Bigr]
& = & \left(\Tr\Bigl[\Delta_L^\dagger\Delta_L^{\phantom{\dagger}}\Bigr]\right)^2
-\dfrac{1}{2}\,
\Tr\Bigl[\Delta_L^{\phantom{\dagger}}\Delta_L^{\phantom{\dagger}}\Bigr]
\Tr\Bigl[\Delta_L^\dagger\Delta_L^\dagger\Bigr]
\;,
\cr
%%%
\Tr\Bigl[\widetilde{\Phi}\widetilde{\Phi}^\dagger\Delta_L^{\phantom{\dagger}}\Delta_L^{\dagger}\Bigr]
& = & \Tr\Bigl[\Phi^\dagger\Phi\Bigr] \Tr\Bigl[\Delta_L^{\dagger}\Delta_L^{\phantom{\dagger}}\Bigr]
-\Tr\Bigl[
\Phi^\dagger\Phi\Delta_L^{\dagger}\Delta_L^{\phantom{\dagger}}
\Bigr]
\;,
\cr
%%%
\Tr\Bigl[\widetilde{\Phi}^\dagger\widetilde{\Phi}\Delta_R^{\phantom{\dagger}}\Delta_R^{\dagger}\Bigr]
& = & \Tr\Bigl[\Phi\Phi^\dagger\Bigr] \Tr\Bigl[\Delta_R^{\dagger}\Delta_R^{\phantom{\dagger}}\Bigr]
-\Tr\Bigl[
\Phi\Phi^\dagger\Delta_R^{\dagger}\Delta_R^{\phantom{\dagger}}
\Bigr]
\;.
\end{eqnarray}
\end{widetext}
%%%
%%%
%%%

%%%%%%%%%%%%%%%%%%%%%%%%%%%%%%%%%%%%%%%%%%%%%%%%%%%%%%%%%%%%%%%%%%%%%%%%%%%%%%%%%%%%%%%%%%%%
%%%%%%%%%%%%%%%%%%%%%%%%%%%%%%%%%%%%%%%%%%%%%%%%%%%%%%%%%%%%%%%%%%%%%%%%%%%%%%%%%%%%%%%%%%%%

\end{document}